\documentclass[fleqn,usenatbib]{mnras}
\usepackage{newtxtext,newtxmath}

\usepackage[T1]{fontenc}
\usepackage{ae,aecompl}
\usepackage{graphicx}	
\usepackage{amsmath}	
\usepackage{amssymb}	
\usepackage{siunitx}
\usepackage{tabularx}
\usepackage{subfigure}
\usepackage{microtype}
\usepackage{multirow}
\usepackage[percent]{overpic}
\usepackage{color}
\DeclareSIUnit \parsec {pc}
\DeclareSIUnit \solarmass {\mbox{M$_{\sun}$}}
\DeclareSIUnit \au {au}
\DeclareSIUnit \year {yr}
\DeclareSIUnit \jansky {Jy}

\title[Modelling SEDs of candidate FHSCs]{What can the SEDs of first hydrostatic core candidates reveal about their nature?}

\author[A. K. Young et al.]{
Alison K. Young,\thanks{E-mail: ayoung@astro.ex.ac.uk (AKY)}
Matthew R. Bate,
Chris F. Mowat,
\newauthor Jennifer Hatchell and
Tim J. Harries
\\
School of Physics and Astronomy, University of Exeter, Stocker Road EX4 4QL, UK
}

\date{Accepted XXX. Received YYY; in original form ZZZ}

\pubyear{2017}

\begin{document}
\label{firstpage}
\pagerange{\pageref{firstpage}--\pageref{lastpage}}
\maketitle

\begin{abstract}
The first hydrostatic core (FHSC) is the first stable object to form in simulations of star formation. This stage has yet to be observed definitively, although several candidate FHSCs have been reported. We have produced synthetic spectral energy distributions (SEDs) from 3D hydrodynamical simulations of pre-stellar cores undergoing gravitational collapse for a variety of initial conditions. Variations in the initial rotation rate, radius and mass lead to differences in the location of the SED peak and far-infrared flux. Secondly, we attempt to fit the SEDs of five FHSC candidates from the literature and five newly identified FHSC candidates located in the Serpens South molecular cloud with simulated SEDs. The most promising FHSC candidates are fitted by a limited number of model SEDs with consistent properties, which suggests the SED can be useful for placing constraints on the age and rotation rate of the source. The sources we consider most likely to be in FHSC phase are B1-bN, CB17-MMS, Aqu-MM1 and Serpens South candidate K242. We were unable to fit SerpS-MM22, Per-Bolo 58 and Chamaeleon-MMS1 with reasonable parameters, which indicates that they are likely to be more evolved. 
   
\end{abstract}

\begin{keywords}
infrared: ISM - hydrodynamics - methods: numerical - radiative transfer  - Stars: formation - submillimetre: ISM 
\end{keywords}



\section{Introduction}
The theory of how stars form via the gravitational collapse of a dense molecular cloud core was introduced in the seminal work of \citet{larson1969}, 
who was the first to predict the existence of the first hydrostatic core (FHSC). Further 1D simulations performed by others, including \citet{masunaga1998}, \mbox{\citet{masunaga2000}}, \citet{vaytet2012} and \mbox{\citet{vaytet2013}} support the initial description while exploring the dependence of pre-stellar evolution on the initial conditions and the detail with which the physical processes are modelled. The extension to three dimensions (e.g. \citealt{bate1998}) and the inclusion of additional physics such as magnetic fields (e.g. \citealt{commercon2012a,tomida2015}) 
 has further refined the initial description of \mbox{\citet{larson1969}}.

A key stage in the early star formation process is the formation of the FHSC. This occurs when the central regions have become sufficiently dense to reabsorb a significant proportion of the thermal radiation and the cooling timescale then exceeds the collapse timescale. This causes the temperature in the centre of the core to rise quickly such that the thermal pressure can support the gas near the centre against gravitational collapse. Outside of the central core, gravitational collapse continues and so the mass of the central core increases. The central temperature of the FHSC eventually increases to $\sim$\SI{2000}{\kelvin}, whereupon molecular hydrogen dissociates and initiates a second collapse to form the second (stellar) core.  Without rotation the typical radius of an FHSC is $\sim$\SI{5}{\au} \citep{larson1969}.  Rotation gives rise to a flattened FHSC which can even have a disc-like morphology if the molecular cloud core had a high initial angular velocity (e.g. \mbox{\citealt{bate1998, bate2011, st2006}}). Magnetic fields may act to slow rotation and therefore reduce the size of the disc \mbox{\citep{tomida2015}}. The additional support from magnetic fields against gravitational collapse can also lead to the formation of a pseudo-disc \citep{commercon2012a}.

Definitive observational evidence to support the theoretical description of early star formation is still required. In particular, the definitive detection of an FHSC remains elusive due to its low luminosity and location deep within a dense molecular cloud core. Attempts have been made both to observe bound objects in star-forming regions and to simulate observations to determine the observational characteristics of the FHSC.

There have been several observations of candidate FHSCs. The class of objects known as very low luminosity objects (VeLLOs) contains submillimetre sources with internal luminosity $<~0.1~\rm{L}_{\odot}$ \citep{young2004}. Several of these sources remain undetected or are faint in \textit{Spitzer Space Telescope} (\textit{Spitzer}) measurements and have been proposed as FHSC candidates. These are generally sources that are designated observationally as less evolved class 0 young stellar objects (YSOs) because they are very faint at $\lambda \leq$~\SI{24}{\micro\metre}, their envelope mass exceeds the mass of the central object and they lack strong, high-velocity CO outflows \citep{andre1993}. This class is likely to contain objects from FHSCs to accreting stellar cores since slow outflows have been detected near some (e.g. \citealt{andre1999,bourke2006}) and others have probably evolved well beyond stellar core formation (e.g. \citealt{dunham2006}).

The expected nature of the FHSC SED has changed little with attempts to simulate spectral energy distributions (SEDs) over the last 20 years. These have explained the effects of inclination \citep{boss1995},  considered different dust models and optical depths \citep{masunaga1998, omukai2007} and investigated differences in the SED of an evolving protostellar core from first collapse to beyond stellar core \citep{young2005}. All have shown the peak of the FHSC SED to be between \SI{100}{\micro\metre} and \SI{500}{\micro\metre} and that the SED is featureless.

\citet{saigo2011} produced synthetic SEDs at various stages throughout the first core phase. Their results show that the SED peak shifts to shorter wavelengths as the FHSC evolves, from $\lambda \simeq$~\SI{200}{\micro\metre} early on to $\lambda \simeq$~\SI{80}{\micro\metre} late in the first core phase. They plotted their SEDs along with that of a VeLLO and found that the FHSC has a lower luminosity and is much dimmer at $\lambda \lesssim$~\SI{100}{\micro\metre}.

\citet{commercon2012a} produced synthetic SEDs for the evolution of a rotating FHSC from a radiation magnetohydrodynamical model. The SED peaks did not change after FHSC formation and they were not able to distinguish the FHSC from the stellar core from the SEDs. Like \citet{saigo2011}, they found the SED peak to be at $\lambda \simeq$~\SI{200}{\micro\metre} early in the FHSC phase. The evolution of the SED was found to be similar for models with very different magnetic field strengths.

Now, in the era of the \textit{Atacama Large Millimeter/submillimeter Array} (ALMA), work has begun to exploit the advance in sensitivity in submillimetre wavelengths (e.g. \citealt{dunham2016}) and it may now be possible to resolve FHSC structures. However, the questions remain of how to positively identify the FHSC observationally and how to better target sources designated as FHSC candidates for further observation. In this paper we explore the differences in the SED of FHSCs at various evolutionary stages and with different properties with the aim of determining whether the SED can shed light on the nature of FHSC candidates. As far as we are aware, there have been no attempts to fit synthetic SEDs to particular observations of candidate FHSCs. \citet{robitaille07} produced a large set of synthetic protostar SEDs to create a fitting tool based on a $\chi^2$ fitting method that observers can use to help interpret their observed SEDs. Here we adopt a similar systematic method to fit the SEDs of five candidates reported in the literature and five newly identified FHSC candidates with model SEDs to test whether the SED can place constraints on their properties. 

\begin{figure*}
\centering
\includegraphics[width=12cm]{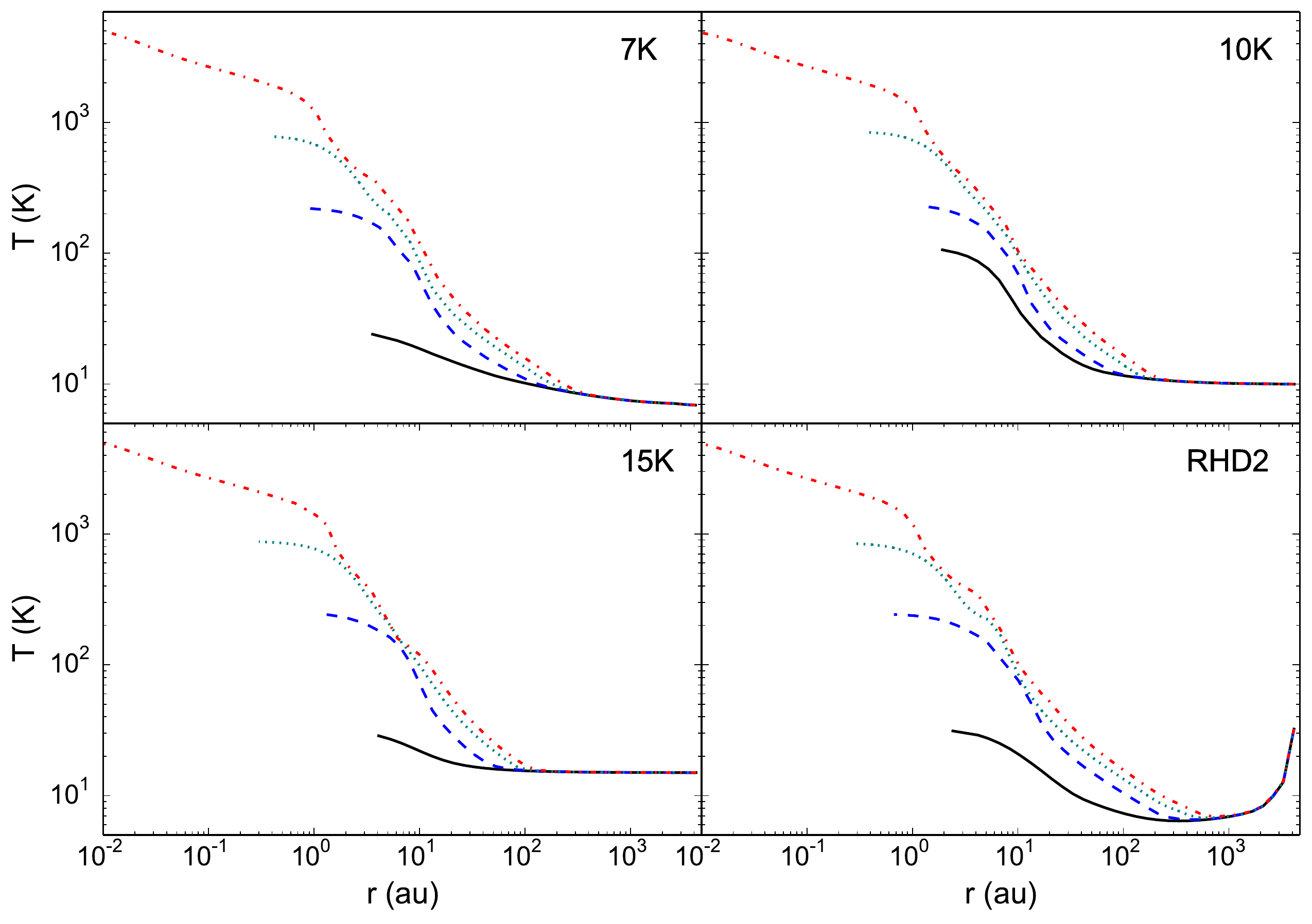}
\caption{Temperature profiles during the collapse of initially-static cores computed using the RHD1 model with initial temperatures \SI{7}{\kelvin}, \SI{10}{\kelvin} and \SI{15}{\kelvin}, and for the RHD2 model.   Snapshots are plotted when the central density is \SI{1e-12}{\gram\per\centi\metre\cubed}, \SI{5e-11}{\gram\per\centi\metre\cubed}, \SI{1e-9}{\gram\per\centi\metre\cubed} and \SI{1e-4}{\gram\per\centi\metre\cubed} in each panel. The temperature rises significantly only in the central $\sim$ \SI{10}{\au} and at $r>$~\SI{200}{\au} the temperature does not change significantly throughout the FHSC and the second collapse stages. For the RHD2 model, note the increase in temperature at radii $>$~\SI{800}{\au} where there is less extinction of the ISRF.}
\label{fig:tempprofiles}
\end{figure*}

\section{Method}
We performed three-dimensional radiation hydrodynamical (RHD) simulations of the collapse of molecular cloud cores to stellar core formation using a smoothed particle hydrodynamics (SPH) code. These simulations are similar to those of \mbox{\citet{bate2011}} and \mbox{\citet{bate2014}} and were performed for cores of different initial masses, rotation rates and radii. A few calculations were performed with magnetic fields, modelled with ideal magnetohydrodynamics (MHD). We also investigated the effect of an alternative method of radiative transfer within the RHD model and of varying the external radiation field. Snapshots from these models were used as input for frequency-dependent radiative transfer modelling to produce synthetic SEDs. In the radiative transfer code we varied the dust grain properties and the viewing inclination.

\subsection{The radiation (magneto-)hydrodynamical code}
\label{sec:methodRHD12}

We performed the protostellar core collapse simulations using the {\sc sphNG} code which originated from \cite{Benz1990}, but has since been extensively developed.  It now includes include individual particle timesteps \cite{Bate1995}, the variable smoothing length formalism of \cite{PriMon2004b, PriMon2007}, radiative transfer in the flux-limited diffusion approximation \cite{WhiBatMon2005, whitehouse2006}, magnetohydrodynamics using the formalism of \cite{PriMon2005} with divergence cleaning \cite{TriPri2012,tricco2016} and artificial resistivity \cite{tricco2013}.  Most recently, the radiative transfer has been combined with a model for the diffuse interstellar medium \citep{bateketo}.

Throughout this paper, we refer to the RHD method developed by \citet{whitehouse2005} and \citet{whitehouse2006} as RHD1. This model treats matter and radiation temperatures separately and performs radiative transfer using the flux-limited diffusion approximation. In these simulations the initial temperature is uniform and the boundary temperature is set equal to the initial temperature. As the core collapses and evolves, the central regions heat faster than the outer regions giving rise to the evolution in temperature profile shown in Fig.~\ref{fig:tempprofiles} for a non-rotating case. The majority of the infalling envelope remains cold throughout the FHSC phase.

We repeated calculations with the RHD model developed by \cite{bateketo} (hereafter RHD2) to compare the resulting SEDs from this more physical model with the simpler RHD1 model. The RHD2 model treats dust, gas and radiation temperatures separately and includes heating from the interstellar radiation field (ISRF), cooling via atomic and molecular line emission and collisional thermal coupling between the gas and dust to model the low density outer parts of the core more accurately. In the RHD2 model the initial gas and dust temperatures are the equilibrium temperatures calculated from these interstellar medium (ISM) heating and cooling processes. Initially, the temperature is lowest at the centre of the core due to extinction of the ISRF. Fig.~\ref{fig:tempprofiles} (lower right) shows the temperature profiles during the subsequent evolution of the core. As core collapse progresses, the centre heats up and warms the inner regions but a cold region remains between the internally heated central region and the outer regions of the core at $r\gtrsim$~\SI{800}{\au} which are heated by the ISRF.

\subsection{TORUS - the radiative transfer code}
We used the TORUS radiative transfer code \citep{harries2000} to simulate SEDs for selected RHD snapshots. TORUS reads in the SPH particle positions, densities and temperatures and maps the particles onto a 3D adaptive mesh refinement (AMR) grid following the method of \cite{rundle2010}. The maximum mass per cell is specified such that if this is exceeded the cell divides to give a higher resolution.

In these collapsing cloud cores the only source of emission is the warm dust. For RHD1 snapshots, the gas temperatures are used to compute the dust emission but for RHD2 snapshots the dust temperatures are calculated separately so these are used instead. An observer is placed at a specified distance and the optical depths through the grid to this observer are calculated. A Monte Carlo method is used to determine the locations and directions of thermal emission and then the path of the photons through the grid to the observer is calculated, including absorption and scattering events. Only direct thermal dust continuum emission makes a significant contribution to the SED when the maximum grain size $a_{\mathrm{max}} \lesssim$ \SI{1}{\micro\meter}. When larger dust grains are included, there is also a contribution from scattered thermal emission.

\section{Hydrodynamical models}
\subsection{Initial conditions}
\label{initcond}

\begin{table*}
\caption{A summary of the radiation hydrodynamical models performed. The parameters used in the hydrodynamical simulations to investigate the effect of changing the initial temperature, incident ISRF, mass, initial rotation, initial radius and magnetic field strength on the SED are listed in the respective rows. (a) The incident ISRF is changed by adding an additional boundary of molecular gas, expressed in terms of the column density of molecular hydrogen. (b) The MHD models employ the radiative transfer treatment of RHD2.}
\label{tab:hydromodels}
\begin{tabular}{|l|ll|llll|}
	\hline
	\multirow{3}{*}{Temperature} & \multicolumn{2}{c|}{\SI{7}{\kelvin}} 	& \multirow{3}{*}{RHD1}	& \multirow{3}{*}{\SI{1}{\solarmass}}	& \multirow{3}{*}{\SI{4700}{\au}} & \multirow{3}{*}{$\beta=0$}  \\
								 & \multicolumn{2}{c|}{\SI{10}{\kelvin}}	&	&	&	&	 \\
								 & \multicolumn{2}{c|}{\SI{15}{\kelvin}}	&	&	&	&	\\
	\hline
	\multirow{4}{*}{ISRF$^\mathrm{a}$} 	& \multicolumn{2}{c|}{none} 								  & \multirow{4}{*}{RHD2} & \multirow{4}{*}{\SI{1}{\solarmass}} & \multirow{4}{*}{\SI{4700}{\au}} & \multirow{4}{*}{$\beta=0$} \\
																& \multicolumn{2}{c|}{\SI{1e20}{\per\centi\meter\squared}} &					&	  & &  \\
																& \multicolumn{2}{c|}{\SI{2e20}{\per\centi\meter\squared}} &					&	  & &  \\
																& \multicolumn{2}{c|}{\SI{5e20}{\per\centi\meter\squared}} &					&	  & &   \\
																& \multicolumn{2}{c|}{\SI{1e21}{\per\centi\meter\squared}} &
																				&     & &   \\
																& \multicolumn{2}{c|}{\SI{5e21}{\per\centi\meter\squared}} &
																				&     & &   \\
	\hline
	\multirow{4}{*}{Mass} & \SI{0.5}{\solarmass} & \SI{3700}{\au} & \multirow{4}{*}{RHD2} & & & \multirow{4}{*}{$\beta=0$} \\
						  & \SI{1}{\solarmass}   & \SI{4700}{\au} & & & & \\
						  & \SI{2}{\solarmass}   & \SI{5900}{\au} & & & & \\
						  & \SI{5}{\solarmass}   & \SI{8000}{\au} & & & & \\
	\hline
	\multirow{4}{*}{Rotation} & \multicolumn{2}{c|}{$\beta=0$} 		& \multirow{4}{*}{\parbox{1.8cm}{RHD1 (10~K) \\ ~~\& \\ RHD2}} & \multirow{4}{*}{\SI{1}{\solarmass}} & \multirow{4}{*}{\SI{4700}{\au}} & \\
							  & \multicolumn{2}{c|}{$\beta=0.01$} 	&						&									  & & \\
							  & \multicolumn{2}{c|}{$\beta=0.05$} 	&						&									  & & \\
							  & \multicolumn{2}{c|}{$\beta=0.09$} 	&						&									  & & \\
	\hline
	\multirow{4}{*}{Radius} & \multicolumn{2}{c|}{\SI{2000}{\au}} & \multirow{4}{*}{RHD2}	&  \multirow{4}{*}{\SI{1}{\solarmass}} & & \multirow{4}{*}{$\beta=0$}  \\
							& \multicolumn{2}{c|}{\SI{6000}{\au}} &							&									   & &  \\
							& \multicolumn{2}{c|}{\SI{10000}{\au}} &						&									   & &  \\
							& \multicolumn{2}{c|}{\SI{11400}{\au}} &						&									   & &  \\
	\hline
	\multirow{2}{*}{Magnetic Field$^\mathrm{b}$} & \multicolumn{2}{c|}{$\mu=5$} & \multirow{2}{*}{RHD2}	&  \multirow{2}{*}{\SI{1}{\solarmass}} & \multirow{2}{*}{\SI{4700}{\au}} & \multirow{2}{*}{$\beta=0.005$} \\
									& \multicolumn{2}{c|}{$\mu=20$} & & & &  \\
	\hline
\end{tabular}
\end{table*}

The simulations begin with a Bonnor-Ebert sphere \mbox{\citep{bonnor1956,ebert1955}} for which the critical ratio of central density to density at the outer boundary is 14.1:1. We use an unstable density ratio of \num{15.1}. Unless otherwise stated, the initial mass was \SI{1}{\solarmass} and the initial radius was \SI{7.0e16}{\centi\metre} (\SI{0.023}{\parsec}, 4680 AU). These calculations were performed with a resolution ranging from \num{260000} to \num{500000} SPH particles.

We modelled the collapse of a molecular cloud core under different initial conditions to investigate differences in the SEDs. To explore the effects of initial temperature, we used RHD1 with uniform initial temperatures of \SI{7}{\kelvin}, \SI{10}{\kelvin} and \SI{15}{\kelvin} and RHD2 for a non-rotating \SI{1}{\solarmass} core. We also varied the column density at the boundary in RHD2 to alter the intensity of the ISRF.

To investigate the effects of mass we modelled the collapse of non-rotating cores of masses \SI{0.5}{\solarmass}, \SI{1}{\solarmass}, \SI{2}{\solarmass} and \SI{5}{\solarmass} and radii \SI{5.56e16}{\centi\metre} (\SI{3700}{\au}), \SI{7.0e16}{\centi\metre} (\SI{4700}{\au}), \SI{8.82e16}{\centi\metre} (\SI{5900}{\au}) and \SI{12.0e16}{\centi\metre} (\SI{8000}{\au}) respectively. The radii were chosen to maintain a constant initial central density of $\rho_{\rm{max}}=$~\SI{1.38e-18}{\gram\per\centi\metre\cubed}. This better reflects conditions in the interstellar medium than keeping a constant radius, since more massive molecular cloud cores have larger radii \citep[e.g.][]{larson1981}. RHD2 was used for these calculations.

To investigate the effects of rotation we modelled the collapse of \SI{1}{\solarmass} cores with RHD2 for different initial rotation rates. The initial angular velocity is set through the ratio of rotational energy to gravitational potential energy $\beta=E_{\rm{rot}}/E_{\rm{grav}}$. Here we used the values $\beta = $~0, 0.01, 0.05 and 0.09. These simulations were also repeated with RHD1 with a uniform initial temperature of \SI{10}{\kelvin} for the comparison with observations and this is described in greater detail in Section~\ref{sec:modselection}.

We also performed collapse simulations of non-rotating \SI{1}{\solarmass} cores with different initial radii, but with the same ratio of central density to density at the outer boundary of 15.1:1, as for the other simulations. Thus, these cores have different mean densities. The additional initial radii chosen were \SI{3.0e16}{\centi\metre} (\SI{2000}{\au}), \SI{9.0e16}{\centi\metre} (\SI{6000}{\au}), \SI{11.0e16}{\centi\metre} (\SI{7400}{\au}), \SI{15.0e16}{\centi\metre} (\SI{10000}{\au}) and \SI{17.0e16}{\centi\metre} (\SI{11400}{\au}). The central density of the cores then ranged from \SI{1.1e-16}{\gram\per\centi\metre\cubed} for the smallest core to \SI{5.9e-19}{\gram\per\centi\metre\cubed} for the \SI{17.0e16}{\centi\metre} core. RHD2 was used for these calculations.

Lastly, we performed radiation magnetohydrodynamical (RMHD) simulations.  The simulations use the same general MHD method as those of \citet{bate2014}, but they use the radiative transfer and ISRF treatment of \citet{bateketo} and they employ the updated artificial resistivity switch of \citet{tricco2013} and divergence cleaning of \citet{tricco2016}. The RMHD simulations were set up with rotation corresponding to $\beta=0.005$ because rotation is necessary for the formation of an outflow. The initial mass-to-flux ratios used were $\mu=5$ and $\mu=20$, with a total core mass of \SI{1}{\solarmass} and initial radius of \SI{7.0e16}{\centi\metre}. These simulations used a resolution of \num{e6} SPH particles.

A summary of the hydrodynamical models is shown in Table~\ref{tab:hydromodels}.

\subsection{Core collapse and FHSC evolution}
\begin{figure}
	\centering
	\includegraphics[width=7.5cm]{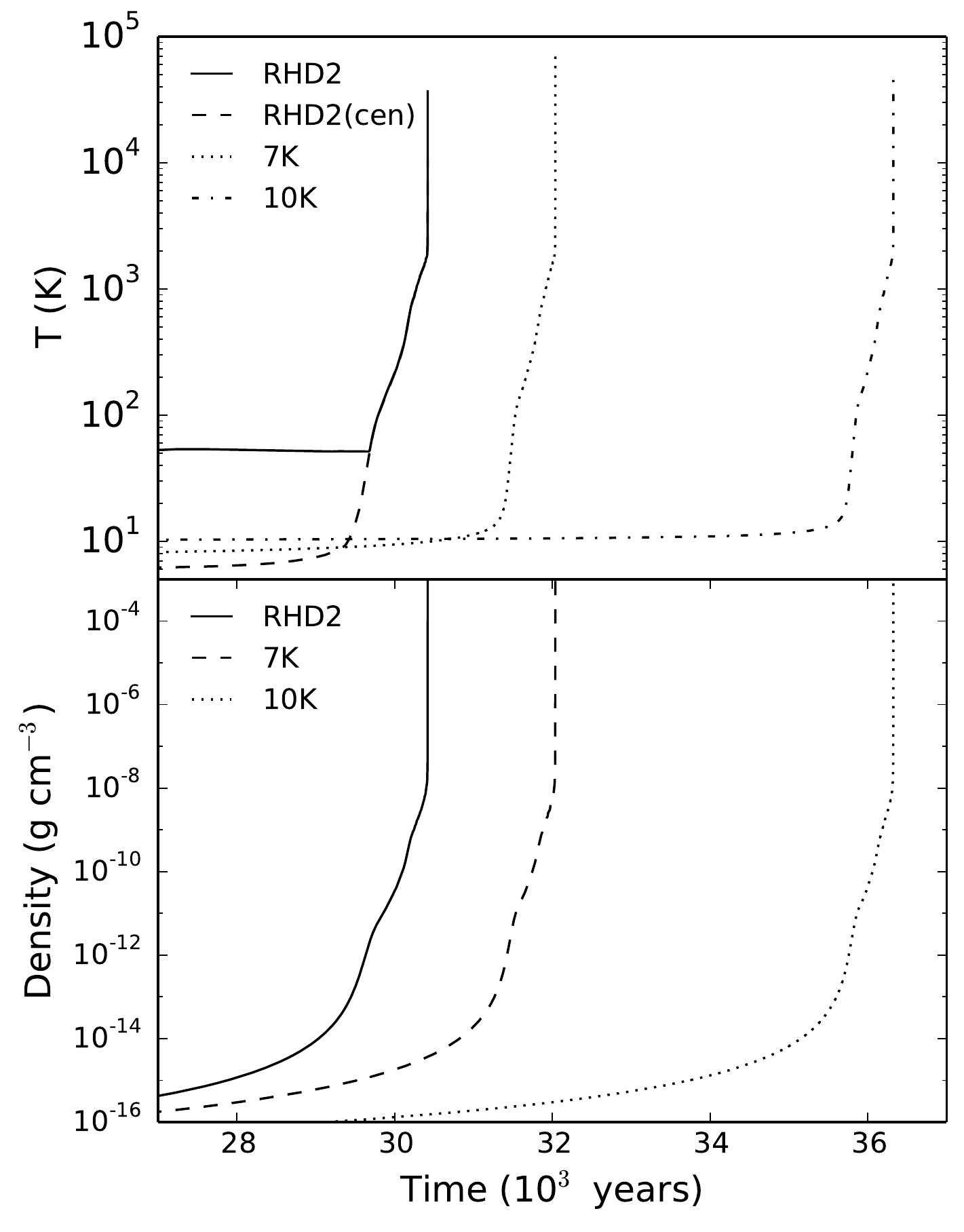}
	\caption{Evolution of the maximum temperature and density of the core for RHD1 models with uniform initial temperatures of \SI{7}{\kelvin} and \SI{10}{\kelvin}, and for the RHD2 model. For the RHD2 models, prior to the formation of the first core, the gas temperature is highest in the outer regions due to heating from the ISRF so we plot both the evolution of the maximum temperature, RHD2, and the temperature at the centre of the core, RHD2(cen). The core collapse proceeds fastest under the RHD2 model and an increase in initial temperature slows the collapse due to the additional thermal pressure. In each case, the FHSC forms and collapses at similar values of maximum density.}
	\label{fig:rhoTevo}
\end{figure}

The evolution of the collapsing core occurs in four stages: first collapse, FHSC phase, second collapse and stellar core formation. The evolution of the maximum density and temperature for cases computed using both RHD1 and RHD2 are shown in Fig.~\ref{fig:rhoTevo}. During first collapse, the maximum density increases slowly by several orders of magnitude while the central temperature increases by just a few Kelvin. The length of the first collapse stage ranges from \num{30000} to \num{36000} years in these simulations because it depends upon the temperature, rotation and magnetic field which provide support against the gravitational collapse as well as the density contrast between the centre and outer regions of the core. The FHSC is formed at a central density of $\sim$~\SI{e-12}{\gram\per\centi\metre\cubed}, when the thermal pressure supports against further collapse. In the \SI{1}{\solarmass} simulations in this paper, the FHSC ranges in size from a radius of \SI{4}{\au} up to \SI{60}{\au} under fast rotation. The central temperature and density of the FHSC both increase as material continues to fall onto it. The collapse takes place more slowly for the initially warmer cores due to the increased pressure support. The collapse is faster for the RHD2 model because the central regions are initially colder.

At $\sim$\SI{2000}{\kelvin} (central density of $\sim$\SI{e-7}{\gram\per\centi\metre\cubed}), molecular hydrogen begins to dissociate and the second collapse begins. The stellar core is formed after second collapse at a central density of $\sim$\SI{e-3}{\gram\per\centi\metre\cubed} but its temperature and density continue to rise rapidly.

\begin{figure}
\centering
	\includegraphics[width=7.5cm]{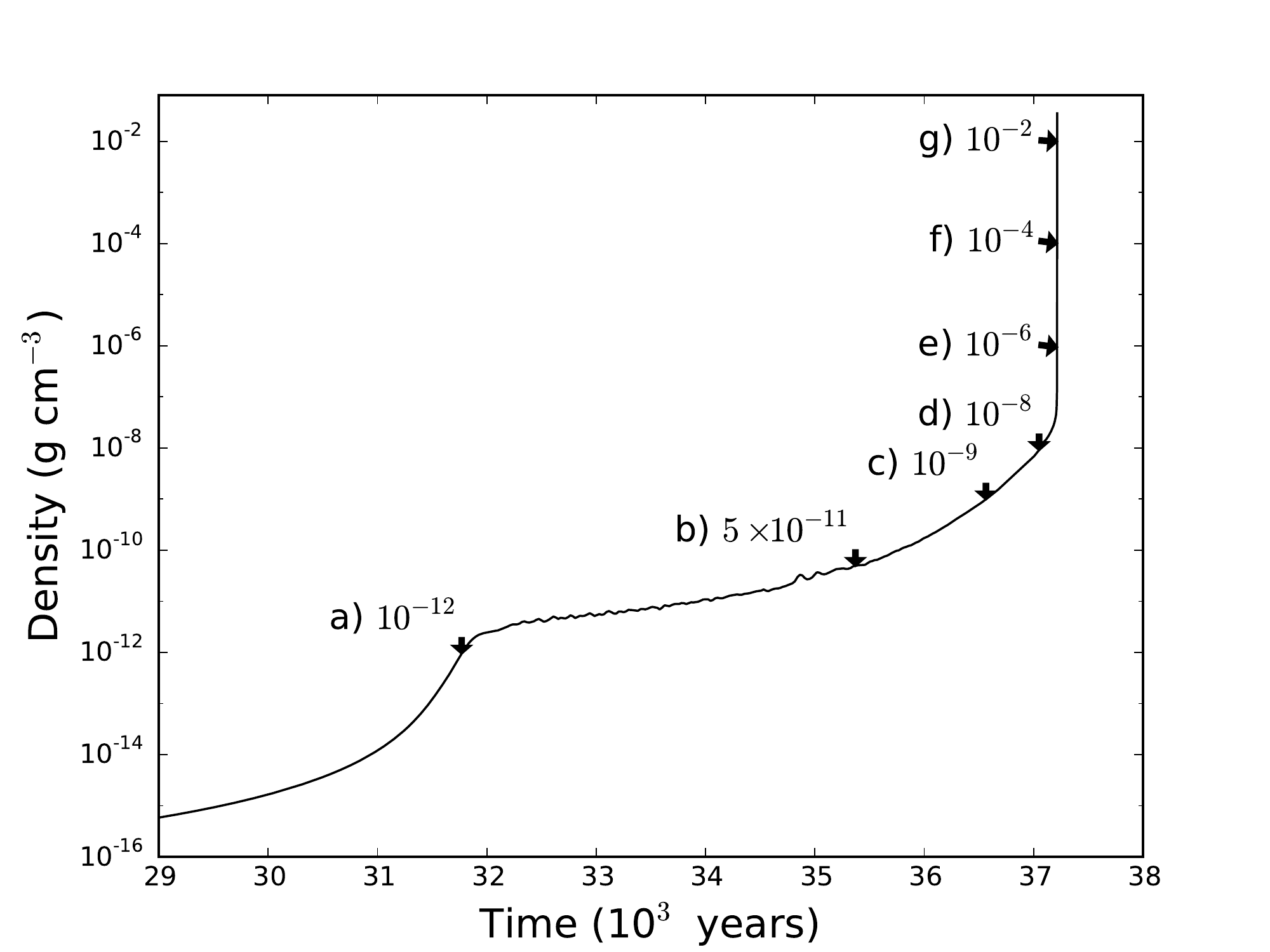}
	\caption{Evolution of the maximum density of a collapsing \SI{1}{\solarmass} core with initial rotation $\beta = 0.09$ from the RHD2 model. The values of the maximum density used to select each snapshot for simulating SEDs are indicated in \si{\gram\per\centi\metre\cubed}. Snapshot (a) occurs just before FHSC formation and snapshot (d) is taken just before the onset of second collapse.}
	\label{fig:B09BK_rhoevo}
\end{figure}

We selected snapshots representing the major phases in FHSC formation and evolution from the RHD simulations for SED modelling. These are defined by the maximum density, as shown in Fig.~\ref{fig:B09BK_rhoevo}, and give six snapshots: before FHSC formation; early-, mid- and late-FHSC phase; and two during second collapse. For some figures, we also show the snapshot just after the formation of the second (stellar) core at $\rho_{\mathrm{max}}=$\SI{e-2}{\gram\per\centi\metre\cubed}.

\section{Modelling SEDs}

Synthetic SEDs were created from the RHD models using TORUS for snapshots of the entire molecular core, including the FHSC and infalling envelope. We altered the dust grain size, dust grain type and inclination of the core relative to the observer to investigate the effects on the SEDs.

The distribution of grain sizes is set via the ISM power law size distribution function \mbox{\citep{mathis1977}} 
\begin{equation} 
\label{eq:graindist}
n(a)\propto a^{-q} 
\end{equation}
with $n(a)$ the number of particles of size $a$. \citet{mathis1977} find $q$ to be in the range $3.3 < q < 3.6$ for various substances including graphite and silicates and so we take the accepted ISM value of $q=3.5$.

The range of dust grain sizes for the distribution was $a_{\mathrm{min}} =$ \SI{0.001}{\micro\metre} to  $a_{\mathrm{max}} = $ \SI{1.0}{\micro\metre}. We then altered the maximum in the range \SIrange{0.5}{1000}{\micro\metre} in separate simulations.

SEDs were simulated for two different grain types: the silicate grain of \citet{drainelee1984} and the amorphous carbon grain type of \citet{zubko1996}.

For all model SEDs a distance of \SI{260}{\parsec} was used because this was the favoured the distance to the Serpens South molecular cloud \citep{Straizys:2003aa}, the region in which the new candidate FHSCs are located, at the time of simulating the SEDs. Rescaling for a different distance is straightforward and we exploit this when fitting to observations.

\subsection{Simulated photometry}
\label{sec:photometry}
It is necessary to simulate photometry from the model SEDs in order to compare them with observations.  This also has the advantage of averaging out the Monte Carlo noise that is present at short wavelengths.  We calculated the monochromatic fluxes for each instrument by convolving the model SEDs with the instrument response functions\footnote{Response functions were taken from the Spanish Virtual Observatory's Filter Profile Service \url{http://svo2.cab.inta-csic.es/theory/fps3/}, \url{http://www.adamgginsburg.com/filters/bolocam_passband.txt}, \url{http://www.eaobservatory.org/jcmt/instrumentation/continuum/scuba-2/filters/} and \url{https://astro.uni-bonn.de/~bertoldi/projects/mambo/manuals.html}.} and redistributing the flux over the assumed SED shape
\footnote{Information on the assumed SED for each instrument were taken from \citet{mayne2012} and the Spanish Virtual Observatory.}
 using the equations presented in Appendix A of \citet{robitaille07}.

\subsection{Fitting model SEDs to observations}
\label{sec:fittingmethod}
We compared observed SEDs to a sample of model SEDs following a similar method to that of \citet{robitaille07}. We selected the models with the smallest $\chi^{2}$ per point for the observed SED, which is calculated using
\begin{equation}
\label{eq:chi2}
	 \chi^{2} = \frac{1}{N}\displaystyle\sum_{i=1}^{N}\left(\frac{F_{\mathrm{i}} - M_{\mathrm{i}}}{\sigma_{\mathrm{i}}}\right)^2,
\end{equation}
$F_{\mathrm{i}}$ and $M_{\mathrm{i}}$ are the observed and model fluxes respectively for each wavelength, $\sigma_{\mathrm{i}}$ is the uncertainty in the observed flux and $N$ is the number of observed wavelengths. For observations where there are upper limits ($U_{\mathrm{i}}$), models with $M_{\mathrm{i}} < U_{\mathrm{i}}$ are assigned a $\chi^2$ contribution of zero. For models with $M_{\mathrm{i}} > U_{\mathrm{i}}$, the $\chi^2$ contribution is
\begin{equation}
\label{eq:upperlimits}
	 \chi^2 = -2 \log{(1- \mathrm{confidence})}
 \end{equation}
  where the confidence is found from the stated observational uncertainty \citep{robitaille07}. Where this is not reported we assume a $2\sigma$ upper limit and hence a confidence of \num{0.9545}. Where possible, all observed SED points are included in the $\chi^2$ calculation. In the cases where there is a detection at a shorter wavelength than available in a given model SED the model SED is discarded because the model flux would be many orders of magnitude lower than required to provide a good fit to the observation. This only affects observations at $\lambda <$ \SI{24}{\micro\metre}.

 The model SEDs must be scaled to allow for differences in the masses and distances of the sources and so each model SED is scaled to the observed SED to minimise its $\chi^{2}$. It is then these final $\chi^{2}$ values that are compared to determine the best fitting models. The scaling is discussed further in Section \ref{sec:scaling}. We do not consider reddening because this effect is negligible at the wavelengths of the cold dust emission of pre-stellar cores. When performing photometry on the observations, the background emission is subtracted from the images and this removes any emission from a wider filament or cloud which means that we are comparing as far as possible only the emission from the core itself with the models.

\begin{figure*}
\centering
\includegraphics[trim={2cm 0 0 0}, clip,width=\textwidth]{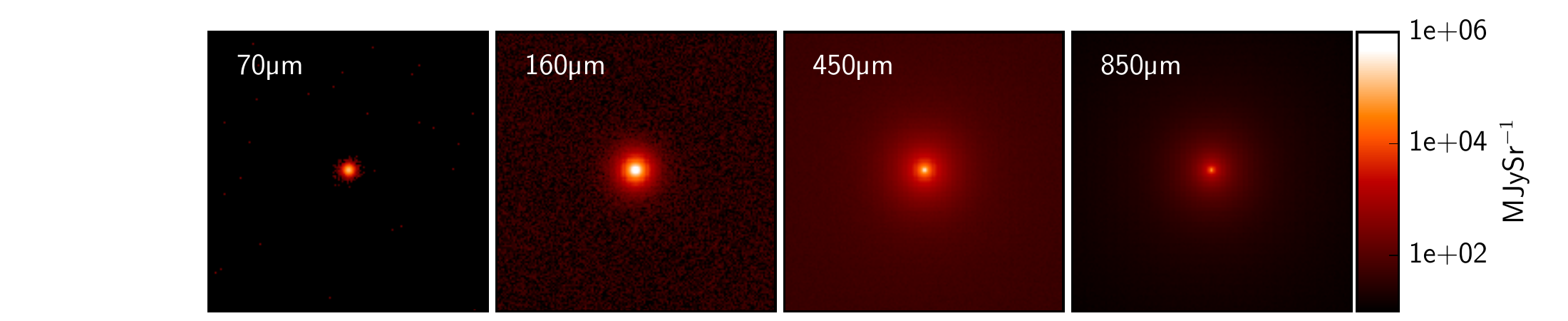}
\caption{Synthetic images of the non-rotating FHSC from RHD2 at four commonly observed wavelengths. The images are each \SI{1.1}{\arcsecond} across which corresponds to \SI{286}{\au} at \SI{260}{\parsec}. The \SI{70}{\micro\metre} flux is emitted only in the warm central regions. The \SI{450}{\micro\metre} and \SI{850}{\micro\metre} images are dominated by emission from close to the FHSC, since the lower opacity allows radiation from deeper inside the core to escape, but there is also a significant contribution from emission from the cold envelope. The \SI{70}{\micro\metre} and \SI{160}{\micro\metre} images are affected by Monte Carlo noise because the emission from the envelope is very low.} 
\label{fig:B0BKimages}
\end{figure*}

\begin{figure}
\centering
\includegraphics[width=7cm]{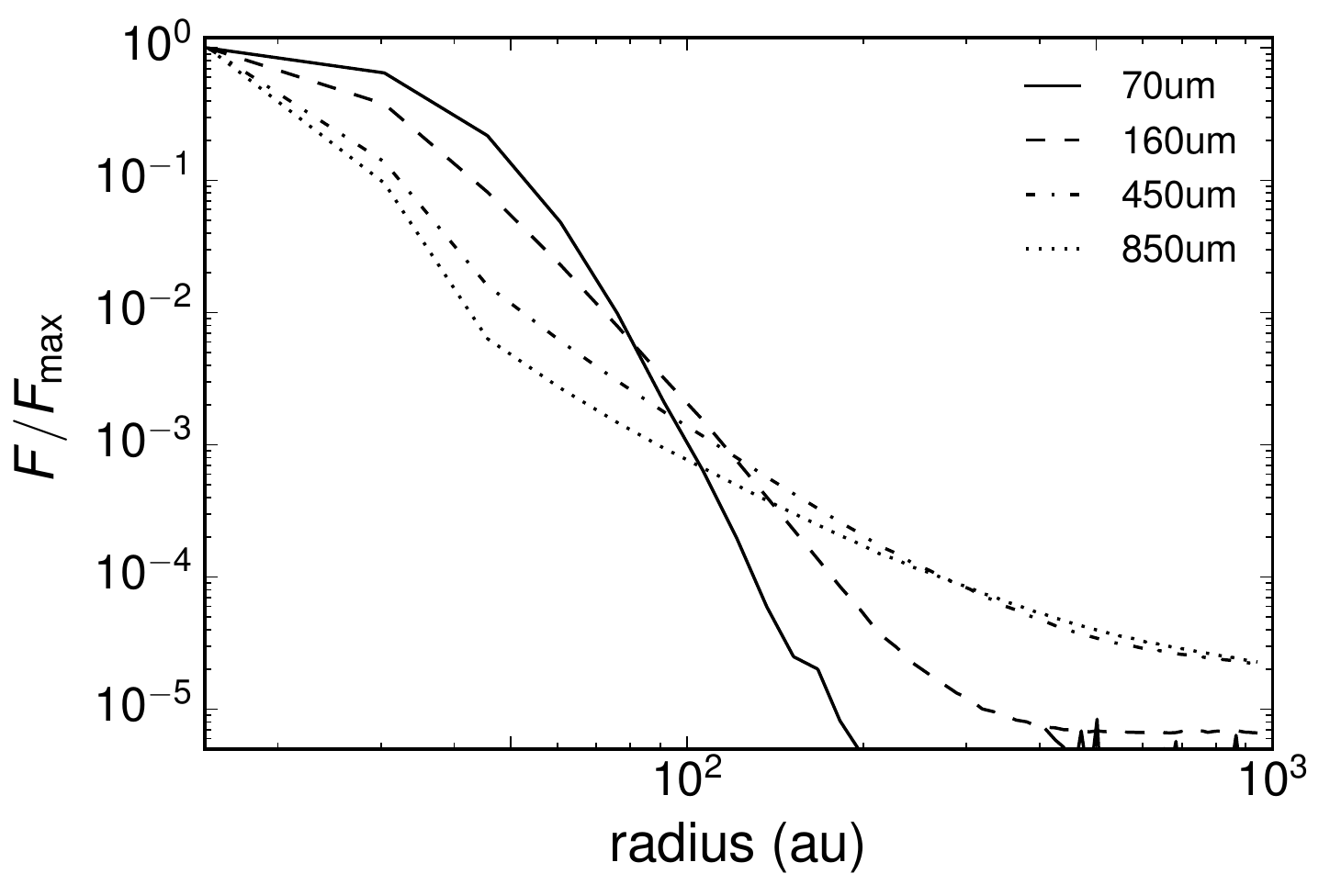}
\caption{Radial average intensity profiles for the images of a non-rotating FHSC from RHD2 shown in Fig.~\ref{fig:B0BKimages} at four commonly observed wavelengths. The flux intensity profiles are normalised to the intensity in the centre of the FHSC image at each wavelength. The flux peaks more sharply at longer wavelengths because more of the observed flux was emitted from deeper within the core. There is also more emission from the cold envelope at the longer wavelengths.}
\label{fig:B0BKradint}
\end{figure}

\begin{figure*}
\centering
\includegraphics[trim{2.5cm 0 0 0},clip,width=\textwidth]{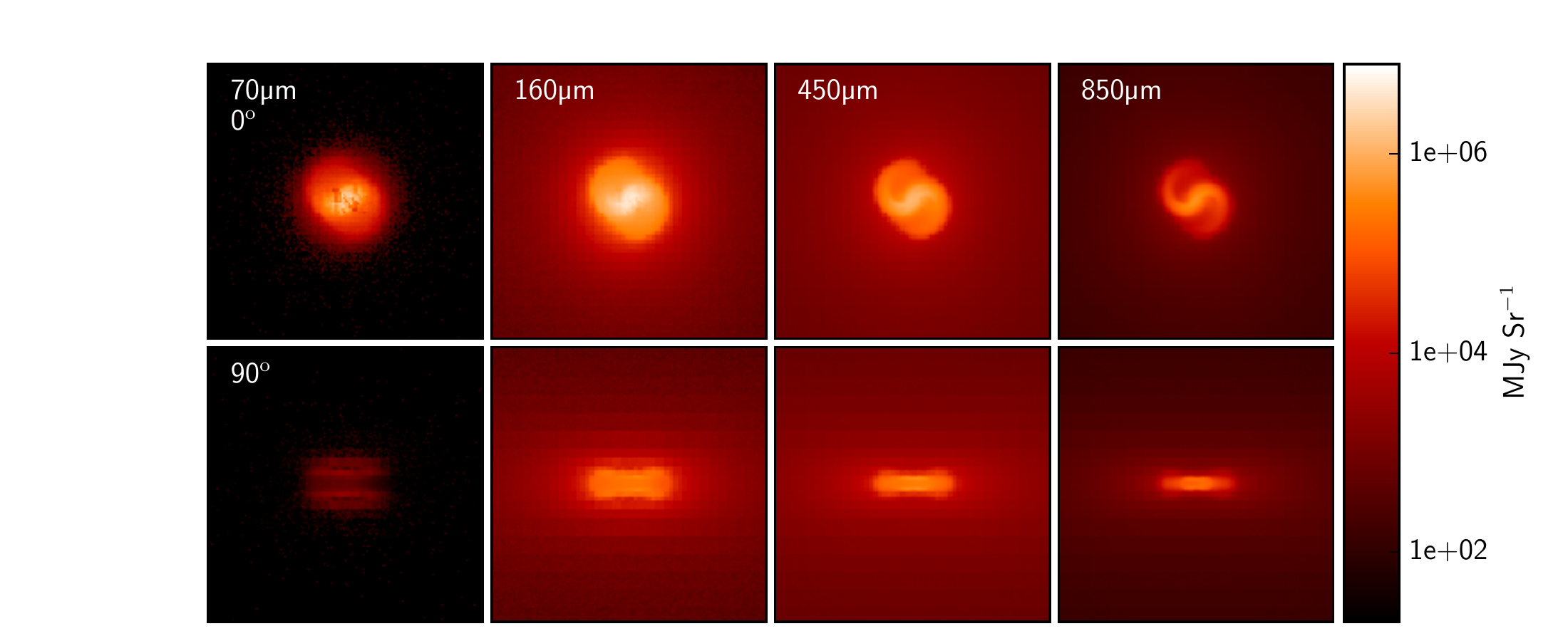}
\caption{Synthetic images of the rotating FHSC from RHD1 with initial temperature of \SI{10}{\kelvin} and initial rotation corresponding to $\beta=0.05$ at four wavelengths with $i=$~\ang{0} (face on) images above and $i=$~\ang{90} images below. The images are each \SI{1.1}{\arcsecond} across which corresponds to \SI{286}{\au} at \SI{260}{\parsec} and logarithmic scaling was used. Longer wavelengths allow us to see deeper into the core and distinguish smaller scale structures. At $i=$~\ang{90} the observed \SI{70}{\micro\metre} flux is emitted in the optically thin regions above and below the disc and the disc itself appears dark. The \SI{70}{\micro\metre} and \SI{160}{\micro\metre} images have a poorer resolution because most of the emission is coming from $>$~\SI{20}{\au} from the centre where the AMR grid resolution is coarser.}
\label{fig:B05images}
\end{figure*}

\section{Results}
\label{sec:SEDresults}
In this section we describe the effects of various initial conditions and parameters on the evolution of the SED. These simulations were performed with no rotation, no magnetic field, a total core mass of \SI{1}{\solarmass} and an initial radius of \SI{7e16}{\centi\metre} (\SI{4700}{\au}) using the RHD2 model unless otherwise stated. The SEDs were produced for a distance of \SI{260}{\parsec} and an inclination of \ang{0}, with silicate type dust grains following the size distribution of $q=3.5$ between $a_{\rm{min}} =$~\SI{0.001}{\micro\metre} and $a_{\rm{max}} =$~\SI{1.0}{\micro\metre} unless otherwise stated. SEDs are plotted with a flux scale cut off approximately an order of magnitude below the typical observational sensitivity of \textit{Spitzer} at \SI{24}{\micro\metre}. 

\subsection{Interpreting the SED}
\label{sec:SEDmeans}

As is the case for many astrophysical objects, flux from a dense pre-stellar core is not emitted from a single `photosphere' at all wavelengths. It is useful to estimate where most of the flux at the observed wavelengths is emitted in order to understand the results in the following sections. The opacity depends upon the density and temperature of the dust, as well as the dust properties such as grain size and composition and the wavelength of the incident radiation. The density and temperature vary by several orders of magnitude and dust grains sublimate at $\sim$\SI{1500}{\kelvin}, causing the opacity to vary significantly as a function of radius and time.

The observed flux will be dominated by radiation from the smallest radii it can `escape' from, i.e. the photosphere. \citet{masunaga1998} and \citet{omukai2007} discuss the locations of the photospheres in a pre-stellar core at different wavelengths in detail based on optical depth calculations. We produced synthetic images at the commonly observed wavelengths of \SI{70}{\micro\metre}, \SI{160}{\micro\metre}, \SI{450}{\micro\metre} and \SI{850}{\micro\metre} from a calculation of a non-rotating, and hence spherically-symmetric, early FHSC using RHD2. These objects are very faint at \SI{24}{\micro\metre} so we did not produce images for that wavelength. The images shown in Fig.~\ref{fig:B0BKimages} show which structures are observed at each wavelength and, hence, which parts of the object contribute to the SED at each wavelength.
We see in Figs.~\ref{fig:B0BKimages} and \ref{fig:B0BKradint} that the \SI{70}{\micro\metre} flux is emitted in the central regions of the core but the flux does not peak as sharply at the centre as for longer wavelengths. The opacity is higher at \SI{70}{\micro\metre} and flux emitted from the very centre is obscured. Consequently,s the FHSC is not directly visible at wavelengths shorter than $\sim$\SI{800}{\micro\metre} since it lies beneath the effective photospheres. The total flux at \SI{850}{\micro\metre} is dominated by emission from the cold envelope but, since the opacity is low, the smaller structures in the warmer regions are distinguishable and the non-rotating source in Fig.~\ref{fig:B0BKimages} appears more compact.

\begin{figure*}
\centering
\includegraphics[width=17cm]{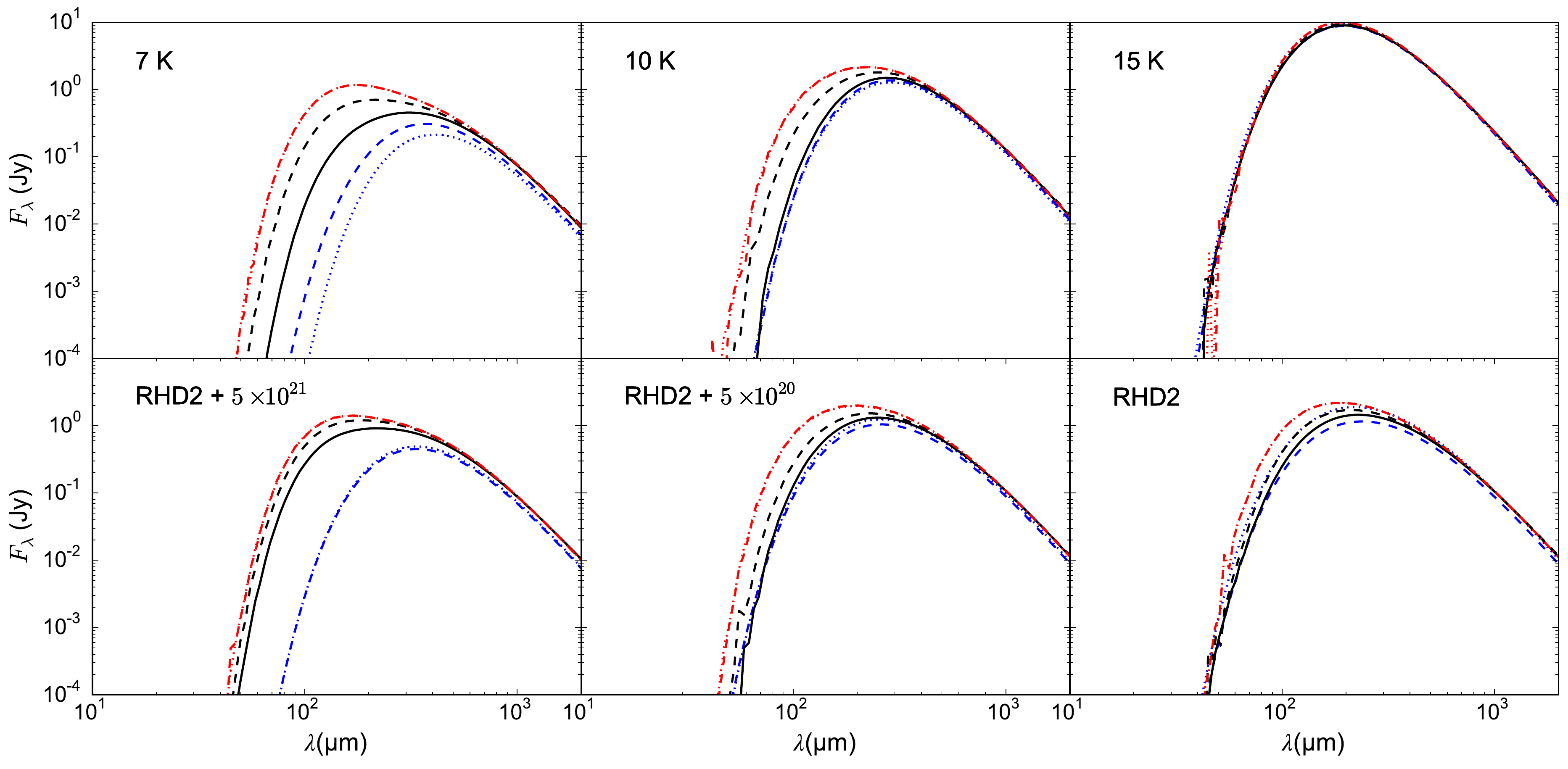}
\caption{SED evolution for the collapse of cores of initial temperatures \SI{7}{\kelvin}, \SI{10}{\kelvin} and \SI{15}{\kelvin} and for the RHD2 model (initial temperatures determined from exposure to the ISRF) with an additional boundary column density of \SI{5e21}{\per\centi\metre\squared}, \SI{5e20}{\per\centi\metre\squared}, and an unattenuated ISRF. SEDs are from snapshots where the central density is \SI{1.4e-18}{\gram\per\centi\metre\cubed} (the initial conditions, blue dotted lines), \SI{1e-12}{\gram\per\centi\metre\cubed} (blue dashed lines), \SI{5e-11}{\gram\per\centi\metre\cubed} (solid black lines), \SI{1e-9}{\gram\per\centi\metre\cubed} (dashed black lines), \SI{1e-4}{\gram\per\centi\metre\cubed} (red dotted lines) and \SI{1e-2}{\gram\per\centi\metre\cubed} (red dashed lines). The SEDs show less evolution for a warmer initial temperature because the initial SED is already brighter. Similarly, the RHD2 SEDs show less evolution with increased exposure to the ISRF.}
\label{fig:tempSEDs}
\end{figure*}

In reality, rotational structures are ubiquitous so we also present synthetic images of an FHSC from the model with initial rotation $\beta=0.05$ using RHD1 and a uniform initial temperature of \SI{10}{\kelvin} in Fig.~\ref{fig:B05images}. This calculation produces a dynamically rotationally unstable FHSC which goes bar-unstable and develops trailing spiral arms \citep[c.f.][]{bate1998}.  With rotation, the optical depth is reduced along the axial direction ($i=$~\ang{0}), which means that the observed flux is emitted from closer to the FHSC boundary than for the non-rotating core and we can see the central spiral morphology. At $i=$~\ang{90} the hot central core is completely obscured at \SI{70}{\micro\metre} and the observed flux at this wavelength is emitted in the lower density regions above and below the disc. At longer wavelengths the opacity decreases and a more compact structure is observed.

In summary, for all observed wavelengths the total flux is dominated by the regions surrounding the FHSC. It is therefore the temperature structure of this warm region outside the FHSC and the wider envelope that affects the shape of the SED which means that temperature differences within the FHSC are to a large extent irrelevant. This is what we consider in the following sections when comparing differences between SEDs of cores formed under different initial conditions.

\subsection{Initial core temperature and external radiation}
\label{sec:resultstemp}

The temperature structure of the core changes as the object evolves and this directly affects the flux at each wavelength reaching an observer. We investigate the SED evolution first with cloud cores of uniform initial temperatures of \SI{7}{\kelvin}, \SI{10}{\kelvin} and \SI{15}{\kelvin} using the RHD1 model and secondly, with a cloud core subjected to an interstellar radiation field (ISRF). We also compared the SED evolution of cores exposed to different ISRF intensities.

Fig.~\ref{fig:tempSEDs} shows the evolution of the SED as core collapse progresses from first collapse ($\rho_{\mathrm{max}}=$\SI{e-12}{\gram\per\centi\meter\cubed}), through the FHSC stage, to second collapse ($\rho_{\mathrm{max}}=$\SI{e-4}{\gram\per\centi\meter\cubed}) with the RHD1 model for the three values of initial temperature, as well as for the RHD2 model. For a higher initial temperature the SED is brighter and peaks at a shorter wavelength early on in the collapse. As the core evolves and the FHSC forms the SED peak shifts to shorter wavelengths (Fig.~\ref{fig:tempSEDs}, top row). For the cores with higher initial temperatures, there is less variation in the SED as the object evolves and we see that for an initial temperature of \SI{15}{\kelvin} (Fig.~\ref{fig:tempSEDs}, lower left panel) the SEDs are identical throughout, even to second collapse. This result indicates that, for warmer pre-stellar molecular cloud cores, no information can be gained from the SED as to its evolutionary state.

The differences in SED evolution between the cores with different initial temperatures suggests there must be differences in their radial temperature profiles. From the radial temperature profiles shown in Fig.~\ref{fig:tempprofiles} we can see that the FHSC has not heated gas beyond a radius of about \SI{300}{\au} above its initial temperature in the \SI{7}{\kelvin} case which means that the envelope remains colder than that of the \SI{15}{\kelvin} core even to the second collapse phase when the central temperature quickly rises by several thousand Kelvin. The higher temperature of the envelope in the \SI{15}{\kelvin} case leads to increased emission from larger radii and the photospheres will then be at a larger radii, where there is less heating from the FHSC. This effect is stronger at shorter wavelengths since the low density gas is still optically thin for $\lambda >$ \SI{700}{\micro\metre}. For a higher initial temperature, the regions probed by shorter wavelengths will therefore have a smaller temperature variation as the core evolves. Because of this and because much of the material is initially hotter, the SED displays less evolution.

There is also a difference in the shape of the FHSC SED for different initial temperatures. The SEDs of the warmer cores are closer to a single-temperature blackbody curve whereas the SEDs of the \SI{7}{\kelvin} case become steeper at shorter wavelengths and flatter to longer wavelengths as the core evolves. \citet{omukai2007} also report SEDs of this shape and explain that it is due to the interplay of increasing optical depth with decreasing wavelength and decreasing temperature at larger radii. The photospheres of the cooler cores are nearer the FHSC where there is a steeper temperature gradient (see Fig.~\ref{fig:tempprofiles}). The photospheres for different wavelengths will be at a spread of radii in all cases but when the envelope is cooler the set of observed wavelengths samples the region nearer the centre and so will sample a broader range of temperatures. The larger the differences in temperatures sampled by each wavelength, the less the SED will resemble that of a single temperature blackbody. We therefore expect the SEDs of colder cores to show the most evolution because the temperature contrast is greater and this contrast increases as the FHSC develops.

A uniform initial temperature is unrealistic (see e.g. \citealt{ward-thompson2002aa,nielbock2012aa,roy2014aa}); therefore we now consider a core exposed to the ISRF and use its equilibrium temperature profile as the initial temperature conditions.

\begin{figure}
\centering
	\includegraphics[width=7cm]{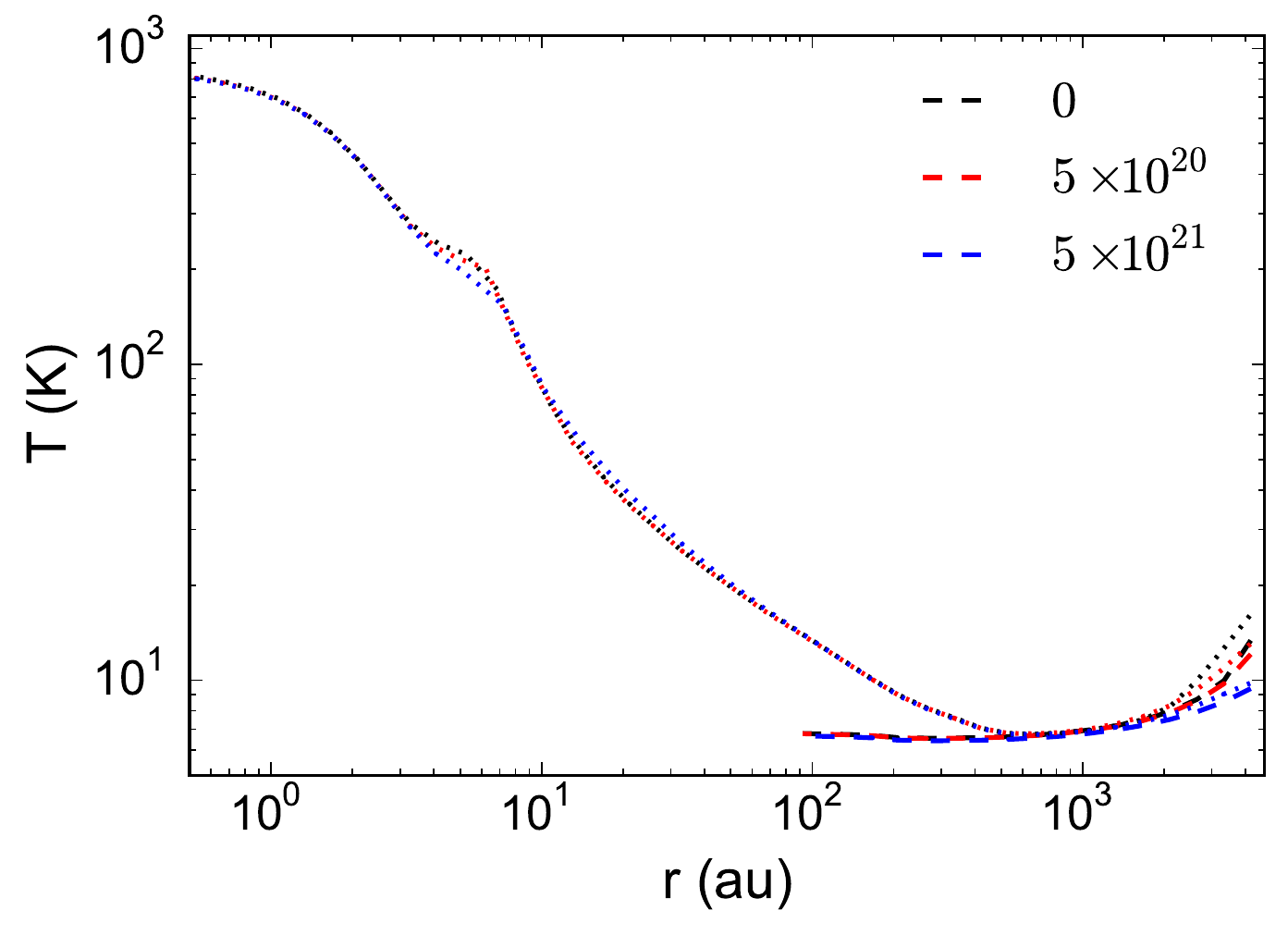}
	\caption{Radial dust temperature profiles for the initial conditions (dashed lines) and at $\rho_{\mathrm{max}}= 10^{-9}$~\si{\gram\per\centi\metre\cubed} (dotted lines) for non-rotating protostellar cores with an additional boundary density, defined by the column number density of molecular hydrogen (shown in legend, \si{\per\centi\metre\squared}), to reduce the incident ISRF.}
	\label{fig:BDtprofs}
\end{figure}

 \begin{figure}
 \centering
	\includegraphics[width=7cm]{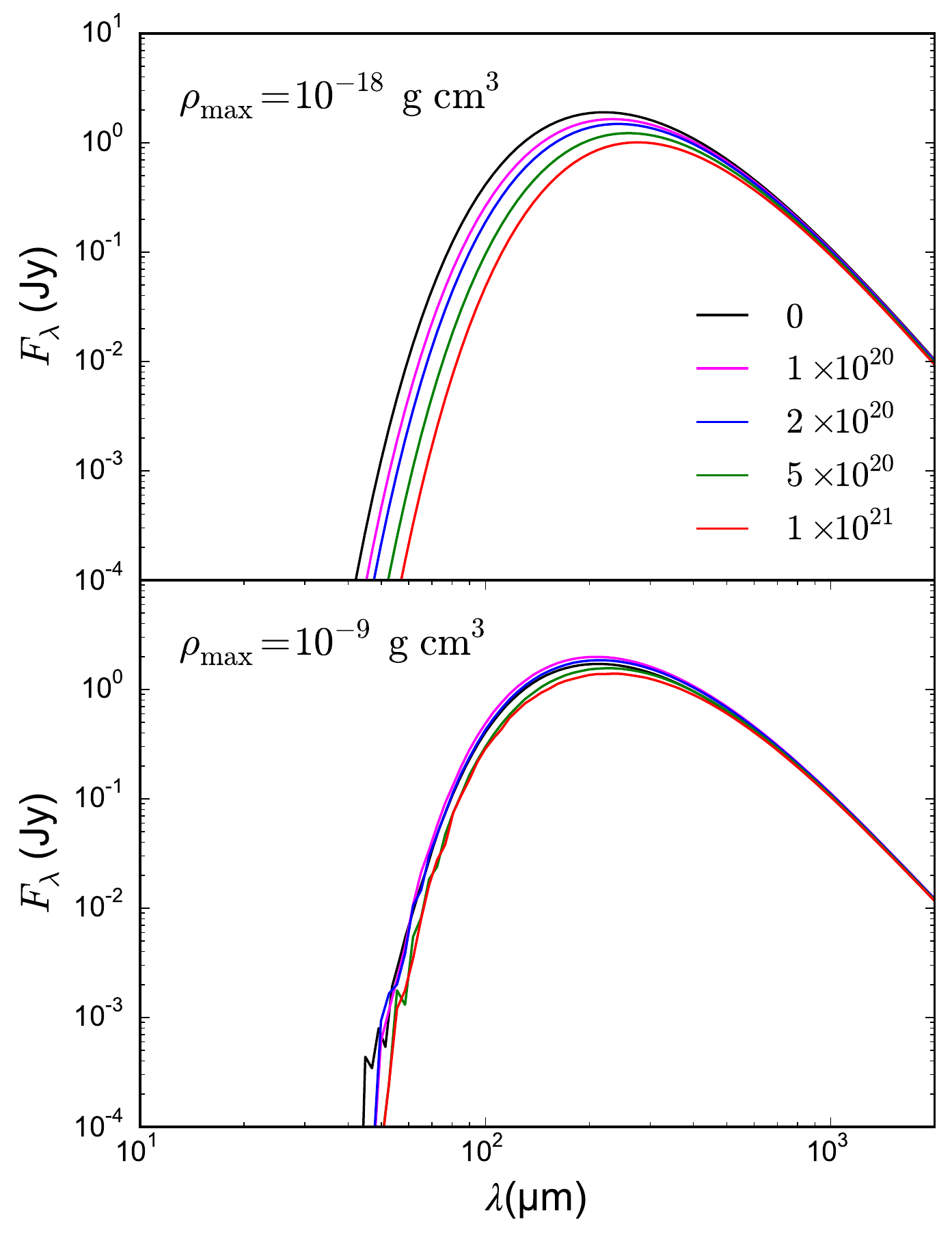}
	\caption{SEDs for the initial conditions (top) and at $\rho_{\mathrm{max}}=10^{-9}$~\si{\gram\per\centi\metre\cubed} for non-rotating protostellar cores with different column densities at the boundary to reduce the ISRF, defined by the column number density of molecular hydrogen (\si{\per\centi\metre\squared}) shown in the legend.}
	\label{fig:bounddens}
\end{figure}

\begin{figure*}
	\centering
	\subfigure[]{\includegraphics[width=7cm]{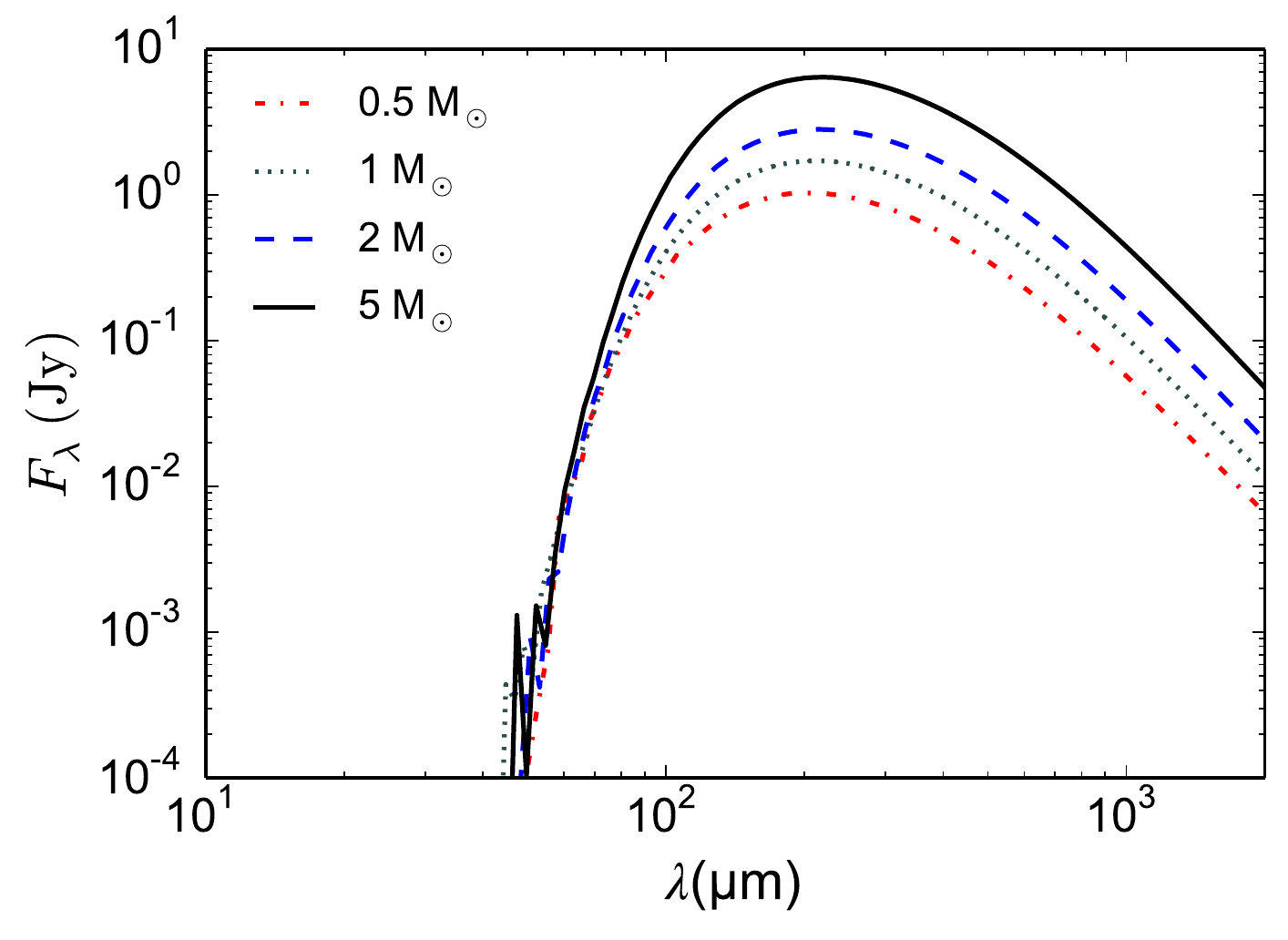}}{\label{fig:DiffMass}}\qquad
	\subfigure[]{\includegraphics[width=7cm]{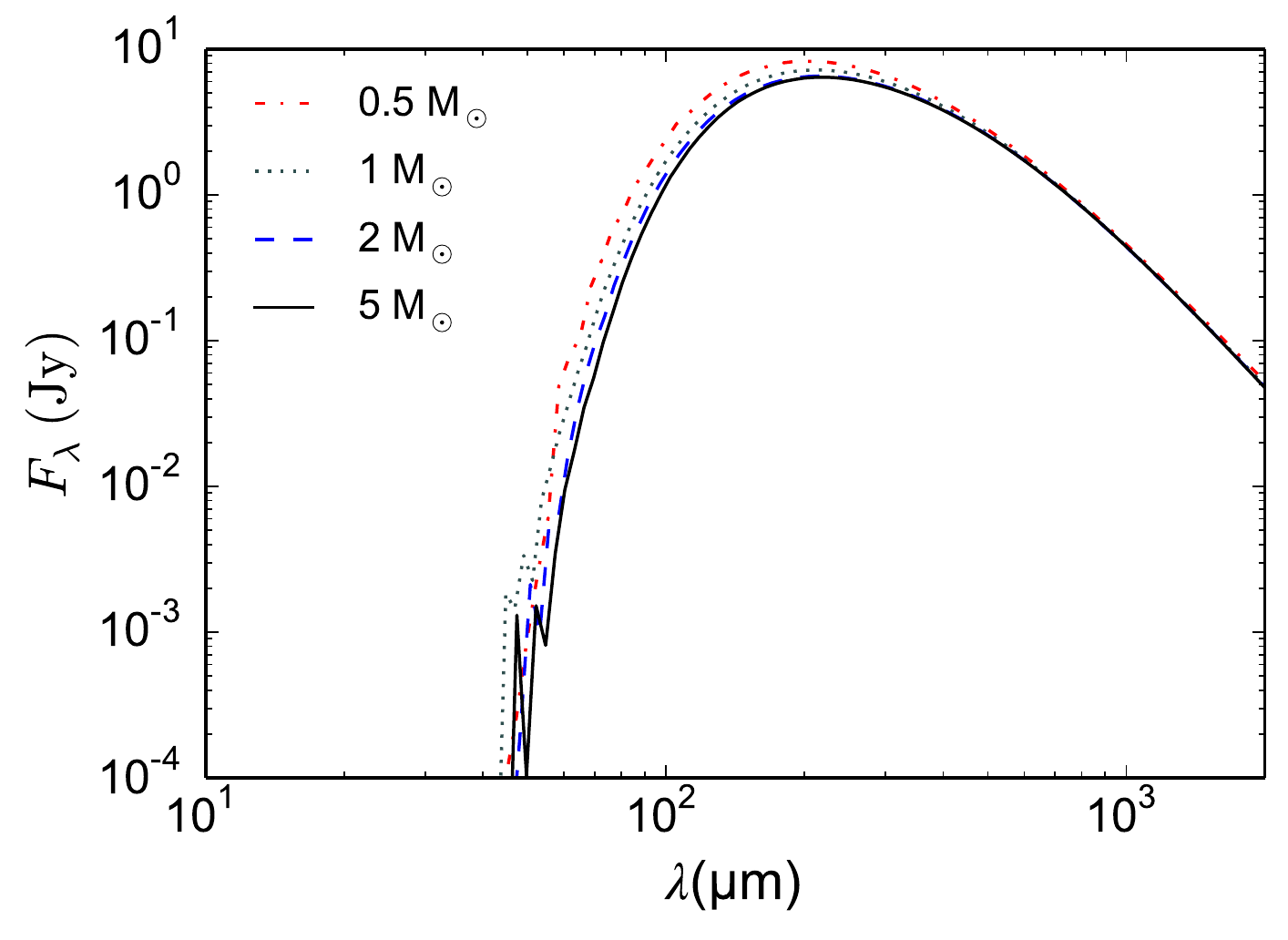}}{\label{fig:scaledmass}}
	\caption{(a) SEDs for collapsing pre-stellar cores of different initial masses for snapshots when $\rho_{\mathrm{max}}=$~\SI{10e-9}{\gram\per\centi\metre\cubed}. (b) SEDs of different mass FHSCs scaled to fit the \SI{5}{\solarmass} FHSC by minimising {$\protect\chi^2$} for points $\lambda >$~\SI{700}{\micro\meter}. Higher mass cores are more luminous, as expected, but darker at \SI{70}{\micro\metre} compared to \SI{160}{\micro\metre}.}
	\label{fig:DiffMasswhole}
\end{figure*}

\begin{figure}
	\centering
	\includegraphics[width=7cm]{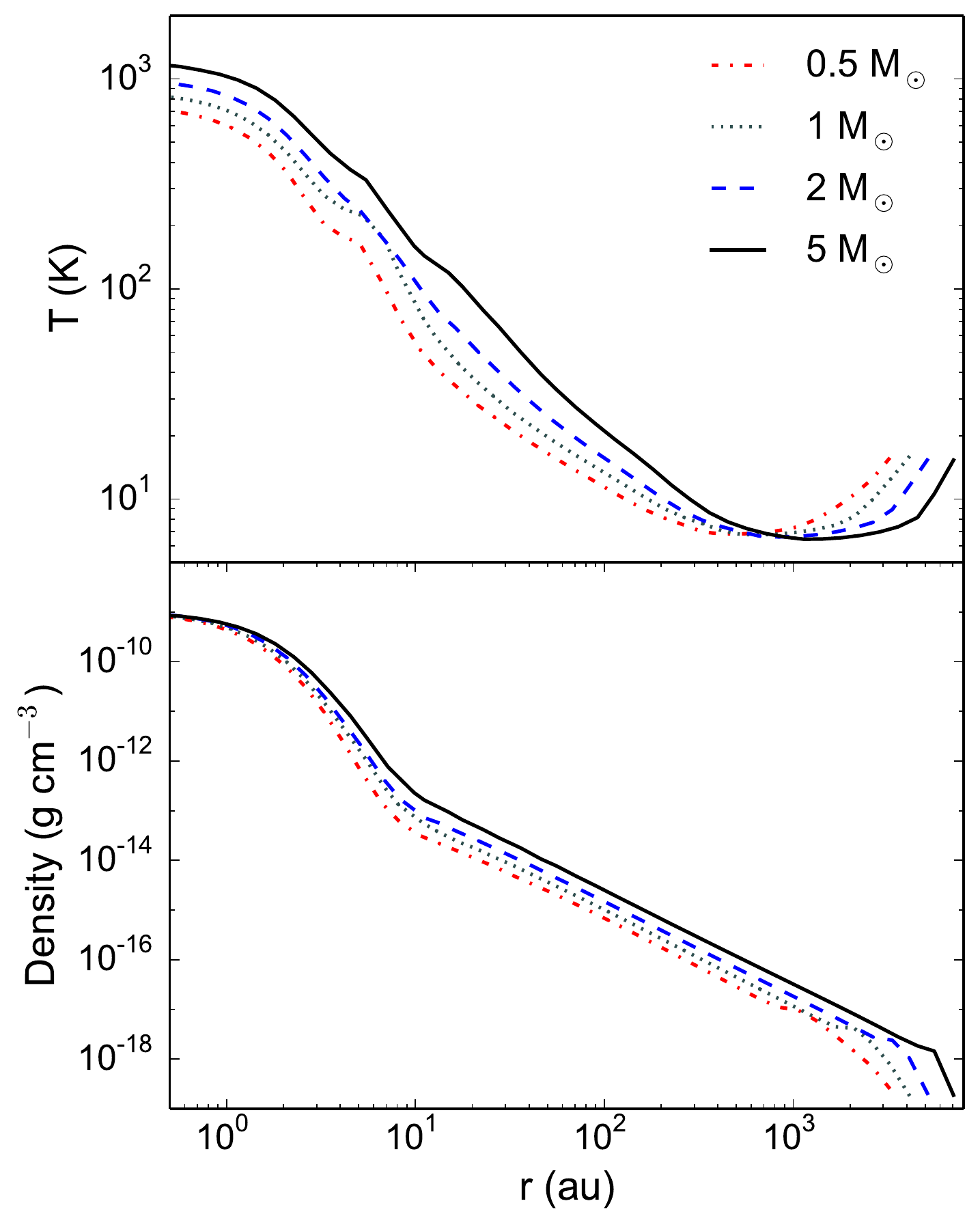}
	\caption{Radial profiles of the dust temperature and density for cores of different masses during the FHSC phase when $\rho_{\mathrm{max}}=$~\SI{10e-9}{\gram\per\centi\metre\cubed}. The FHSC extends to similar radii in all cases, as shown by the change in the slope of the density profile at $\sim$~\SI{1e14}{\centi\metre}.}
	\label{fig:MassProfs}
\end{figure}

The SED evolution with the RHD2 model is presented in Fig.~\ref{fig:tempSEDs} (bottom row). The RHD2 SEDs display a smaller increase in flux at shorter wavelengths as the core evolves than the RHD1 SEDs. The RHD2 SED of the early FHSC stage peaks at a shorter wavelength compared to those of the \SI{7}{\kelvin} and \SI{10}{\kelvin} RHD1 cores due to the heating of dust by the ISRF at $r>$~\SI{2700}{\au} to more than \SI{10}{\kelvin}, as shown in Fig.~\ref{fig:tempprofiles} (lower right). This region contains around a third of the total mass, which means that the ISRF heats a greater proportion of the core than the FHSC does during its lifetime. The FHSC heating only makes a significant difference to the SED later on when the region surrounding the FHSC ($20\lesssim r \lesssim 30$~\si{\au}) is heated to $>$~\SI{20}{\kelvin}. The SED is dominated by flux from the ISRF-heated region through to second collapse phase so we do not see the strong progression to shorter wavelengths as the core evolves that was the case for the cold RHD1 SEDs, as well as for the SEDs of \citet{saigo2011}, \mbox{\citet{commercon2012a}} and \mbox{\citet{young2005}}. There is an increase in flux at \SI{70}{\micro\meter} and \SI{160}{\micro\meter}, while the flux at longer wavelengths remains nearly constant and this could be useful for estimating the evolutionary stage of a source, even though the SEDs of the FHSC phase do not appear to be clearly distinct from those of the first and second collapse stages.

In the RHD1 models, gas and dust are assumed to have the same temperature, but using RHD2 the gas and dust temperatures may differ substantially.  Dust emission is the main source of continuum emission and so the dust temperatures calculated in the RHD2 model are used to produce the SEDs. The dust and gas temperatures only diverge significantly in the low density region at the edge of the core where they are not well coupled. We find there is, however, a significant difference in the SEDs if we use the gas temperature rather than the dust temperature because of this difference in the outer parts of the core. For detailed SED modelling it is therefore important to model the outer dust temperatures well. Given the differences in SED evolution between the RHD1 and RHD2 models, it appears beneficial to model the core collapse with the more physical radiative transfer method.

The ISRF incident on the protostellar core was varied by adding an additional column density of molecular material (gas and dust) at the boundary. This additional density is not included in the SED calculations so its only effect is to reduce the external radiation incident on the cloud core. 
In an observational study of dense cores, \citet{kim2016} report that there is a temperature increase of 3-6~K between the central and outer regions of the core with the central temperature being 7-8~K. The temperature profiles in Fig.~\ref{fig:BDtprofs} show that the initial conditions of all of the RHD2 models fall within these ranges.

The SED evolution with RHD2 with boundary column density values of \SI{5e20}{\per\centi\metre\squared} and \SI{5e21}{\per\centi\metre\squared} of molecular hydrogen is\ shown in Fig.~\ref{fig:tempSEDs} (bottom row).
For the initial conditions ($t=0$), $\lambda_{\mathrm{max}}$ shifts to longer wavelengths for increased boundary density, as shown by Fig.~\ref{fig:bounddens}, because heating from the ISRF is attenuated leading to lower dust temperatures in the outer regions of the core (see Fig.~\ref{fig:BDtprofs}). However, once collapse begins, the increasingly dense central core becomes the dominant heat source and so there is much less of a difference between the SEDs of FHSCs from models with different values of the boundary column density after FHSC formation. We do also see greater evolution of the SED with a reduced ISRF because the infrared flux is less dominated by emission from the outer regions and is more sensitive to internal temperature changes.

\begin{figure*}
	\centering
	\includegraphics[trim={0 0.4cm 0 0}, clip, width=\textwidth]{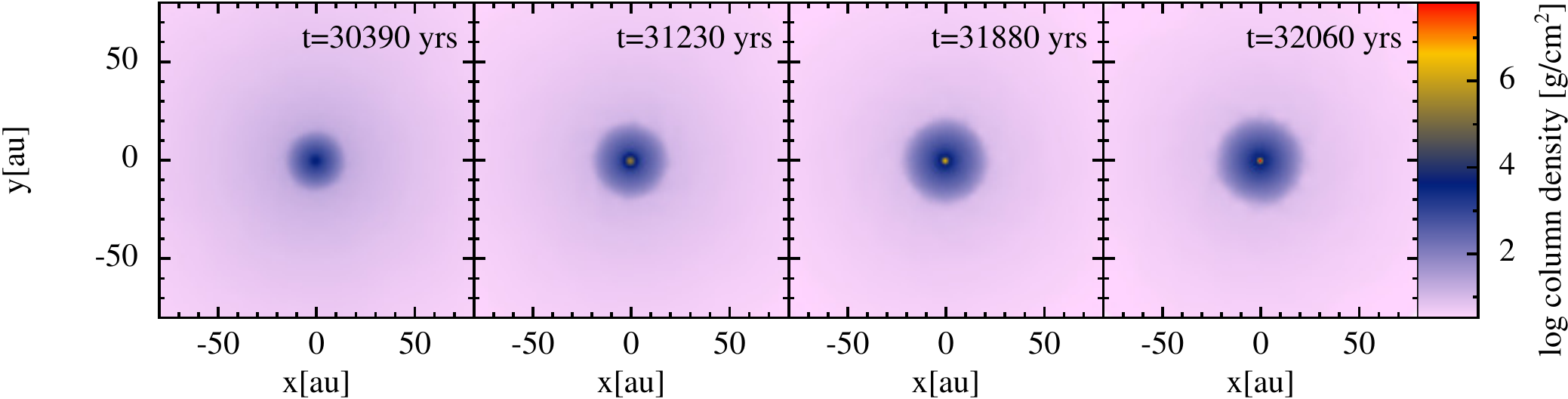}
	\includegraphics[trim={0 0.4cm 0 0}, clip, width=\textwidth]{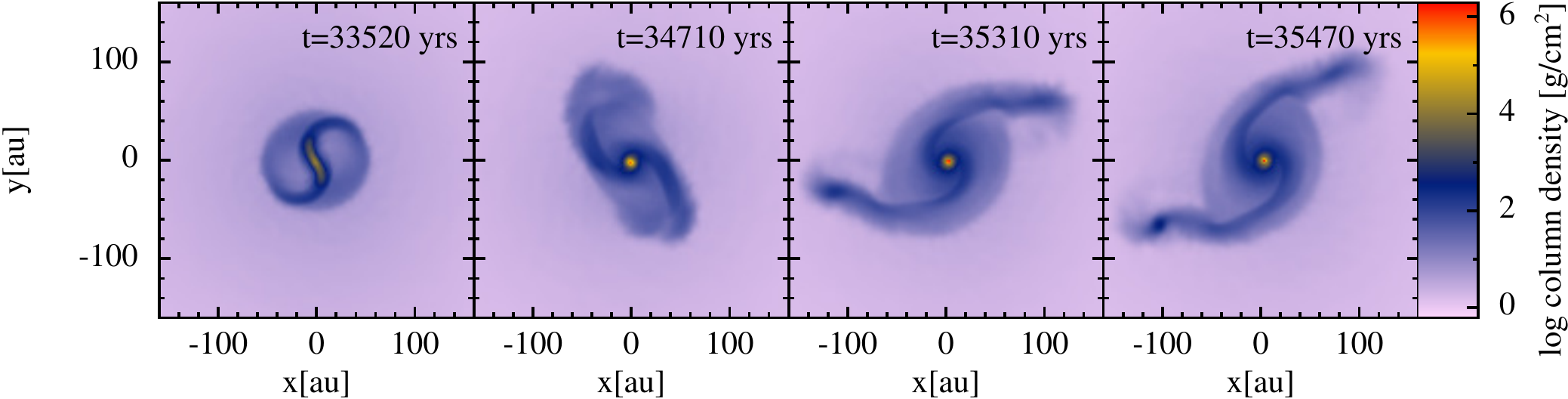}
	\includegraphics[trim={0 0 0 0}, clip, width=\textwidth]{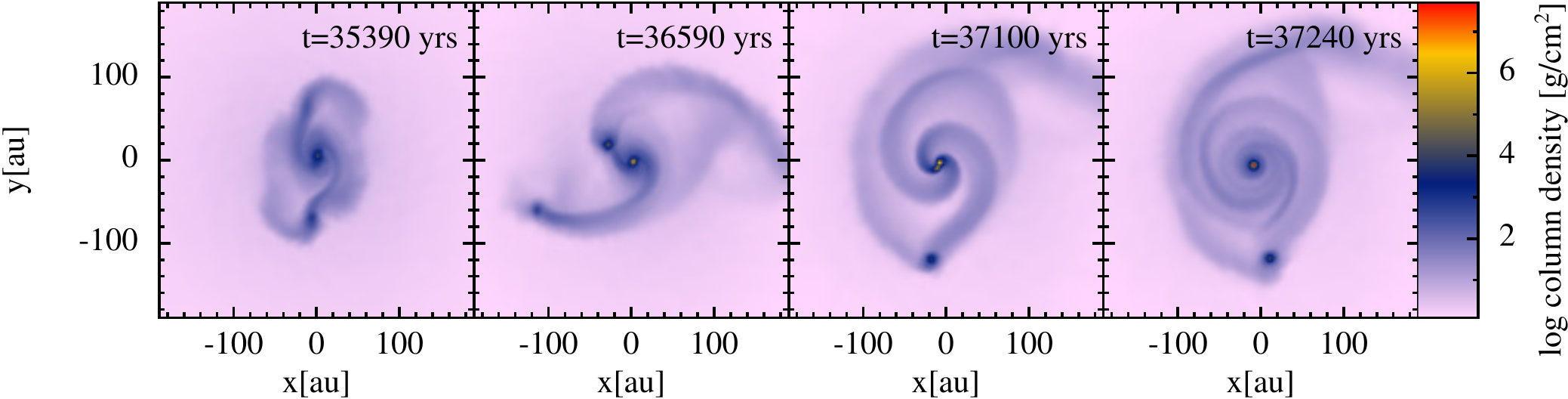}
	\caption{Snapshots of the column density for RHD2 simulations with initial rotation $\beta=0.01$ (top row), $\beta=0.05$ (middle row) and $\beta=0.09$ (bottom row). These are some of the snapshots from which SEDs were simulated. The maximum density, $\rho_{\rm{max}}$ in each panel is \SI{5e-11}{\gram\per\cubic\centi\meter}, \SI{e-9}{\gram\per\cubic\centi\meter}, \SI{e-8}{\gram\per\cubic\centi\meter} and \SI{e-6}{\gram\per\cubic\centi\meter} from left to right. The FHSC is an oblate spheroid in the $\beta=0.01$ simulation. The faster rotating cores form spiral arms and fragments. The two inner fragments formed in the $\beta=0.09$ simulation fall back together before the onset of second collapse. At $t=0$ the peak column density is $<$~\SI{0.1}{\gram\per\centi\metre\squared}.}
	\label{fig:rot_dens}
\end{figure*}

\begin{figure}
	\centering
	\includegraphics[width=7cm]{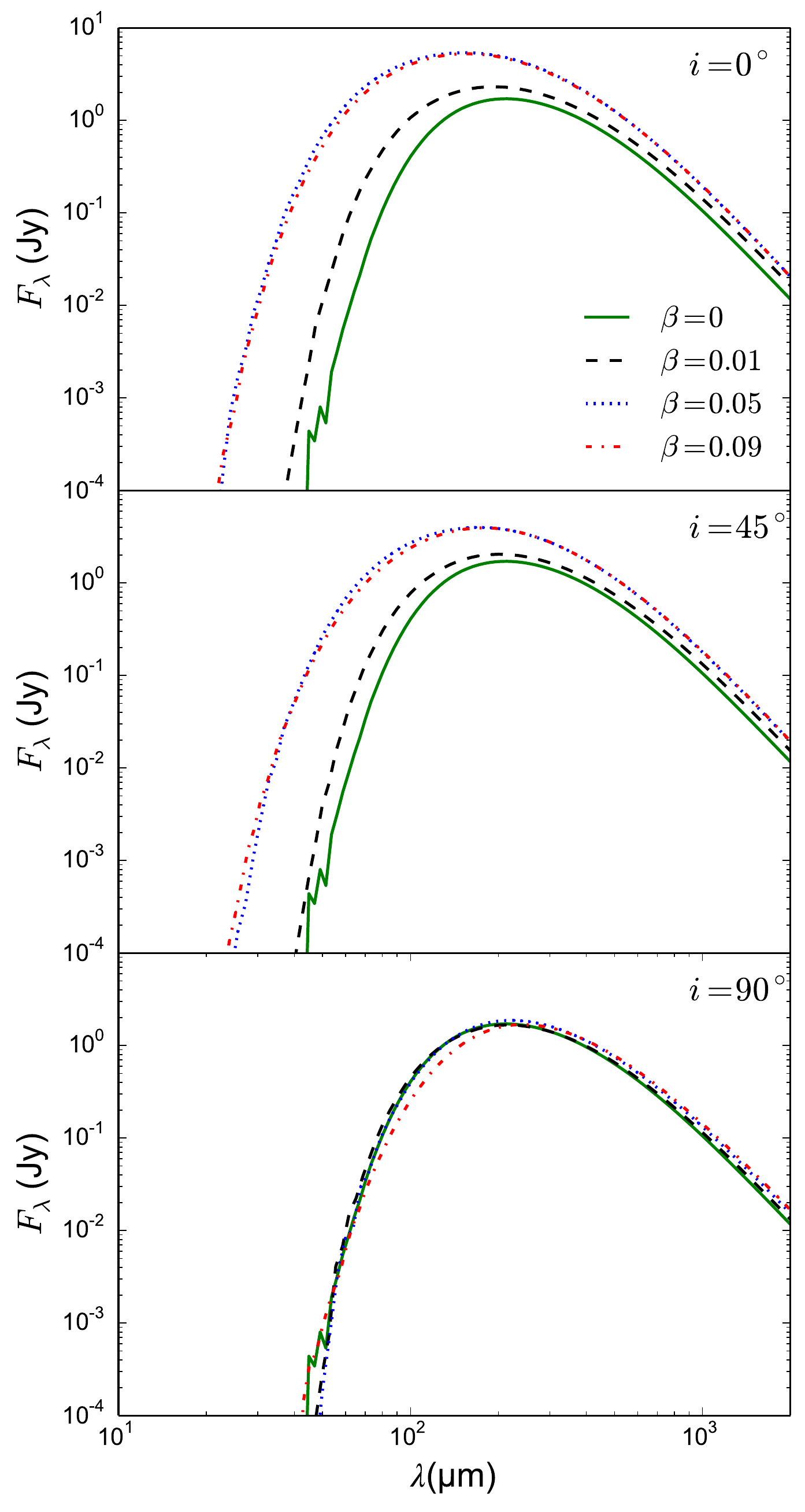}
	\caption{SEDs for collapsing pre-stellar cores with different initial rotation rates for snapshots from the RHD2 model when $\rho_{\mathrm{max}}= 10^{-9}$~\si{\gram\per\centi\metre\cubed}. Rotation rate is defined via $\beta=E_{\mathrm{rot}}/E_{\mathrm{grav}}$. At faster rotation rates, the core is brighter in the far infrared when viewed face-on. At \ang{90} the SEDs of cores with different rotation rates are very similar. The peak of the SED of the $\beta=0.09$ FHSC is at a slightly longer wavelength because the fragmented FHSC lies within a disc so the opacity perpendicular to the rotation axis is greater than the opacity in the non-rotating case.}
	\label{fig:DiffRot}
\end{figure}

In all cases except for the core with a uniform initial temperature of \SI{15}{\kelvin}, the SED peak shifts to shorter wavelengths and the total flux increases as the temperature increases in the central core. The SED begins to deviate from that of a blackbody because the envelope remains cold as the centre heats up. This change in the SED shape may allow more evolved cores to be distinguished from starless cores. However, protostellar cores with initial temperatures of $\gtrsim$~\SI{15}{\kelvin} or exposed to a strong ISRF are not expected to show much evolution so it is unlikely much information about the central object can be obtained from the SED. In this section we have studied only spherically-symmetric cases. These conclusions do not necessarily hold for rotating protostellar cores and we explore those cases further in Section \ref{sec:resultsrotation}.

\subsection{Mass}
\label{sec:resultsmass}

The total luminosity of a dense core should be greater for a larger core mass but the fluxes at different wavelengths will not necessarily scale equally. We therefore simulated the collapse of \SI{0.5}{\solarmass}, \SI{1}{\solarmass}, \SI{2}{\solarmass} and \SI{5}{\solarmass} cores to investigate the variation in SED shape with mass and to determine how to scale model SEDs to observations. The initial radii were increased along with the mass so as to retain a constant initial central density of $\rho_{\mathrm{max}}=$~\SI{1.38e-18}{\gram\per\centi\metre\cubed}. The initial radii ranged from \SI{3700}{\au} for \SI{0.5}{\solarmass} to \SI{8000}{\au} for \SI{5}{\solarmass}. These simulations were performed with the RHD2 model.

The SEDs of cores of different masses in Fig.~\ref{fig:DiffMasswhole} (left) show that increasing the total mass of the protostellar core results in an overall increase in flux, as expected, because there is more radiating material.

To compare the differences in SEDs further we scaled the SEDS of cores of different masses to the SED of the \SI{5}{\solarmass} core (Fig.~\ref{fig:DiffMasswhole}, right). Flux at wavelengths $\lambda > $~\SI{700}{\micro\metre} depends upon the mass, temperature and dust emissivity of the source. Here, the dust properties are the same for each case and the temperatures are very similar, which means the effects of these factors on the flux at $\lambda > $~\SI{700}{\micro\metre} are negligible and we can consider the flux here as being proportional only to the mass. To implement the scaling, we therefore multiplied each SED by a scale factor which was increased until the $\chi^2$ value for wavelengths $\lambda > $~\SI{700}{\micro\metre} was minimised.

When the SEDs are scaled in this way it is apparent that the lower mass cores are brighter at wavelengths $\lambda <$~\SI{200}{\micro\meter} compared to longer wavelengths, an effect also found by \citet{tomida2010dec}. We see from the density profile in Fig.~\ref{fig:MassProfs} that in each case the FHSC has the same radius and that the density only deviates towards the boundary so we would expect a similar flux at $\lambda<$~\SI{100}{\micro\metre} from the hot FHSC. However, the radiation from the FHSC is absorbed by the cooler surrounding gas and most of the observed flux is emitted by the cooler gas further out. For cores of a greater total mass, the FHSC is surrounded by a more massive envelope which leads to disproportionately more longer wavelength emission and increased absorption of shorter wavelengths by the envelope.

The collapse occurs much more quickly for the more massive cores. For the \SI{0.5}{\solarmass} core, the $\rho_{\rm{max}}=$~\SI{e-9}{\gram\per\centi\metre\cubed} snapshot is taken at \SI{35600}{\year} and for the \SI{5}{\solarmass} core it is much sooner at \SI{25700}{\year}. Since the collapse of the \SI{5}{\solarmass} core happens much more quickly, there is less time for the heat to disperse, leading to the hotter central temperature. Despite this, there is not much more infrared flux for the more massive cores because the SED is still dominated by emission from the cooler dust in the envelope.

When comparing to observations, it is useful to be able to scale synthetic SEDs since the sources will have various masses and distances, which may not be known precisely. We find the SED of an FHSC does vary with mass so there will be limits to scaling synthetic SEDs of a core of one mass to observations of a source of a different mass. This is considered further in Section \ref{sec:scaling}.

\subsection{Initial rotation}
\label{sec:resultsrotation}

\begin{figure}
	\centering
	\includegraphics[width=7cm]{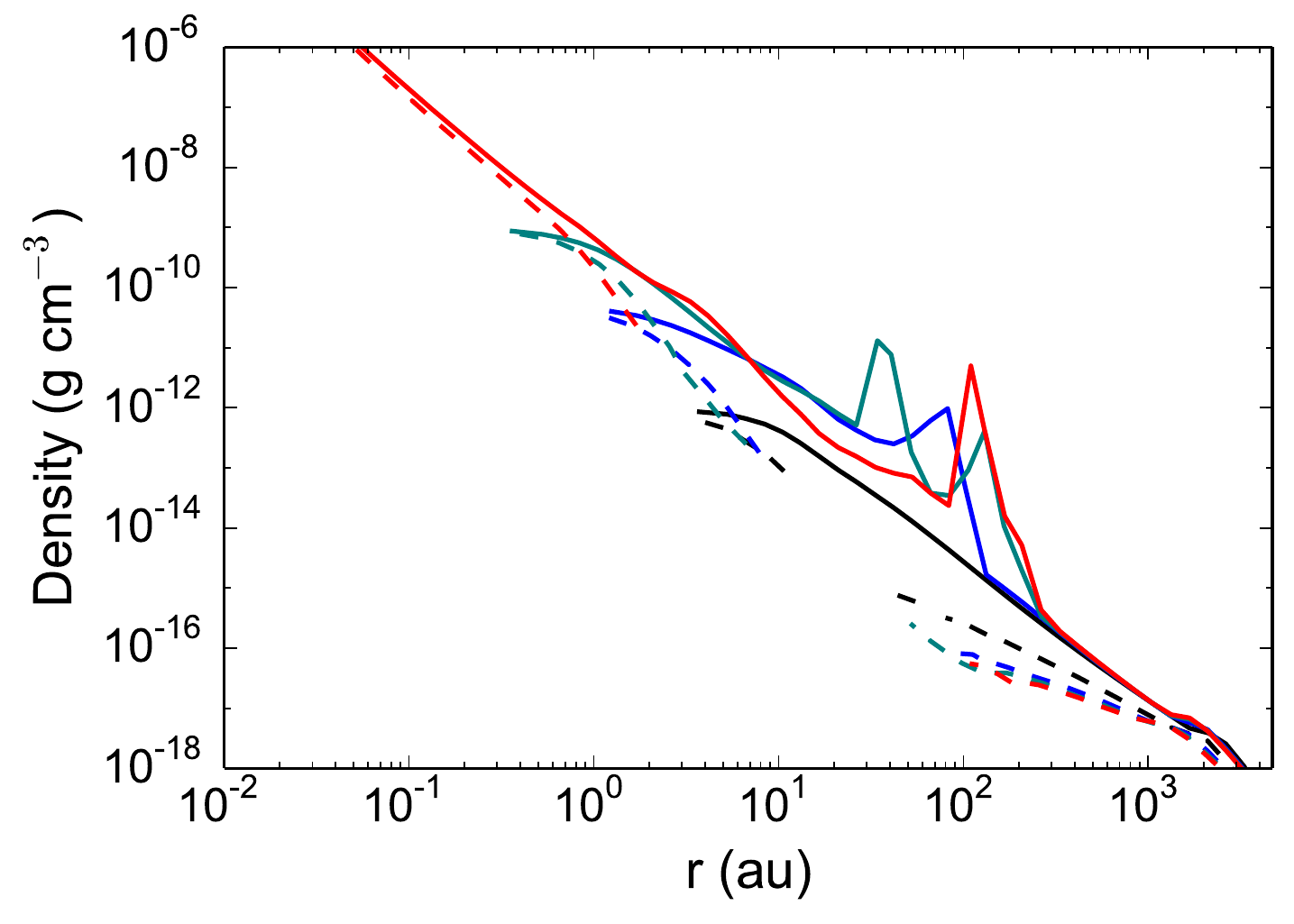}
	\caption{Radial density profiles of the core of initial rotation $\beta=0.09$ at snapshots from the RHD2 model when the maximum density is \SI{1e-12}{\gram\per\centi\metre\cubed} (black), \SI{5e-11}{\gram\per\centi\metre\cubed} (blue),\SI{1e-9}{\gram\per\centi\metre\cubed} (green) and \SI{1e-6}{\gram\per\centi\metre\cubed} (red). Dashed lines show the vertical radial profile (i.e. along the rotation axis) and solid lines show the horizontal radial profile. There is a greater density in the horizontal direction, that is in the plane of the disc structure, and this difference increases as the object evolves. This difference in the density profiles in the horizontal and vertical directions is responsible for the difference between SEDs at different inclinations. The `spike' in the density is due the presence of a fragment and the gaps in the vertical profiles are where there are insufficient particles to calculate an average density.}
	\label{fig:B0910Krho}
\end{figure}

 \begin{figure*}
	\centering
	\includegraphics[width=17cm]{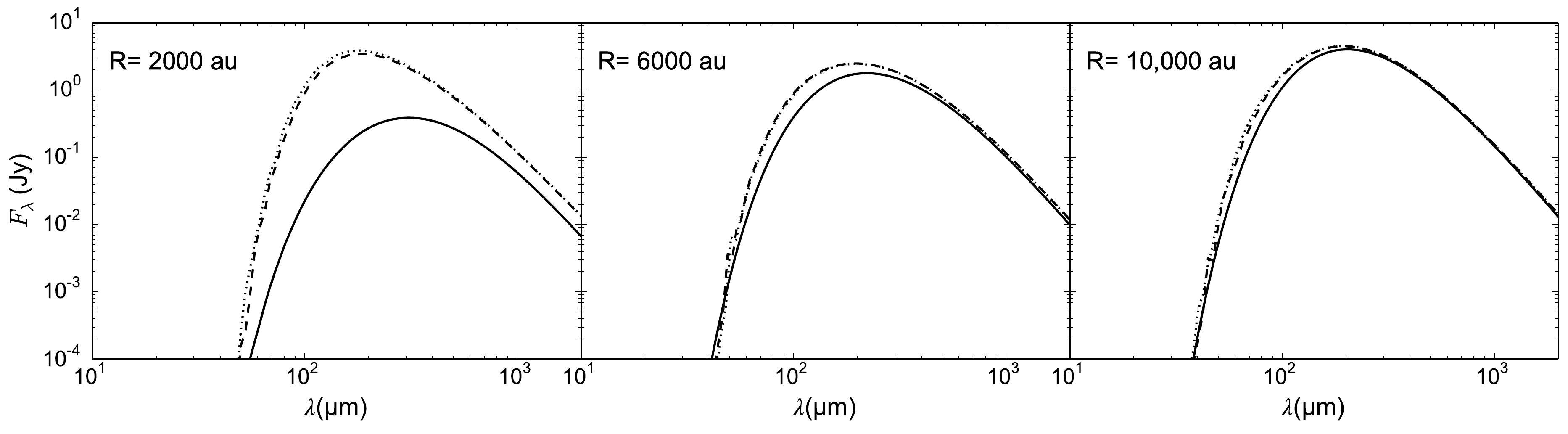}
	\caption{The evolution of SEDs for the collapse of \SI{1}{\solarmass} non-rotating pre-stellar cores with initial radii of \SI{2000}{\au}, \SI{6000}{\au} and \SI{10000}{\au} from left to right. The SEDs are calculated from snapshots at values of maximum density of \SI{1e-12}{\gram\per\centi\metre\cubed} (first collapse, solid lines), \SI{1e-8}{\gram\per\centi\metre\cubed} (late FHSC stage, dashed lines) and \SI{1e-4}{\gram\per\centi\metre\cubed} (second collapse, dotted lines). There is less variation of the SED as the core evolves for cores with larger initial radii. For an initial radius of \SI{10000}{\au} the second collapse SED is observationally indistinguishable from the SED of the early FHSC.}
	\label{fig:DiffRadii}
\end{figure*}

Differing initial rotation rates give rise to various structures. In the non-rotating case the FHSC is spherically symmetric but becomes increasingly oblate with faster rotation. For faster rotation rates, the FHSC may be dynamically rotationally unstable, developing spiral arms and perhaps fragmenting (e.g. \citealt{bate1998,bate2011,st2006}). If these structures lead to differences in the SED then this would provide a probe for the internal structures of pre-stellar cores without requiring high resolution imaging. 

Dense cores are observed to have typical rotation rates corresponding to $\beta=0.02$ \citep{goodman1993} and range between $\sim 10^{-4}$ and $0.07$ \citep{caselli2002}. We simulated SEDs for $\beta=$~0, 0.01, 0.05 and 0.09 to cover the range of observed values, noting that for $\beta<0.01$ the structure is similar to that of a non-rotating ($\beta=0$) core. The SEDs in this section were simulated using RHD2 but we also produced SEDs for rotating cores with RHD1 with an initial temperature of $T_{\rm{initial}}=$ \SI{10}{\kelvin} for comparison with observations. The column density snapshots from the RHD2 simulations in Fig.~\ref{fig:rot_dens} show the structure in each case. The $\beta=0$ case is spherically-symmetric, the $\beta=0.01$ case gives an oblate FHSC, $\beta=0.05$ forms spiral arms and a small fragment before stellar core formation, and $\beta=0.09$ forms spiral arms than fragments into three FHSCs, two of which merge before stellar core formation.
	
The initial rotation rate affects the shape of the SED as the object evolves. From Fig.~\ref{fig:DiffRot} we see that the higher rotation rate leads to an SED with a peak at a shorter wavelength and a shallower slope to shorter wavelengths, at low inclinations (face on). With a faster rotating core there is less obscuring material between the core and observer at low inclinations. This is illustrated by plotting the radial density profiles along the rotation axis and perpendicular to the rotation axis, as shown in Fig.~\ref{fig:B0910Krho}. At radii less than \SI{1000}{\au} there is a clear difference in density at positions in the parallel and perpendicular directions due to the presence of a disc. This difference increases as the object evolves and more gas accretes onto the disc. As shown in Section \ref{sec:SEDmeans}, flux at $\lambda <$~\SI{100}{\micro\meter} is emitted by the warm dust surrounding the FHSC. A lower density outside this region will reduce the opacity and therefore more flux will reach the observer. Hence at faster rotation rates, with a more oblate structure, the SED will peak at shorter wavelengths and there will be more flux at $\lambda <$~\SI{100}{\micro\meter} and detectable flux even at \SI{24}{\micro\meter} in many cases, when viewed at low inclinations (face-on), as we see in Fig.~\ref{fig:DiffRot}~(top).

The reverse is true for observations at high inclinations. The density increases perpendicular to the rotation axis as the disc structure forms, the opacity increases, and less radiation from the regions heated by the FHSC reaches the observer. This results in much less evolution of the SED. Beyond $\beta=0.05$ the pattern is different because the FHSC fragments. The flux reaching an observer will therefore also depend on the location of the fragments because the structure is not axisymmetric.

The SEDs of cores with different rotation rates do have different properties, so it is possible for an SED to provide some information as to the structure of the central regions of a collapsing pre-stellar core. At high inclinations the SEDs of cores with different rotation rates are very similar. In addition, the edge-on SEDs are very similar to those of less evolved cores in the first collapse or early FHSC phase. This means that any differences due to rotation may be difficult to interpret when the inclination is not known. SEDs would need to be combined with other forms of observation to infer the structure.

\begin{figure*}
\centering
	\includegraphics [width=17.5cm]{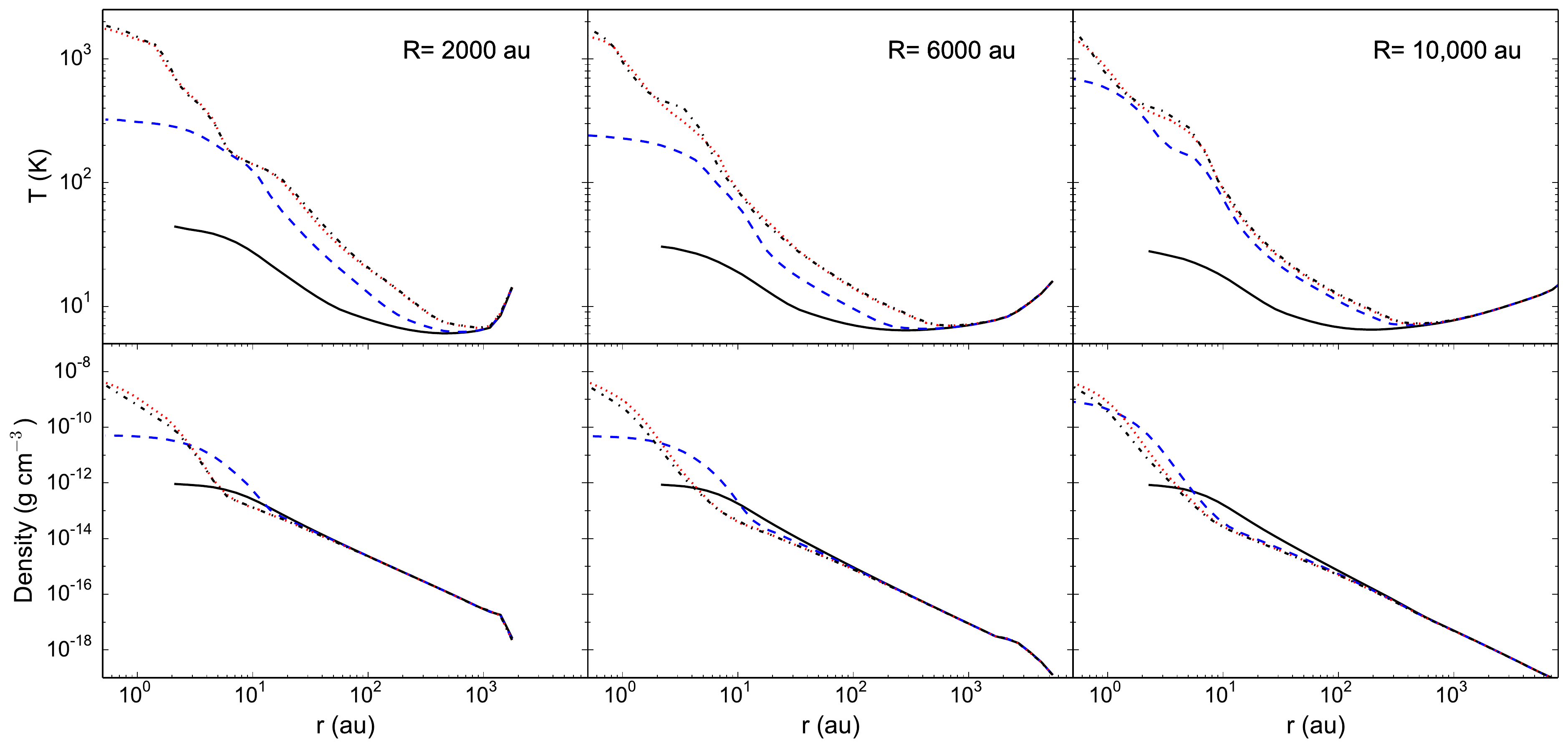}
	\caption{Radial density and temperature profiles during the collapse of \SI{1}{\solarmass} non-rotating pre-stellar cores with initial radii of \SI{2000}{\au}, \SI{6000}{\au} and \SI{10000}{\au} for snapshots when the maximum density was \SI{e-12}{\gram\per\centi\metre\cubed} (solid black lines), \SI{5e-11}{\gram\per\centi\metre\cubed} (blue dashed lines), \SI{1e-8}{\gram\per\centi\metre\cubed} (black dot-dashed line) and \SI{e-4}{\gram\per\centi\metre\cubed} (red dotted lines). The differences in the temperature structures and the extent of the envelopes give rise to differences in their SEDs as seen in Fig.~\ref{fig:DiffRadii}.}
	\label{fig:radiiprofiles}
\end{figure*}

\subsection{Initial radius}
\label{sec:resultsradius}

Cores of different initial radii but equal masses will necessarily have different density structures, which in turn will affect the temperature and opacity. The simulations for exploring the effects of varying the initial radius were performed using the RHD2 model for \SI{1}{\solarmass} cores, with heating from the ISRF.

It is clear from the SEDs of evolving cores with different initial radii shown in Fig.~\ref{fig:DiffRadii} that there is much more variation as the object evolves for the core with the smallest initial radius. As for the case of an initial radius of $r_{\rm{ini}}=$ \SI{4700}{\au} shown in Section~\ref{sec:resultstemp}, the SEDs of the core of $r_{\rm{ini}}=$ \SI{6000}{au} do not show much evolution, with just a small increase in flux between \SI{50}{\micro\meter} and \SI{300}{\micro\metre} and no significant difference between the SEDs of late FHSC and second collapse stages. When the initial radius is increased further there is even less variation of the SED with age of the core. As the initial radius of the core increases, it becomes less unstable to gravitational collapse so the collapse progresses mores slowly. The core of radius $r_{\rm{ini}}=$ \SI{11400}{\au} did not collapse.

The radial density and temperature profiles of each core are shown in Fig.~\ref{fig:radiiprofiles} and we find that the FHSC has the same radius in each case, as was reported by \mbox{\citet{vaytet2017}}. The first collapse SED is coldest for the smallest core. The smallest core is more dense so the ISRF is attenuated more strongly and does not heat the core as deeply as it does for the larger cores. The effect of this is apparent in Fig.~\ref{fig:radiiprofiles} as the temperature falls more sharply near the boundary of the smaller core and the core has a lower minimum temperature. A greater fraction of the mass of the smaller core is cold, giving rise to a colder SED during the early FHSC stage for a smaller core.

The smallest core also has a brighter and more sharply peaked second collapse SED. In all cases, the FHSC heats the envelope out to $\sim$~\SI{700}{\au} so for a smaller core, a larger fraction of the total mass is heated, leading to a warmer SED.

For larger cores we find that the evolution of the central object is almost completely obscured. Unless the core is rotating sufficiently to reduce the opacity significantly along the rotation axis, the application of SEDs to determine the evolutionary state of pre-stellar cores is limited to smaller cores.

\subsection{Magnetic field}
\label{sec:resultsmagfield}

\begin{figure*}
	\centering
	\includegraphics[width=17cm]{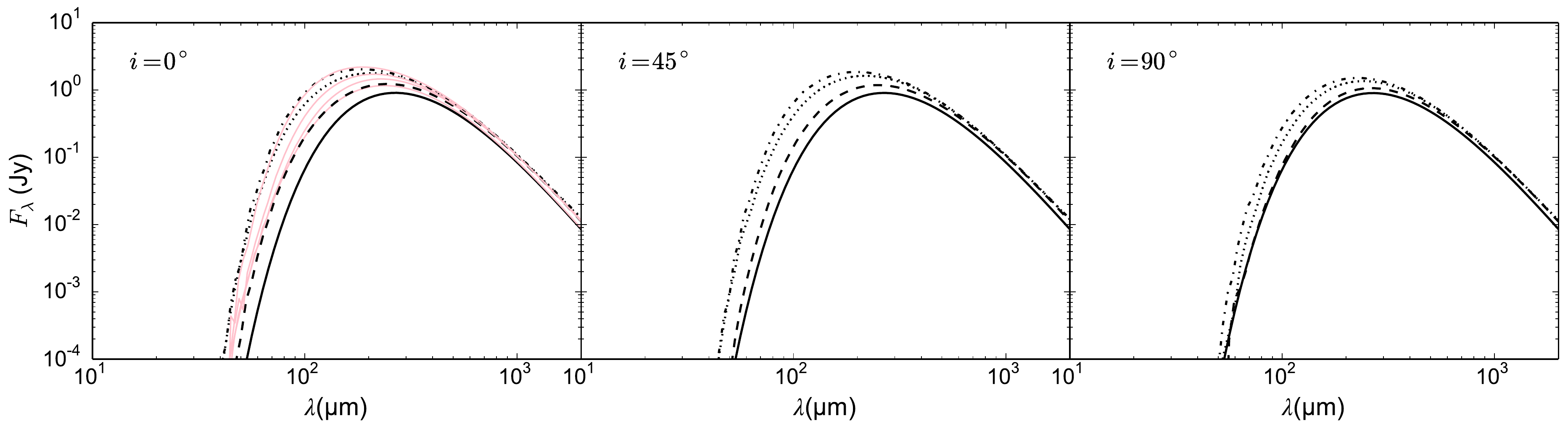}
	\caption{Evolution of the SED at three inclinations for a collapsing \SI{1}{\solarmass} core with the MHD model with initial rotation $\beta=0.005$ and $\mu=5$. The SEDs were calculated from snapshots when the maximum density was \SI{e-12}{\gram\per\centi\meter\cubed}, \SI{5e-11}{\gram\per\centi\meter\cubed}, \SI{1e-9}{\gram\per\centi\meter\cubed} and \SI{e-4}{\gram\per\centi\meter\cubed} from right to left. The SED evolution from the non-rotating RHD2 model are shown in pink in the $i=$~\ang{0} panel. The SED becomes bluer and brighter as the object evolves, however the presence of an outflow does not lead to any additional increase in infrared flux.}
	\label{fig:MuSEDs}
\end{figure*}

\begin{figure*}
	\includegraphics[height=4.8cm]{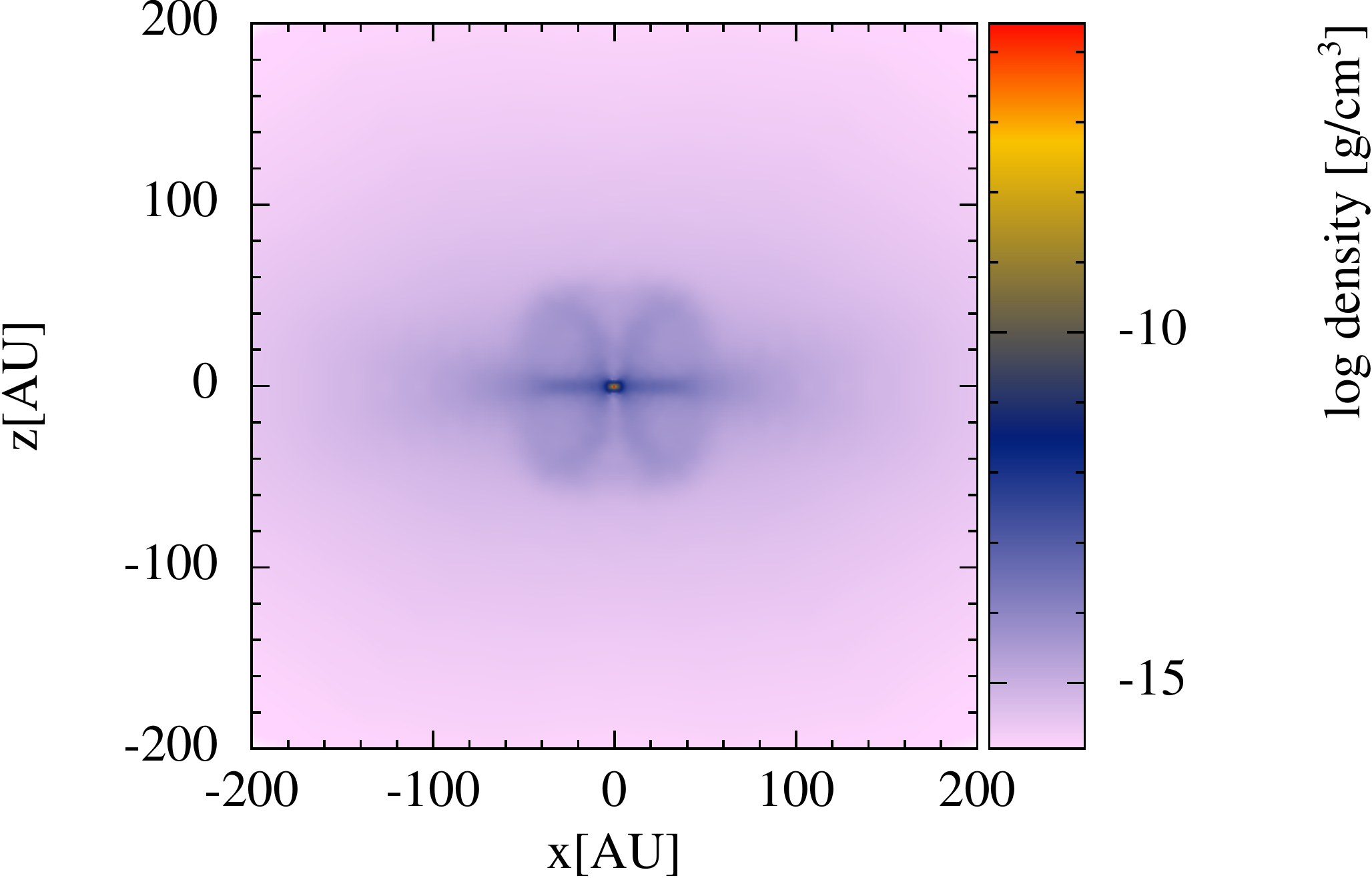}\qquad
	\includegraphics[height=4.8cm]{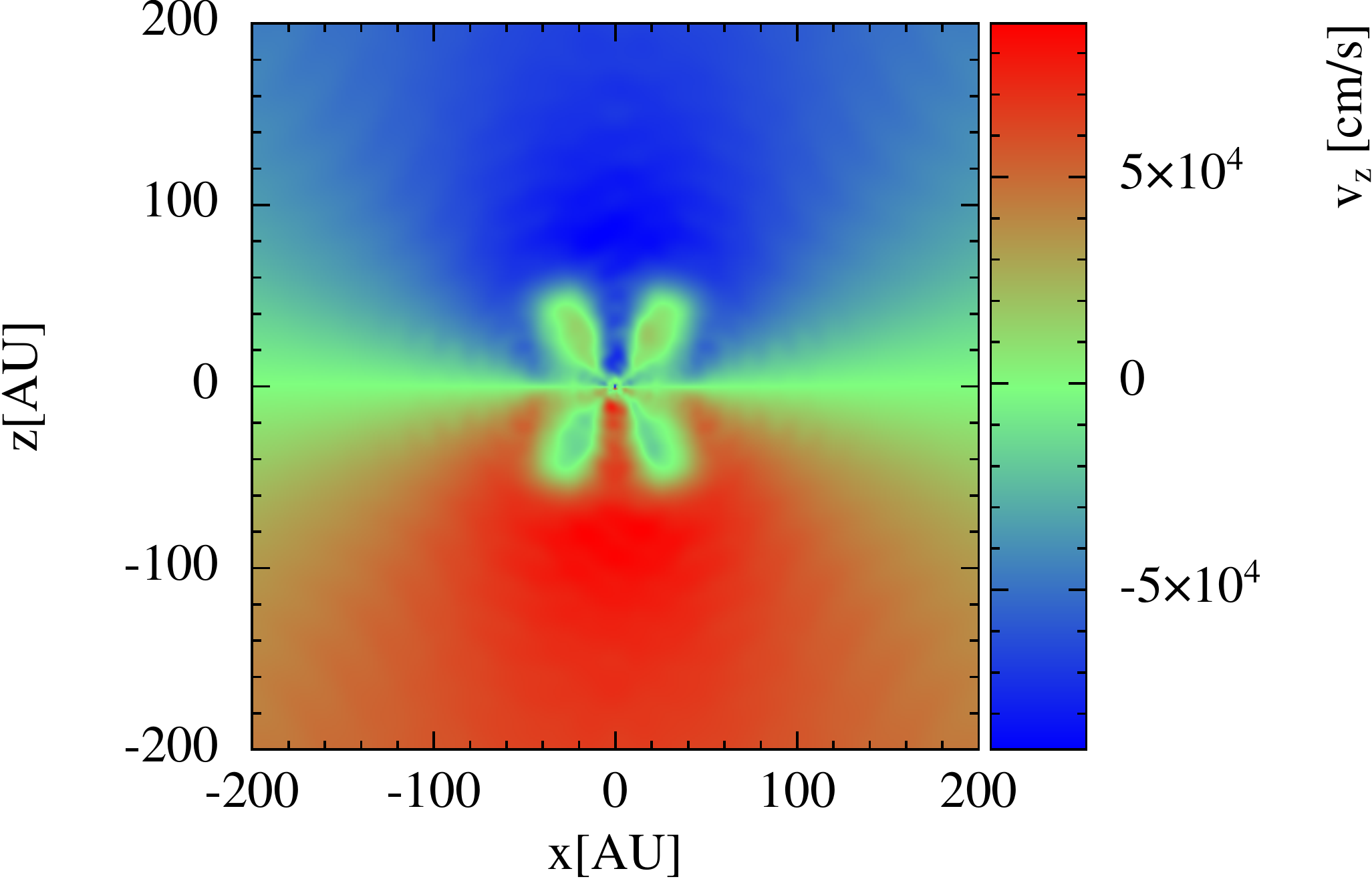}
	\caption{Density slice (left) and vertical velocity, $v_z$, slice (right) for the MHD snapshot used to produce an SED during the second collapse phase when the central density was \SI{e-4}{\gram\per\centi\metre\cubed} (at \SI{30960}{\year}). The initial rotation was $\beta=0.005$ and the initial mass-to-flux ratio was $\mu=5$. At this stage, the outflow has not penetrated far into the cloud and we see that this does not reduce the density along the direction of the outflow and the infall motions are still much greater than outflow motions. Since the outflow affects only a small proportion of the core, the SED is similar to those of non-MHD simulations. The former FHSC extends to a radius of $\sim$~\SI{7}{\au} in this snapshot.}
	\label{fig:MUfigs}
\end{figure*}

We performed MHD simulations of the collapse of a \SI{1}{\solarmass} core with the radiative transfer treatment of RHD2 (see Section \ref{sec:methodRHD12}). These simulations use $\beta=0.005$ and initial mass-to-flux ratios of $\mu=5$ and $\mu=20$, aligned with the rotation axis. Some observed FHSC candidates have substantial short wavelength emission ($\lambda <$~\SI{100}{\micro\metre}), and we wondered whether a magnetically-driven outflow from an FHSC core may be able to boost the observed short wavelength emission compared to that produced by an unmagnetised model. The magnetic field in the $\mu=20$ case was too weak to produce an outflow and so we produced SEDs from snapshots from the $\mu=5$ simulation in which an outflow develops late on in the FHSC phase (c.f. \citealt{bate2014}).

The SED evolution with the MHD simulation is displayed in Fig.~\ref{fig:MuSEDs} and in the $i=$~\ang{0} panel we also plot the SED evolution for RHD2 with $\beta=0$. We see that the SED from the MHD model is initially colder, as was the case with the RHD2 model with the additional boundary column density, but subsequent SEDs are very similar. It is apparent that the presence of an outflow does not increase the flux at short wavelengths. Although the outflow contains a central region of reduced density, the outflow at this early stage is still sufficiently dense to absorb much of the flux from the core. \citet{commercon2012a} also report this effect: looking along the outflow of a magnetised FHSC there was no flux at $\lambda <$~\SI{30}{\micro\meter}. The vertical density slice through the core and the vertical velocities of the gas during the second collapse phase are shown in Fig.~\ref{fig:MUfigs}. These figures show that during the FHSC and second collapse phases, the outflow is still small so only a small proportion of the core is influenced by the outflow and the majority of the envelope structure is similar to the purely hydrodynamical case. It is, therefore, not surprising that the SEDs are similar for models with and without magnetic fields.

\citet{Lewis2015} and \citet{lewis2017} performed MHD simulations for a wide range of initial magnetic field magnitudes and orientations which all produced early protostellar outflows with velocities of $<$~\SI{8}{\kilo\meter\per\second}.  Such slow outflows extended to less than \SI{100}{\au} from the protostar before the stellar core forms (Fig.~\ref{fig:MUfigs}; \citealt{bate2014}). Since both we and \citet{commercon2012a} find that MHD models of FHSCs do not produce SEDs that are significantly different to RHD models, we did not explore SED variation with magnetic field further.

\subsection{Dust grain properties}
\label{sec:resultsgrainsize}

\begin{figure}
\centering
	\includegraphics[width=7cm]{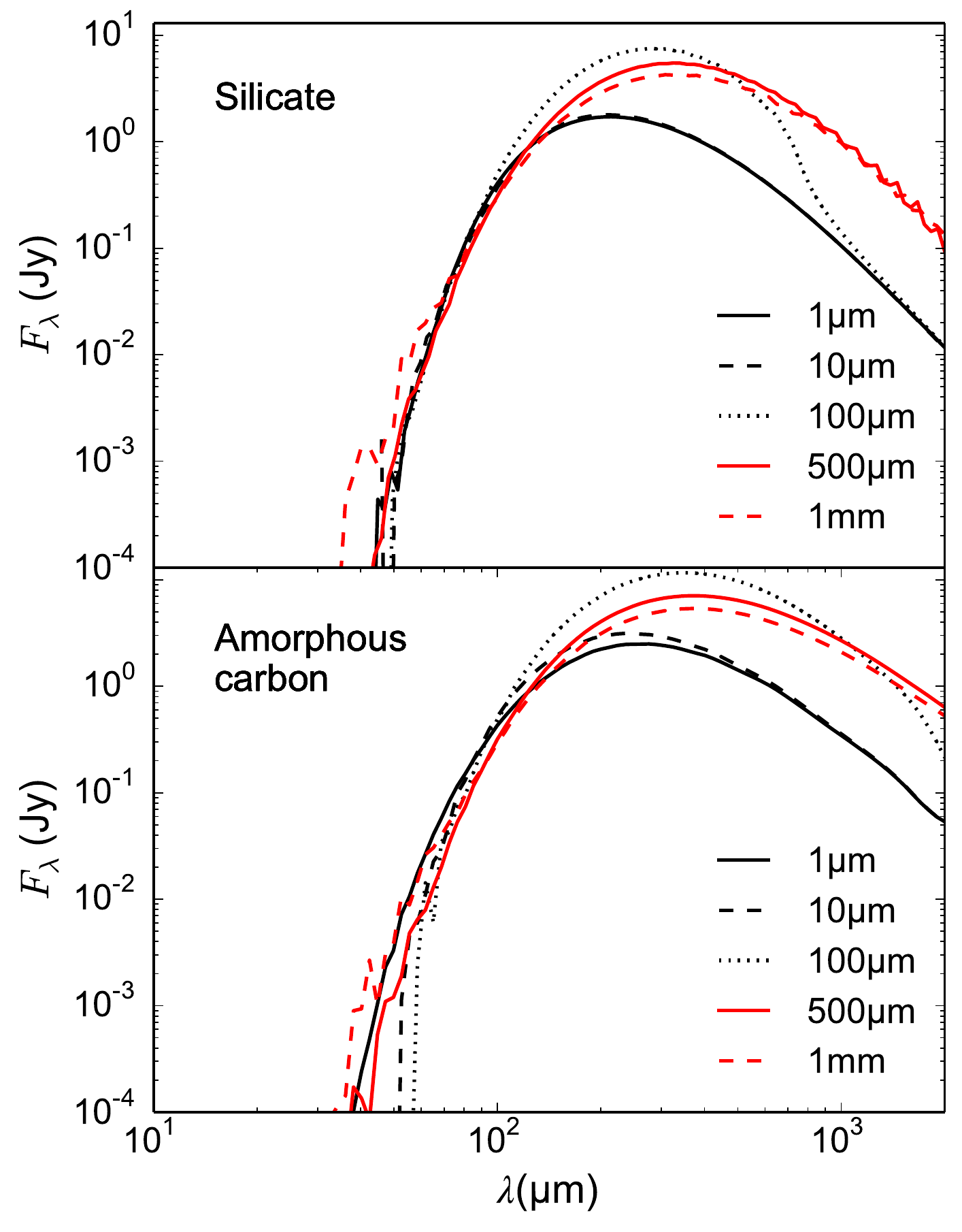}
	\caption{SEDs for a non-rotating FHSC with different maximum dust grain sizes ($a_{\mathrm{max}}$) using the RHD2 model. Top: silicate grain model. Bottom: amorphous carbon grain model.}
	\label{fig:maxgrainsize}
\end{figure}

\begin{table*}
\centering
\caption{Parameters used to produce a range of model SEDs for both the RHD1, with initial temperature \SI{10}{\kelvin}, and RHD2 models. The maximum density $\rho_{\mathrm{max}}$ is used to select specific evolutionary stages as shown in Fig.~\ref{fig:B09BK_rhoevo}. $\beta$ is the initial ratio of rotational energy to gravitational potential energy and a$_{\mathrm{max}}$ is the maximum dust grain size. The combinations of parameters give a total of 1440 synthetic SEDs, noting that $\beta=0$ only has one possible inclination.}
\label{tab:SEDparameters}
\begin{tabular}{lccccc}
\hline
	RHD Model & $\beta$ & $\rho_{\mathrm{max}}$ (\si{\gram\per\centi\meter\cubed}) & Inclination & a$_{\mathrm{max}}$ (\si{\micro\metre}) & Grain Type\\
	\hline
	RHD1 (10K) & 0 & \num{e-12} & \ang{0} & \num{1} & silicates\\
	RHD2 & 0.01 & \num{5e-11} & \ang{45} &\num{10} & graphite\\
	& 0.05 & \num{e-9} & \ang{90} & \num{100} &\\
	& 0.09 & \num{e-8} && \num{200} &\\
	  & & \num{e-6} & & \num{500} &\\
	&	& \num{e-4} & & \num{1000} &\\
	\hline
\end{tabular}
\end{table*}

There is currently significant uncertainty as to the nature of interstellar dust grains. These grains are usually assumed to be no larger than \SI{1}{\micro\metre}, however there have been several suggestions that there may be significant grain growth in dense regions of the ISM (e.g. \citealt{ormel2009, butler2009}). \citet{draine2006} points out that there are regional differences in the wavelength dependence of interstellar extinction which suggests that the grain size distribution does vary.

\mbox{\citet{zubko2004}} compared a variety of interstellar dust models to observational constraints and found that any linear combination of polycyclic aromatic hydrocarbon (PAH) molecules, silicate and amorphous carbon or graphite grains provide good fits to observational constraints. Here we simulate SEDs first with silicate grains and then with amorphous carbon grains. PAH molecules only provide a significant contribution to thermal emission for $\lambda <$~\SI{40}{\micro\metre} \citep{zubko2004}, which would only affect the \SI{24}{\micro\metre} observational data points, for which there is little flux for these cool objects.

We assumed a maximum grain size $a_{\mathrm{max}}=$~\SI{1}{\micro\meter} when simulating the SEDs in the previous sections but in this section we present results for when this maximum grain size is varied.

Decreasing $a_{\mathrm{max}} $ to below \SI{1}{\micro\metre} has no effect on the SED, a result that is supported by the finding of \cite{miyake1993} that opacities for $a_{\mathrm{max}} <$~\SI{10}{\micro\meter} are the same as for a distribution of just submicrometre dust particles. The top panel of Fig.~\ref{fig:maxgrainsize} shows the effect on the SED of increasing the maximum dust grain size. For particles with \SI{10}{\micro\meter}~$\lesssim a_{\mathrm{max}}\lesssim$~\SI{500}{\micro\meter} there is a very small decrease in flux at \SI{50}{\micro\meter}~$ \lesssim \lambda \lesssim $~\SI{100}{\micro\meter} but the flux at $\lambda > $~\SI{200}{\micro\meter} substantially increases for $a_{\mathrm{max}} \gtrsim $~\SI{100}{\micro\meter}. For $a_{\mathrm{max}} \gtrsim$~\SI{500}{\micro\meter}, there is a small increase in flux for  $\lambda < $~\SI{70}{\micro\meter} and, again, a substantial increase in flux at $\lambda > $~\SI{200}{\micro\meter}. Increasing $a_{\mathrm{max}}$ has the effect of increasing the millimetre dust opacity, as long as $a_{\mathrm{max}} \lesssim$~\SI{1}{\milli\metre} \citep{dalessio2001}, which could explain the increased flux at the longer wavelengths with the larger grains. At the same time, for grain sizes over \SI{10}{\micro\meter} the albedo, $\omega_{\nu}$, increases quickly from 0 to $\sim$~0.9 for \SI{1}{\centi\metre} grains \citep{dalessio2001} and so there is an additional contribution from scattered light when there are larger grains present. For larger $a_{\mathrm{max}}$, the opacity is reduced for $\lambda \lesssim$~\SI{100}{\micro\metre} \citep{miyake1993}, which allows more flux from the core to reach the observer.

Varying the minimum dust grain size, $a_{\mathrm{min}}$, between \SI{0.001}{\micro\meter} and \SI{0.05}{\micro\meter} with $a_{\mathrm{max}}$ fixed at \SI{1}{\micro\meter} had no effect on the SED. \cite{draine2006} points out that for a grain size distribution of $p=3.5$ most of the mass is contained in the large grains and so the numbers of large grains is affected little by $a_{\mathrm{min}}$.

The results presented so far use silicate dust grains and next we compare these with SEDs simulated with the amorphous carbon grain model of \citet{zubko1996}, shown in Fig.~\ref{fig:maxgrainsize} (bottom). Compared to the SED produced using silicate grain properties with $a_{\mathrm{max}}=$~\SI{1}{\micro\metre}, the flux drops off less steeply either side of the peak with amorphous carbon with the result that there is more flux at both the shortest and longest detectable wavelengths. For $a_{\mathrm{max}}=$~\SI{10}{\micro\metre}, the flux at $\lambda < $~\SI{100}{\micro\meter} decreases. Flux at $\lambda > $~\SI{100}{\micro\meter} increases significantly for grains of $a_{\mathrm{max}}=$~\SI{100}{\micro\metre} and larger, with the $\lambda < $~\SI{100}{\micro\meter} flux returning to similar values as for $a_{\mathrm{max}}=$~\SI{1}{\micro\metre} when $a_{\mathrm{max}}>$~\SI{500}{\micro\metre}.

The SED is very sensitive to grain size in the presence of large dust grains. The exact composition and size distribution of dust grains in pre-stellar cores is still largely unknown, therefore we consider a range of maximum sizes of both silicate and amorphous carbon grains when comparing to observed SEDs.

\section{Comparison to observations}
\label{sec:resultscomparison}

In this section we describe the set of model SEDs we have produced to compare to the observations of FHSC candidates. We then discuss the ten FHSC candidates and the results of fitting models to those observed SEDs.

\subsection{Selection of models}
\label{sec:modselection}

Simulations were performed for both RHD1 and RHD2 for four different initial rotation rates. Both RHD1 and RHD2 models were included to ascertain whether the more complex treatment of radiative transfer would provide better fits to observations. For RHD1, a uniform initial temperature of \SI{10}{\kelvin} was used. Fig.~\ref{fig:B09BK_rhoevo} shows the evolution of the maximum density from first to second collapse. The FHSC phase lasts from $\sim$~\SI{4e-12}{\gram\per\centi\meter\cubed} to $\sim$~\SI{5e-8}{\gram\per\centi\meter\cubed}, but may commence at a higher maximum density for lower rotation rates. Six snapshots were selected from each RHD model as indicated in Fig.~\ref{fig:B09BK_rhoevo} to allow comparison of the SED as the core evolves and also between SEDs of different cores at similar evolutionary stages.

Dust properties vary across different regions so we chose to simulate SEDs for a selection of different properties. Model SEDs which use amorphous carbon grains peak at longer wavelengths than the equivalent SED using silicate grains, which more closely matches the locations of some of the observed SED peaks. We chose the smallest value of $a_{\mathrm{max}}=$~\SI{1}{\micro\metre} because reducing it below 1um had no effect and current observations indicate a maximum interstellar grain size of at least \SI{1}{\micro\metre}. We included SED models with dust grains $a_{\mathrm{max}}>$~\SI{1}{\micro\metre} to try to reproduce the `wider' SED of SerpS-MM22 and the infrared-bright SEDs of Per-Bolo 58 and Chamaeleon-MMS1.

The range of parameters for which we simulated SEDs is outlined in Table~\ref{tab:SEDparameters} and this gives a total of 1440 synthetic SEDs which we compared with observations.

\begin{figure}
\centering
\includegraphics[width=7cm]{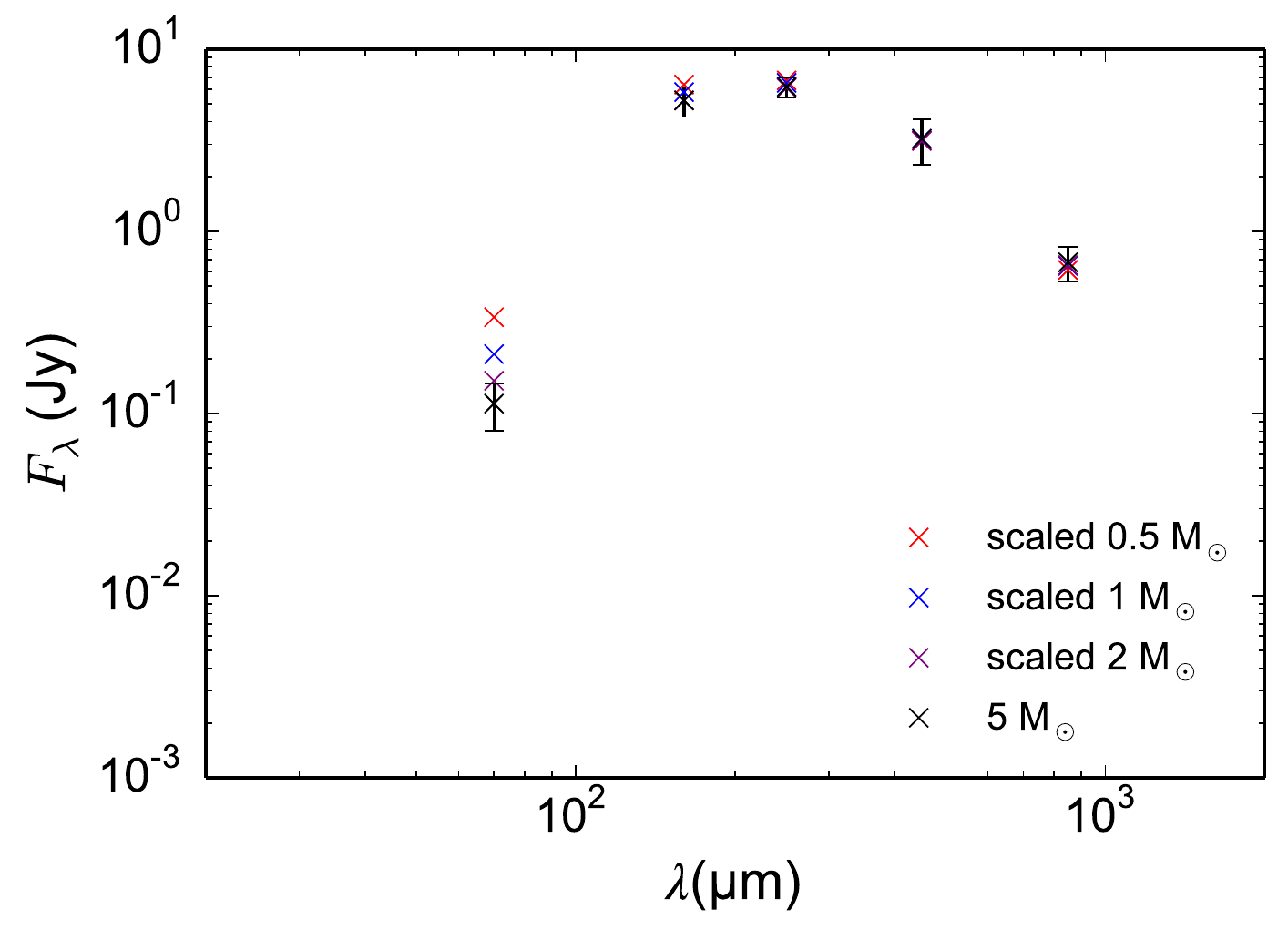}
\caption{Monochromatic fluxes calculated from the SEDs of different mass FHSC models scaled to fit the SED of the \SI{5}{\solarmass} FHSC via the {$\protect\chi^2$} method described in Section~\ref{sec:fittingmethod}. Error bars for the mean error of a selection of real, observed sources are plotted at the corresponding wavelengths. At \SI{70}{\micro\metre} the scaled model fluxes lie outside the error bar which indicates that there is a limit to the scaling of a factor of $\sim2$, beyond which the SEDs would be observationally distinguishable at $\lambda \lesssim$~\SI{100}{\micro\metre}.}
\label{fig:masserr}
\end{figure} 

\begin{figure}
\centering
\includegraphics[width=7cm]{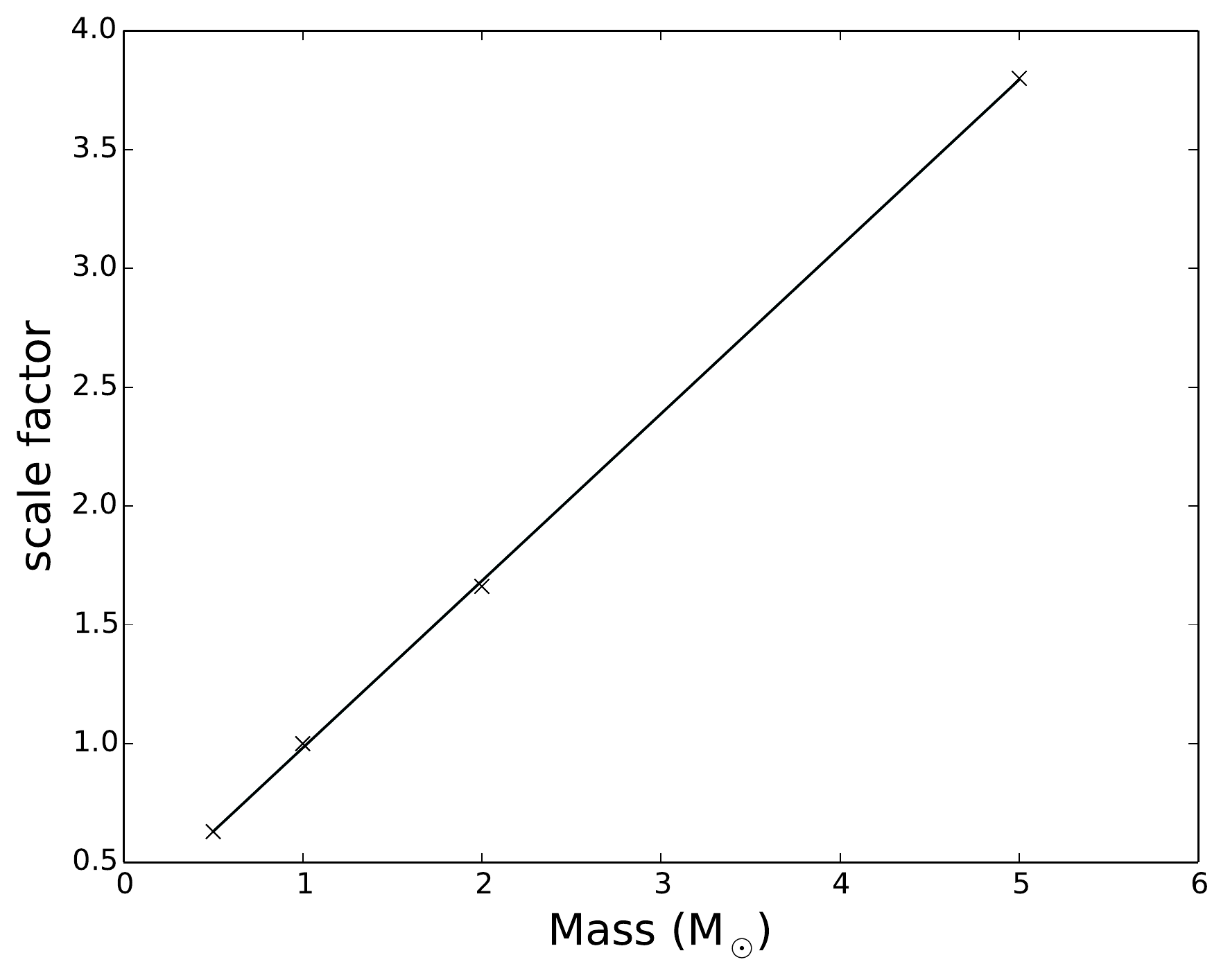}
\caption{Scale factors necessary to scale the \SI{1}{\solarmass} SED to the SEDs of three cores of different masses. Here we assume the same values of fractional uncertainty as for Fig.~\ref{fig:masserr}. A linear fit is calculated and plotted alongside.}
\label{fig:linscalefit}
\end{figure}

\subsection{Scaling model SEDs}
\label{sec:scaling}

The difference between SEDs of varying mass cores described in \ref{sec:resultsmass} indicates that there are restrictions to using scaled SEDs to fit observations. To assess the validity of using scaled SEDs we applied the $\chi^2$ fitting method to scale the SEDs of \SI{0.5}{\solarmass}, \SI{1}{\solarmass} and \SI{2}{\solarmass} cores during FHSC phase to the SED of a \SI{5}{\solarmass} core also during FHSC phase. The monochromatic fluxes derived from the model SEDs of different mass cores at the wavelengths of the Serpens South observations (except \SI{24}{\micro\meter}) were multiplied by an increasing scale factor until the $\chi^2$ value was minimised. The scaled monochromatic fluxes are plotted with the monochromatic fluxes of the \SI{5}{\solarmass} SED in Fig.~\ref{fig:masserr}. The average errors of the observations of eight sources at these wavelengths are plotted with the synthetic \SI{5}{\solarmass} fluxes.

The flux varies most between the different mass FHSCs at \SI{70}{\micro\meter}. At this point the scaled \SI{2}{\solarmass} SED lies just outside the observational uncertainty, indicating that a mass difference of up to a factor of $\sim2$ is indistinguishable observationally when scaled. From this we conclude that it is acceptable to scale \SI{1}{\solarmass} SEDs to fit observations of different sources as long as the mass of the object lies in the range \SI{0.5}{\solarmass}~$\lesssim M \lesssim $~\SI{2}{\solarmass}.  Models fitted with scale factors outside this range must be treated with caution, particularly at $\lambda \lesssim$~\SI{100}{\micro\metre}.

There is also a distance contribution to the scaling factor which follows the inverse square relation for luminosity and distance. The distance contribution for the scaling of the SEDs simulated at a distance of \SI{260}{\parsec} to observations can be estimated using
\begin{equation}
\label{eq:distscaling}
	\mathrm{scaling~factor} = \Bigg(\frac{260~\mathrm{pc}}{D}\Bigg)^2
\end{equation}
where $D$ is the distance to the source. The scale factor will be $>1$ for sources close than \SI{260}{\parsec} and $<1$ for sources further than \SI{260}{\parsec}.

The approximate distance to observed FHSC candidates is usually known. It is therefore possible to separate the mass and distance contributions to the scale factor. The model SEDs used for fitting to observations are all from \SI{1}{\solarmass} cores and equation \ref{eq:distscaling} will give the contribution of the distance to the scaling factor. Dividing the scale factor by this distance contribution will then provide the approximate contribution from the mass.

An estimate of the relation between mass and scale factor was obtained from scaling the \SI{1}{\solarmass} SED to SEDs of three different values of core mass. The optimum scale factors for each value of mass are fitted by a linear relation of $\mathrm{scale ~factor} = 0.7M+0.3$ as shown in Fig.~\ref{fig:linscalefit}. It is the mass obtained from this relation that needs to lie in the range \SI{0.5}{\solarmass}~$\lesssim M \lesssim $~\SI{2}{\solarmass} for the fitting to be robust. There are often large uncertainties in the distance measurement so these mass values are only very approximate. We also note that it would be difficult to compare these values to observational estimates of core masses, which may include material contained within a smaller or larger radius.

\subsection{Observations of FHSC candidates}
\label{sec:FHSCobs}

We attempt to fit model SEDs to the observed SEDs of B1-bN and B1-bS \citep{pezzuto2012,hirano2014}, Per-Bolo 58 \citep{hatchell2005, enoch2006, schnee2010,enoch2010}, Chamaeleon-MMS1 \mbox{\citep{belloche2006,tsitali2013,vaisala2014}} and CB17-MMS \citep{chen2012}. Per-Bolo 58 and Chamaeleon-MMS1 are brighter in the infrared than the other FHSC candidates while their SEDs peak at longer wavelengths. The B1-b sources and CB17-MMS are undetected at \SI{24}{\micro\meter} and \SI{70}{\micro\meter}.

We also attempt to fit the SEDs of five newly identified FHSC candidates. These are compact submillimetre, infrared-faint sources in Serpens South, a region discovered in the observations of \citet{Gutermuth:2008aa} with the \textit{Spitzer} Space Telescope. Further observations (e.g. \citealt{Maury:2011aa,Konyves:2015aa,Nakamura:2011aa,Kirk:2013aa}) have revealed Serpens South to be a dense filamentary cloud, with a large number of protostars and a high star formation rate (SFR) of \SI{23}{\solarmass\per\mega\year}, which makes Serpens South a promising place to look for FHSC candidates. Four of the sources, Aqu-MM2, SerpS-MM22, SerpS-MM19 and Aqu-MM1 are listed in \citet{Maury:2011aa}. The fifth source, HGBS J182941.1-021339, we refer to as K242, after its identification in the catalogue of \citet{Konyves:2015aa}.

The SEDs of the Serpens South FHSC candidates are constructed from \SI{24}{\micro\meter} \textit{Spitzer} MIPS~1, \SI{70}{\micro\metre} and \SI{160}{\micro\metre} (\textit{Herschel} PACS), \SI{250}{\micro\metre} (\textit{Herschel} SPIRE) and \SI{450}{\micro\metre} and \SI{850}{\micro\metre} (SCUBA-2) fluxes. The observations and data reduction are described in Appendix \mbox{\ref{sec:SSobs}} and the positions and monochromatic fluxes of these sources are given in Table~\mbox{\ref{tab:fhsc_fluxes}}. We also discuss the measurements of the distance of Serpens South in  Appendix \mbox{\ref{sec:SSobs}}. The objects are all faint at \SI{24}{\micro\meter}, with only SerpS-MM22 detected. B1-bN, CB17-MMS, SerpS-MM19 and K242 are also undetected at \SI{70}{\micro\metre}. While the four other candidates have SEDs similar to those of cold gas clumps, SerpS-MM22 is unusual in that it is brighter than the others in the infrared while the SED peak is still around \mbox{\SI{200}{\micro\metre}}. 

\subsection{Results of SED fitting}

In this section we describe the results of fitting model SEDs to the ten sources. We define the model SEDs with a $\chi^2$ within a factor of two of the lowest value as providing a `good' fit to give an indication of the variety of models providing a similiar quality of fit to the best fitting model and to account for the minimum $\chi^2$ values of the sources differing by up to two orders of magnitude. We then look for consistencies among those best fitting models. First we discuss five FHSC candidates from the literature and then the sources in Serpens South.

\begin{figure}
	\centering
	\includegraphics[width=6.5cm]{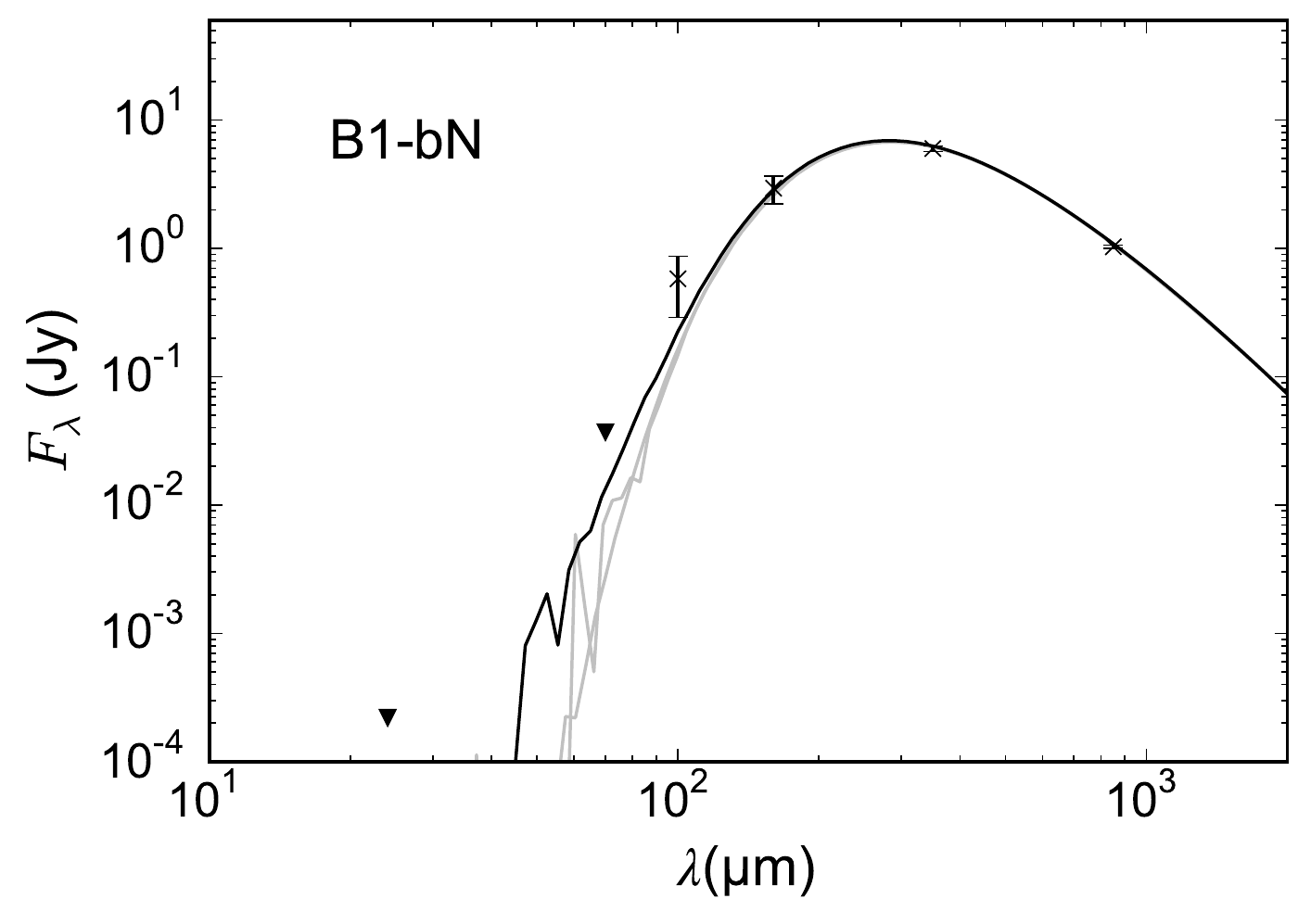}\\
	\includegraphics[width=6.5cm]{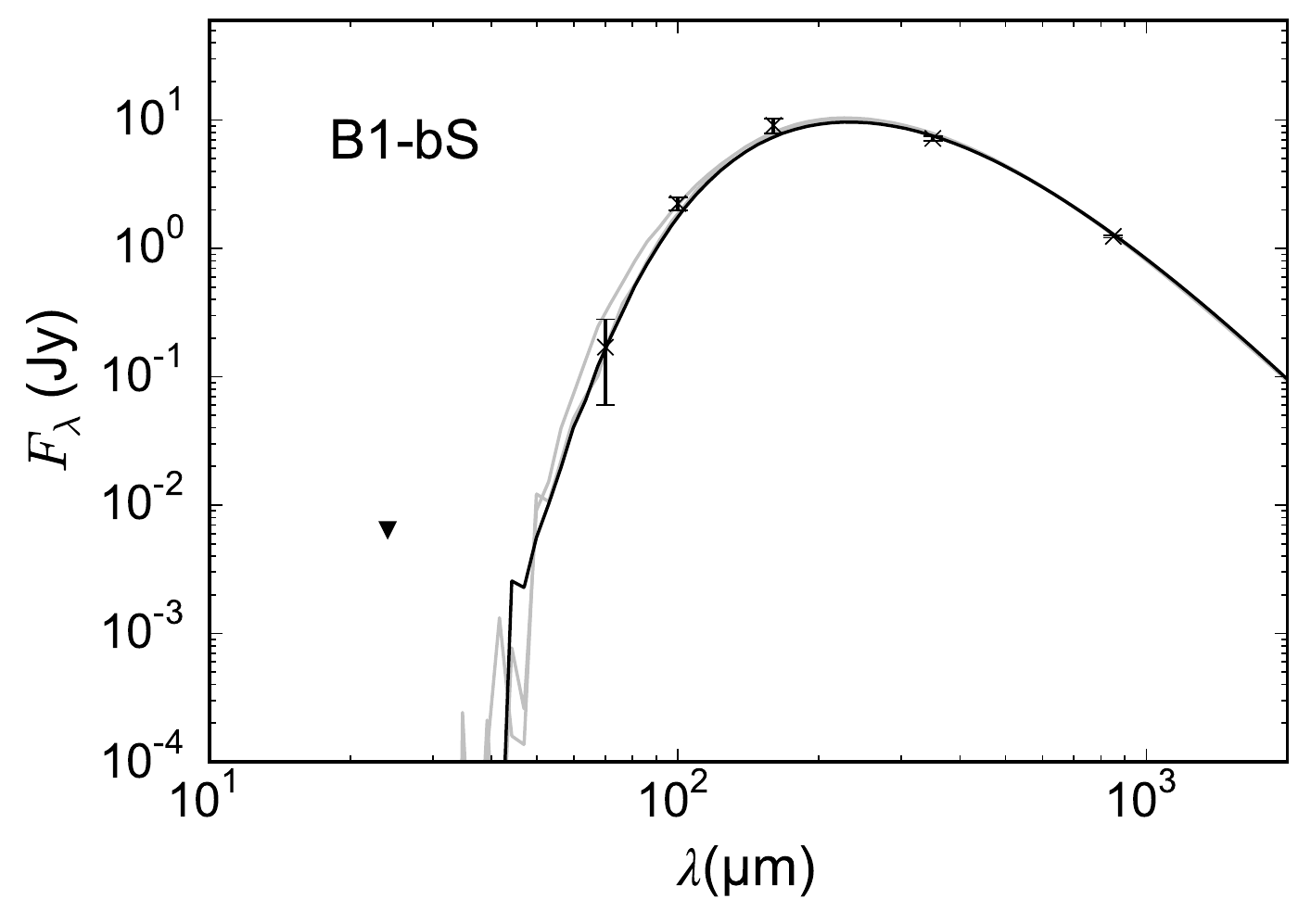}
	\caption{B1-bN (top) and B1-bS (bottom); arrows denote upper limits.  Observational data from \citet{hirano2014} and \citet{pezzuto2012}. B1-bN: Three model SEDs fell within a factor of two of the $\chi^2$ value of the best fitting SED. The best fitting model was from RHD1, $\beta=0.09$, early FHSC, silicate grains, ${a_\mathrm{max}}=~$\SI{10}{\micro\metre} and $i=$~\ang{90}. B1-bS: 3 models fell within a factor of two of the $\chi^2$ value of the best fitting SED. The best fitting model was from RHD2, $\beta=0.05$, early FHSC phase, silicate grains,  ${a_\mathrm{max}}=$~\SI{1}{\micro\metre} and $i=$~\ang{90}.}
	\label{fig:B1bN}
\end{figure}

\begin{figure}
	\centering
	\includegraphics[width=6.5cm]{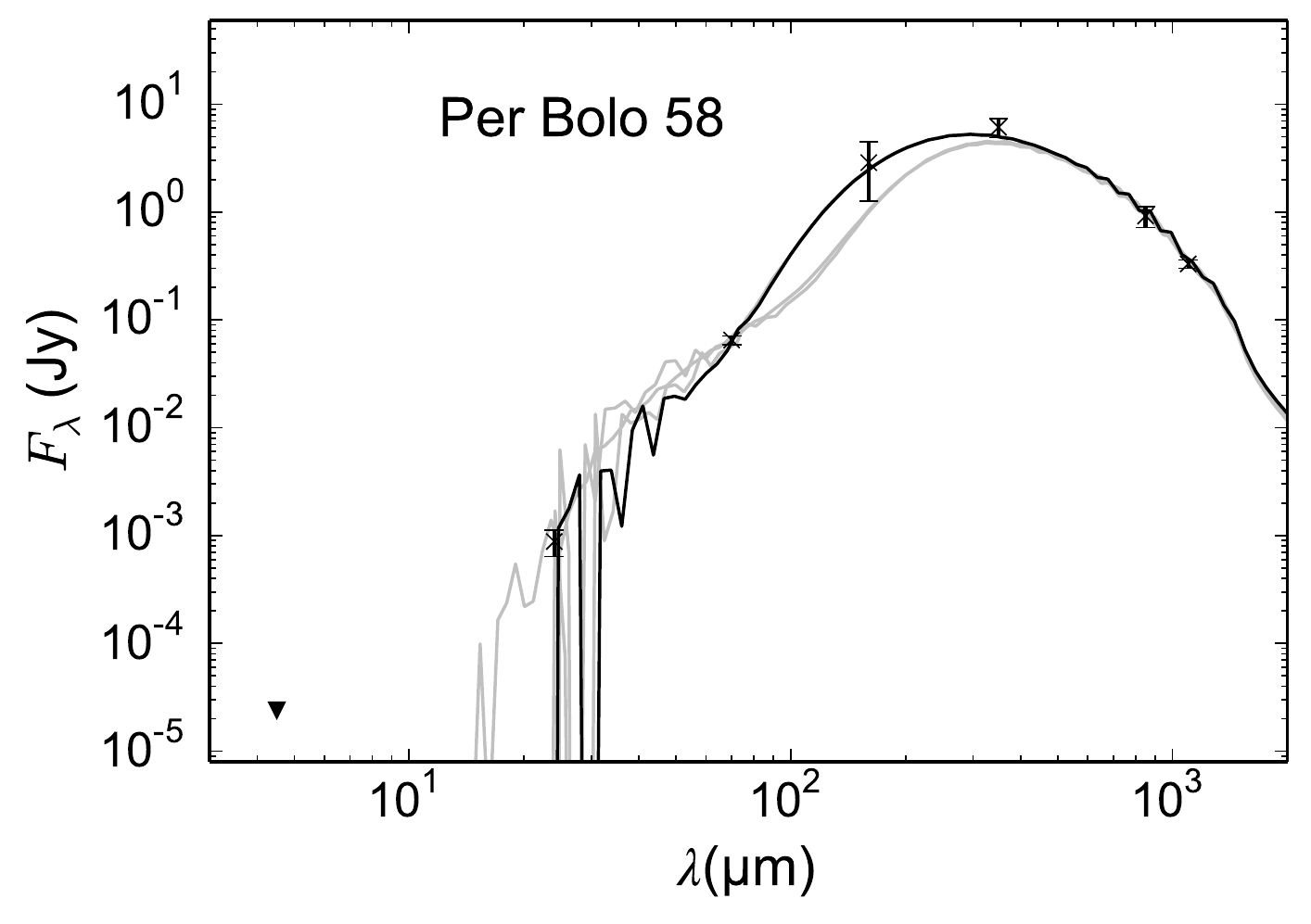}
	\caption{Per-Bolo 58; arrows denote upper limits.  Observational data from \citet{enoch2010}. Just four model SEDs fell within a factor of two of the $\chi^2$ value of the best fitting SED. The best fitting model was from RHD2, $\beta=0.09$, second collapse, silicate grains,  ${a_\mathrm{max}}=$~\SI{200}{\micro\metre} and $i=$~\ang{90}.}
	\label{fig:perbolo58}
\end{figure}

\subsubsection{B1-bN and B1-bS}
We attempted to fit the SEDs of FHSC B1-bN and B1-bS candidates using the photometry presented by \citet{pezzuto2012} and \citet{hirano2014}
\footnote{We use the \SI{70}{\micro\meter} colour corrected Herschel PACS fluxes from \citet{hirano2014} and discard the SPIRE fluxes because the beam sizes are larger than the separation between the sources.}. 
These sources are thought to comprise a wide binary system and there is no evidence that either source contains multiple components \citep{tobin2016}. The SEDs and best fitting models are shown in Fig.~\ref{fig:B1bN}.

B1-bN is fitted equally well by three of the model SEDs, which have minimum $\chi^2$ values between \num{0.2} and \num{0.3}. These are all from the $\beta=0.09$ RHD1 model at a \ang{90} inclination. The three model SEDs are from the FHSC phase and use silicate grains of a maximum size of either \SI{1}{\micro\metre} or \SI{10}{\micro\metre}. The optimum scale factor was \num{5.0} which gives a mass of $\sim$~\SI{5.5}{\solarmass} at the distance of \SI{235}{\parsec} \citep{pezzuto2012} and the effective dust opacity at \SI{850}{\micro\metre} of the best fitting model was \SI{4.0e-3}{\centi\metre\squared\per\gram}. B1-bN therefore looks to be a rotating young FHSC viewed at a high inclination, but is likely to be more massive than the \SI{1}{\solarmass} cores used in our simulations.

B1-bS was well fitted by 3 model SEDs and the best fitting model has a $\chi^2$ of \num{1.2}. All of the best fitting models were from the RHD2 $\beta=0.05$ simulation and viewed at a high inclination.  They were snapshots from the FHSC phase with silicate grains of a maximum size of ${a_\mathrm{max}}=$~\SI{1}{\micro\metre} or  \SI{10}{\micro\metre}. The optimum scale factor was $\sim$\num{6} which gives a mass of $\sim$~\SI{6.6}{\solarmass} at \SI{235}{\parsec} and the \SI{850}{\micro\metre} dust opacity was \SI{3.9e-3}{\centi\metre\squared\per\gram}. Like B1-bN, B1-bS appears to be a rotating FHSC viewed at high inclination located in a more massive core than those used in our simulations.

Slow molecular outflows have been detected for B1-bN and B1-bS \citep{hirano2014, gerin2015} and evidence of pseudo-disc rotation has been reported for B1-bS \citep{fuente2017aa} which are consistent with the sources being FHSCs. The B1-bS outflow is more extended, indicating that B1-bS could be more evolved. In Section \ref{sec:resultsmagfield} we find the SED does not change when there is an outflow from an FHSC so we would expect the results of the SED fitting to be similar had we included MHD models with outflows. \citet{IHsiuLi2017} find no evidence for grain growth in this core which would rule out the \SI{10}{\micro\metre} model SEDs. They suggest that an alternative explanation for this SED shape (in the ~\SI{800}{\micro\metre} to ~\SI{8}{\milli\metre} range) is that the temperature or density increases towards the centre of the core. This is consistent with the structure of the FHSC and envelope that we model here and the corresponding SED appears to fit the fluxes well with a maximum grain size of \SI{1}{\micro\metre}.

\subsubsection{Per-Bolo 58}
Next, we attempted to fit the SED FHSC candidate Per-Bolo 58 in the Perseus molecular cloud \citep{hatchell2005, enoch2006, enoch2010,  schnee2010} and the results are shown in Fig.~\ref{fig:perbolo58}. The optimum scale factor was \num{0.75}, giving a mass of \SI{0.6}{\solarmass} at \SI{260}{\parsec} \citep{enoch2010} and the \SI{850}{\micro\metre} dust opacity was \SI{4.0e-2}{\centi\metre\squared\per\gram}. Per-Bolo 58 was fitted well by just four models, two from RHD1 and two from RHD2, and the minimum $\chi^2$ value was \num{0.67}. These were fast rotating cores at high inclination, in FHSC or second collapse phase. All four used silicate grains of ${a_\mathrm{max}}=$ \SI{200}{\micro\metre}, which was necessary to reproduce the combination of a peak between \SI{200}{\micro\metre} and \SI{300}{\micro\metre} and detectable flux at \SI{24}{\micro\metre} and \SI{700}{\micro\metre}.

\citet{enoch2010} used a distribution of dust grains with ${a_\mathrm{max}}=$ \SI{0.5}{\micro\metre} and were able to produce an SED of this shape with either a wide outflow or a cavity in the envelope to allow the \SI{24}{\micro\metre} to escape. As our models have shown, the outflow from an FHSC is unlikely to reduce the opacity to this extent. Similarly, we do not expect a cavity to be formed since in the lifetime of the FHSC the slow molecular outflow only reaches the innermost regions of the envelope. A slow bipolar outflow was discovered by \citet{dunham2011} with properties that may be consistent with those of either FHSCs or protostars. Since we can only reproduce the SED shape with very large dust grains, this source is likely to be an FHSC only if such grains are present in collapsing pre-stellar cores.It is likely then that Per-Bolo 58 is more evolved than an FHSC. Another possibility is that the source is an unresolved multiple system, however this is ruled out by the multiplicity survey of \citet{tobin2016}. Alternatively, Per-Bolo 58 could well be a very young protostar, perhaps not long after stellar core formation, with a disc and viewed edge-on such that much of the emission from the protostar is reprocessed.

\begin{figure}
	\centering
	\includegraphics[width=6.5cm]{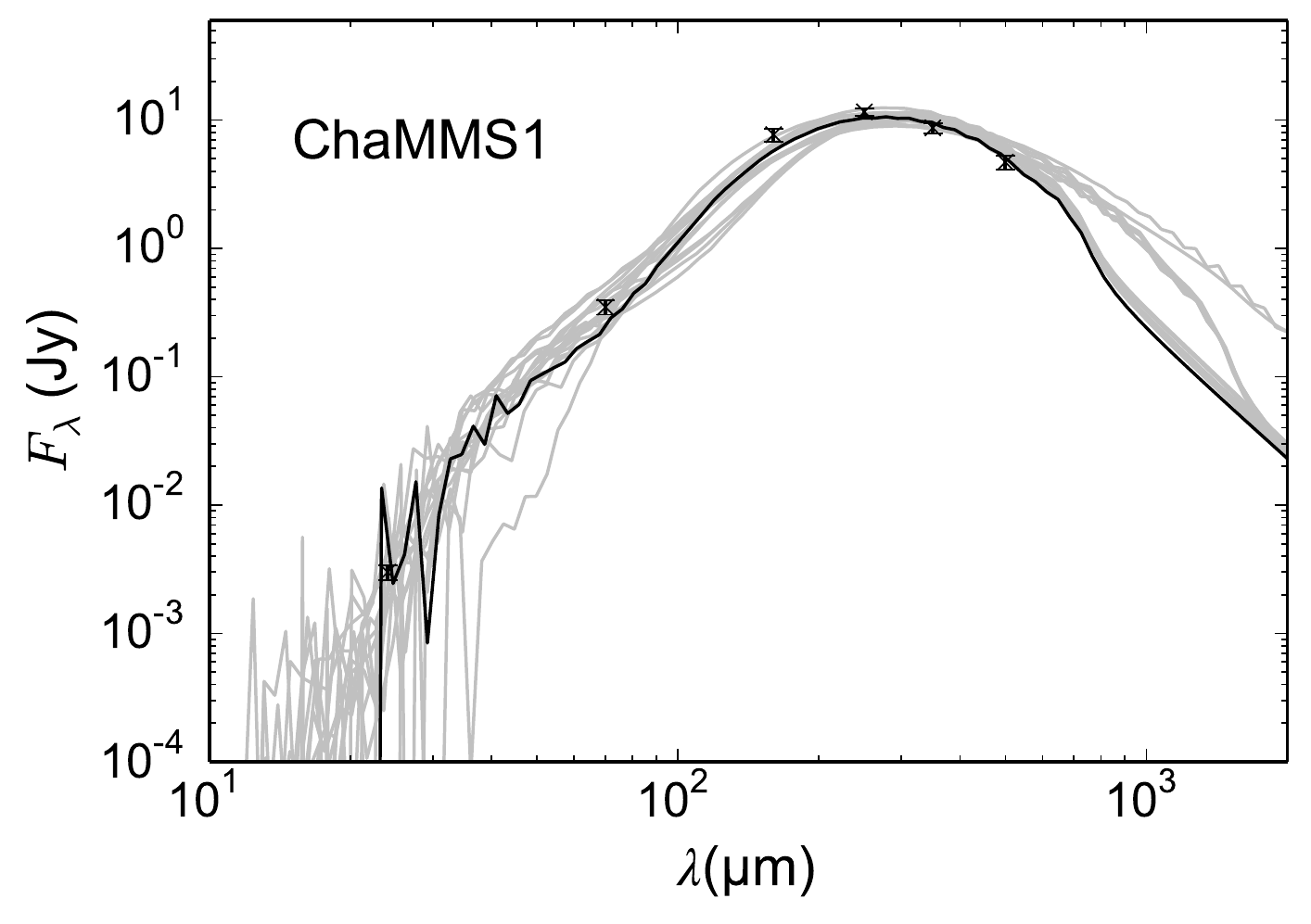}
	\caption{Chamaeleon-MMS1.  Observational data from \citet{vaisala2014}. After SED fits which did not use all six data points were discarded, 17 models fell within a factor of two of the $\chi^2$ value of the best fitting SED. The best fitting model was from RHD2, $\beta=0.09$, late FHSC, silicate grains,  ${a_\mathrm{max}}=$~\SI{100}{\micro\metre} and $i=$~\ang{90}.}
	\label{fig:chamms1}
\end{figure}

\begin{figure}
\centering
	\includegraphics[width=6.5cm]{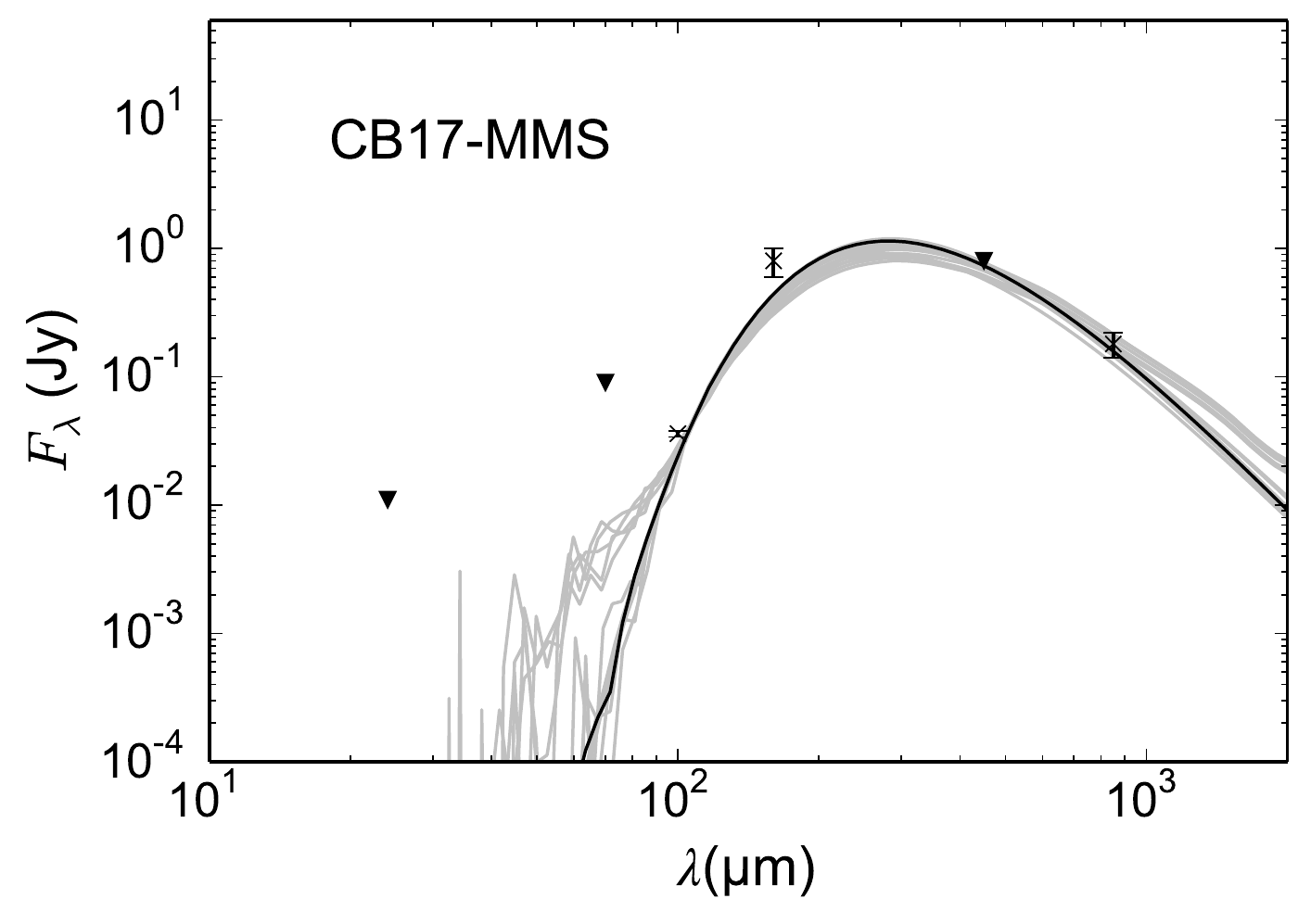}
	\caption{CB17-MMS; arrows denote upper limits.  Observational data from \citet{chen2012}. 22 models fell within a factor of two of the $\chi^2$ value of the best fitting SED. The best fitting model was from RHD1, $\beta=0$, first collapse stage with silicate grains of ${a_\mathrm{max}}=$~\SI{10}{\micro\metre}.}
	\label{fig:CB17MMS}
\end{figure}

\subsubsection{Chamaeleon-MMS1}
The results of the SED fitting for Chamaeleon-MMS1 using the monochromatic fluxes presented in \citet{vaisala2014} and \mbox{\cite{tsitali2013}} are presented in Fig.~\ref{fig:chamms1}. This source was fitted well by 17 models which fall within a factor of two of the best $\chi^2$ value of \num{3.4}. The optimum scale factor was \num{1.3}, giving a mass of \SI{0.3}{\solarmass} at \SI{160}{\parsec} \citep{vaisala2014} and the dust opacity at \SI{850}{\micro\metre} was \SI{7.0e-3}{\centi\metre\squared\per\gram}. All but four of these were from the RHD2. The best fitting models had initial rotation rates of $\beta=0.05$ and $\beta=0.09$ and a \ang{45} inclination was most favoured. There was no clear preference for either FHSC or second collapse phase but first collapse is ruled out. Except for one model with ${a_\mathrm{max}}=$ \SI{1}{\micro\metre}, all the models used \SI{100}{\micro\metre}~$\leq{a_\mathrm{max}}\leq$~\SI{500}{\micro\metre}. These results indicate that Cha-MMS1 is very likely to be an evolved FHSC, rotating quickly and at moderately high inclination.

\citet{vaisala2014} attempted to fit the SED of Cha-MMS1 using the protostar SED fitting tool of \citet{robitaille2006,robitaille07}. The failure to produce a good fit led them to conclude that Cha-MMS1 is more likely to be a second (stellar) core, that is newly formed and accreting, and this is consistent with the results of our fitting which point to an object at least as evolved as a late FHSC. \mbox{\citet{vaisala2014}} also derive the ratio of rotational to gravitation energy to be $\beta=0.07$, which lies midway between our two favoured rotation rates. We find that the best fitting models include large (${a_\mathrm{max}}=$~\SI{100}{\micro\metre} ) dust grains to fit the near straight slope between \SI{24}{\micro\metre} and \SI{160}{\micro\metre}. It may be that a more evolved object with a disc has an SED of a similar shape while assuming only smaller grains as was suggested for Per-Bolo 58.

\subsubsection{CB17-MMS}
From its SED, shown in Fig.~\ref{fig:CB17MMS}, CB17-MMS \citep{chen2012} looks cooler and younger than the previous two sources, with a peak between $\sim$~\SI{150}{\micro\metre} and $\sim$~\SI{250}{\micro\metre}. The minimum $\chi^2$ value was \num{0.44} and 22 models fell within a factor of two of this value. All of these best fitting models were from RHD1, more than half of them used silicate dust grains and most used grains of ${a_\mathrm{max}}=$~\SI{10}{\micro\metre}. Faster rotation rates are favoured at $i=$~\ang{90}, although non-rotating and slower rotating cores at lower inclinations are also among the best fitting models. Half of the best fitting model SEDs are from the first collapse stage. The remaining, more evolved, models are all viewed at $i=$~\ang{90}. The optimum scale factor was $\sim$~0.8 for the models using silicate grains and $\sim$~0.3 for those using carbon grains. This gives a mass of $<$~\SI{0.6}{\solarmass} at \SI{250}{\parsec}. The smaller scale factor leads to a mass contribution below the range studied here. The dust opacity at \SI{850}{\micro\metre} of the best fitting model was \SI{3.9e-3}{\centi\metre\squared\per\gram}.

The SED fitting appears strongly constrained by the \SI{450}{\micro\metre} upper limit, which leads to a poorer fit of the \SI{100}{\micro\metre} and \SI{160}{\micro\metre} detections. The fitting results indicate that CB17-MMS is indeed likely to be an FHSC but could be an even younger collapsing prestellar core without a central object although the \SI{100}{\micro\metre} detection would most likely indicate the former. The model SED scaling indicates a low mass object and, as we show in \ref{sec:scaling}, the SEDs of lower mass cores are relatively brighter at wavelengths $<$~\SI{200}{\micro\metre} and this could explain why the source is detected at \SI{100}{\micro\metre} but not at \SI{450}{\micro\metre}. Kinematic data with a high spatial resolution would determine whether there is a disc and thus whether this is a more evolved, rotating object viewed edge-on or a symmetrical collapsing prestellar core. \cite{chen2012} present CO observations which may show evidence of a slow outflow. If this is correct, the collapsing core models can be ruled out and CB17-MMS remains a likely FHSC.

\begin{figure*}
	\centering
	\includegraphics[width=6cm]{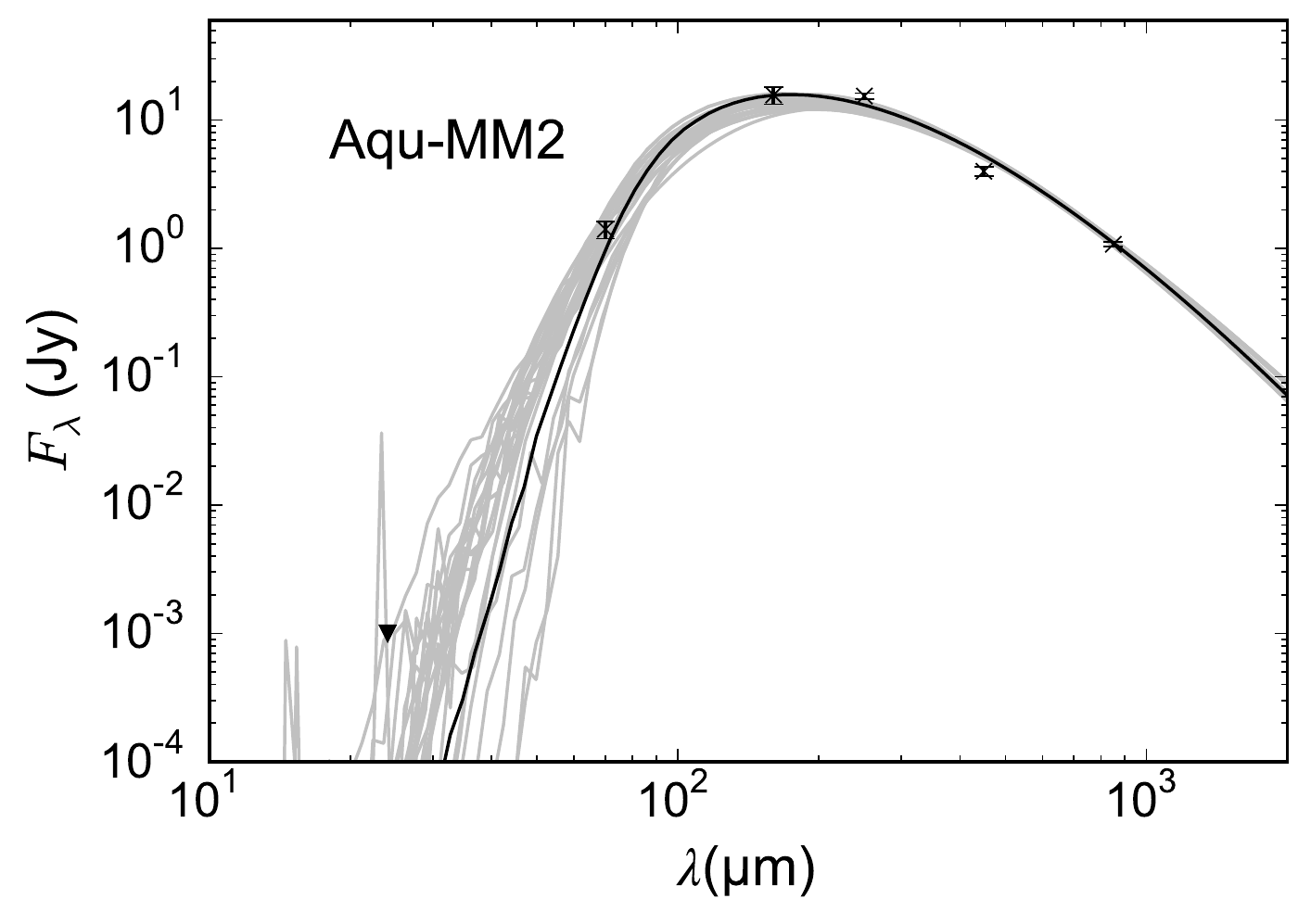}\qquad
	\includegraphics[width=6cm]{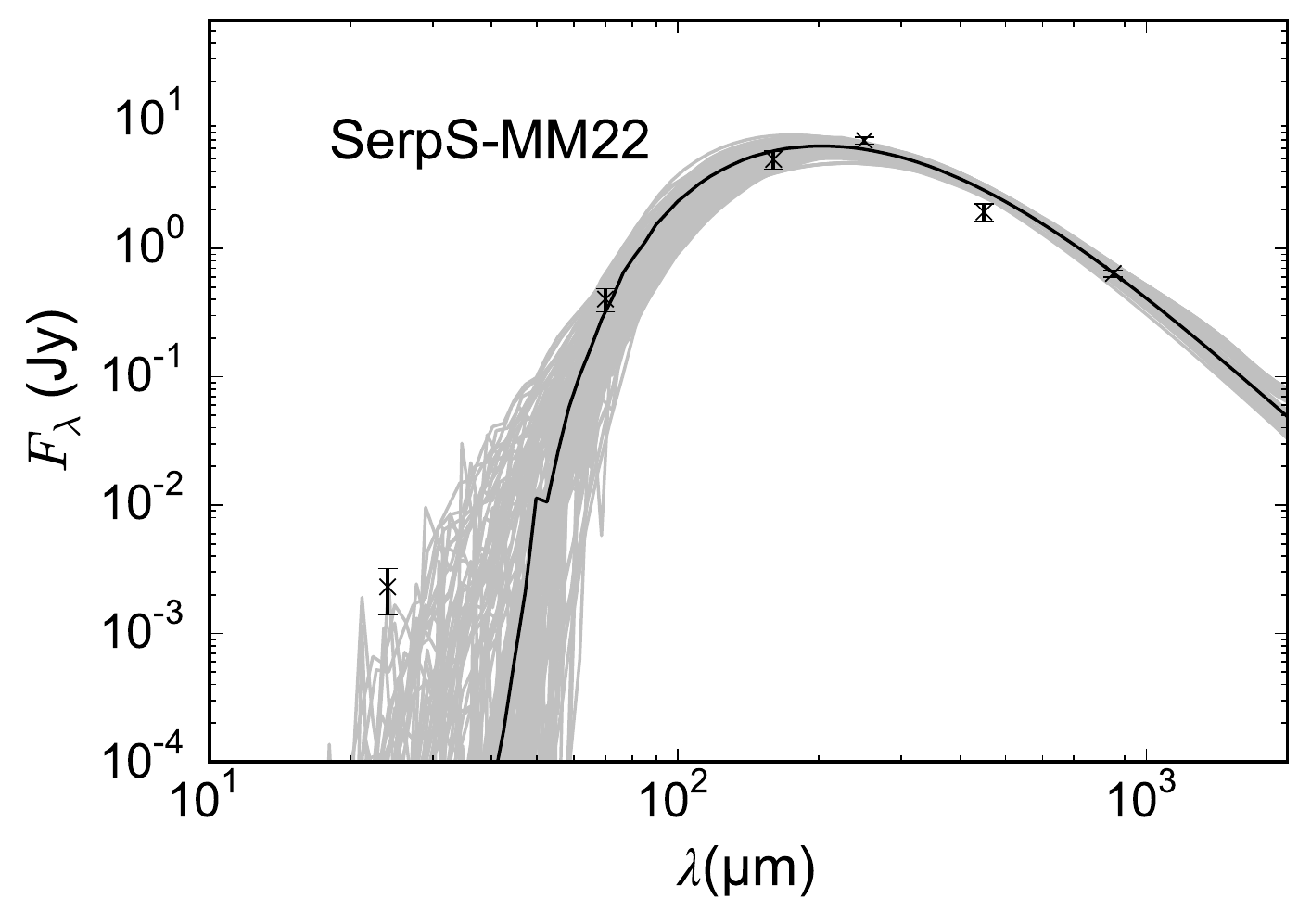}
	\includegraphics[width=6cm]{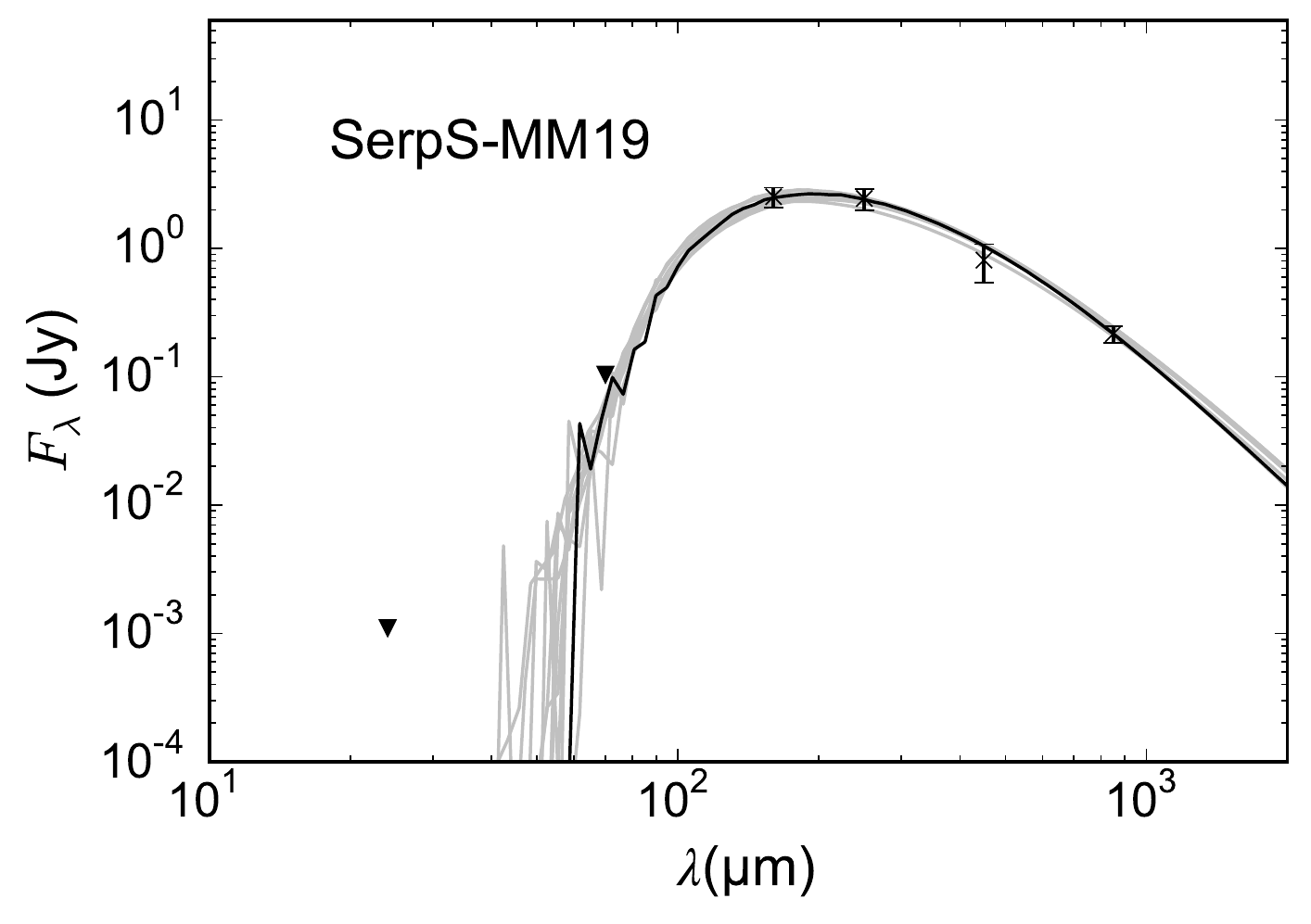}\qquad
	\includegraphics[width=6cm]{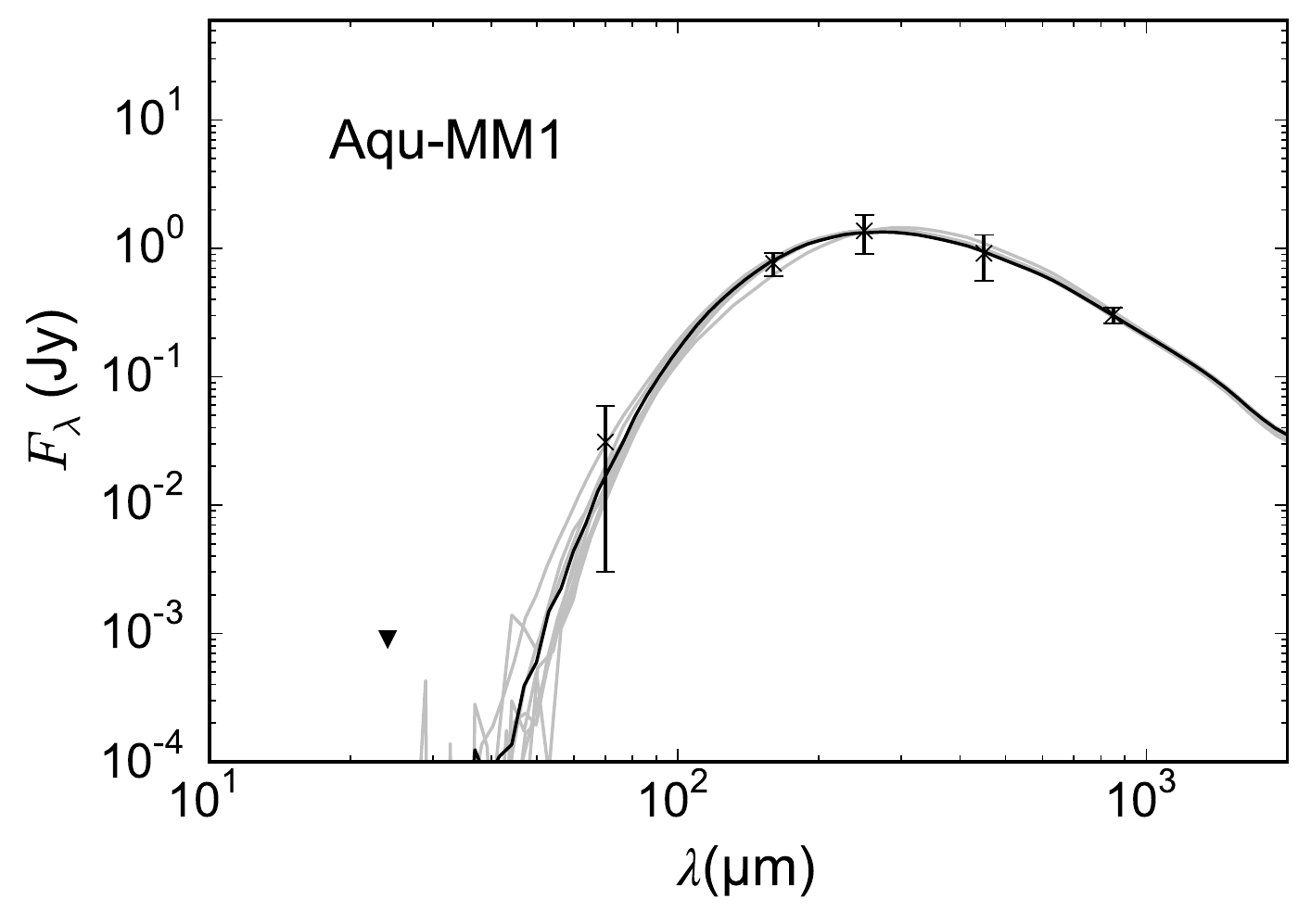}
	\includegraphics[width=6cm]{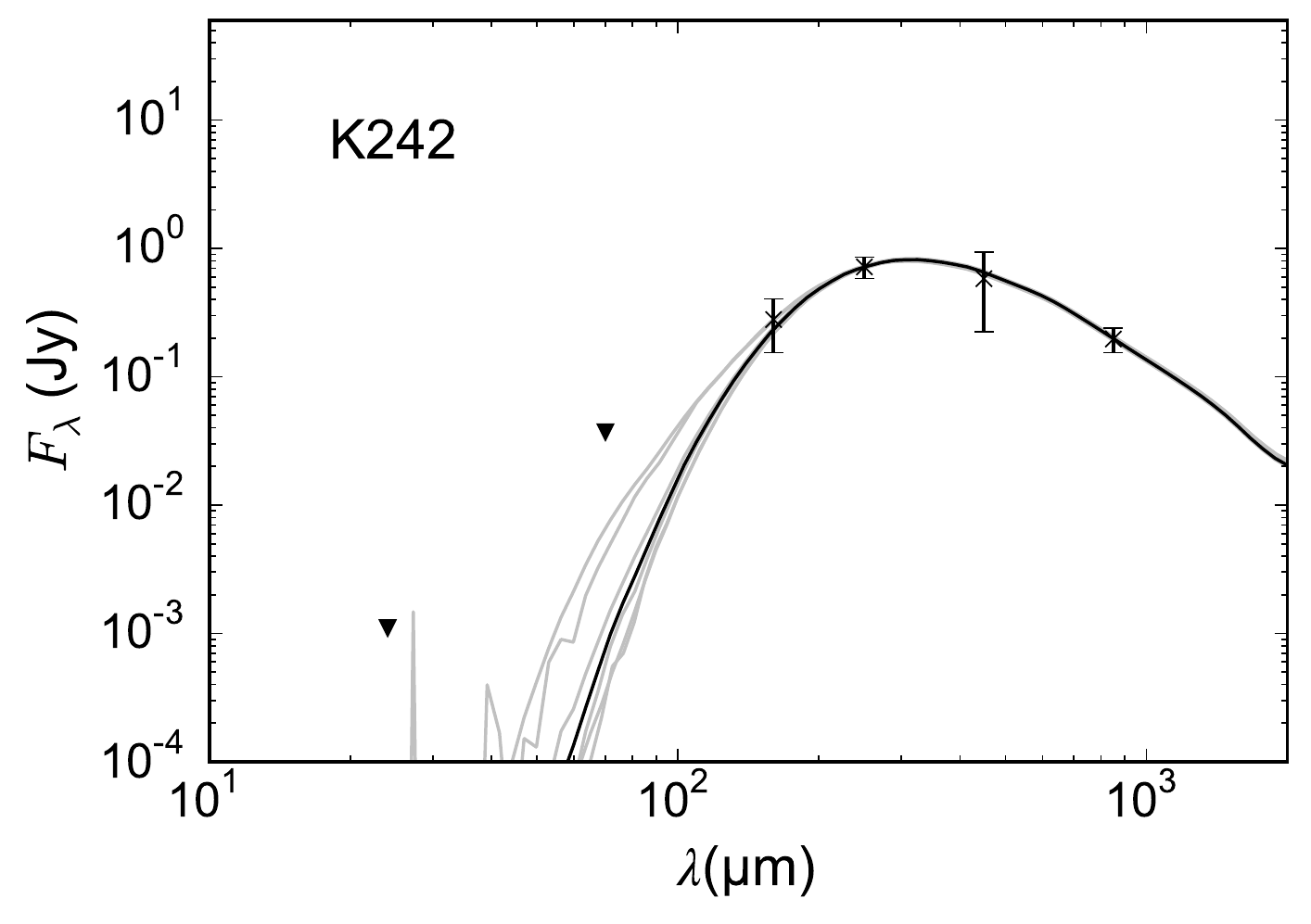}

	\caption{FHSC candidates from observations of the Serpens South region (see Section \ref{sec:FHSCobs} and Appendix \ref{sec:SSobs}); arrows denote upper limits. The best fitting model SEDs are plotted in black and the other model SEDs with $\chi^2$ values within a factor of two of the best fitting model are plotted in grey. The  best fitting model SEDs are: (Aqu-MM2) RHD2, $\beta=0.05$, silicate grains, ${a_\mathrm{max}}=$ \SI{10}{\micro\metre}, $i=$\ang{0}, early FHSC. (SerpS-MM22) RHD2, $\beta=0.09$, carbon grains, ${a_\mathrm{max}}=$ \SI{10}{\micro\metre}, $i=$\ang{0} FHSC. (SerpS-MM19) RHD2, $\beta=0.0$, silicate grains, ${a_\mathrm{max}}=$ \SI{10}{\micro\metre}, $i=$\ang{0} and late FHSC. (Aqu-MM1) RHD2, $\beta=0.05$, carbon grains, ${a_\mathrm{max}}=$ \SI{1}{\micro\metre}, $i=$\ang{90}, early FHSC. (K242) RHD1, $\beta=0.01$, carbon grains, ${a_\mathrm{max}}=$~\SI{1}{\micro\metre}, $i=$\ang{90}, early FHSC.    }
	\label{fig:serpensseds}
\end{figure*}

\subsubsection{Aqu-MM2}
For Aqu-MM2, the minimum $\chi^2$ was \num{4.29} and 28 models out of 1440 fitted well. Slightly more of these model SEDs were produced from the RHD2 model than the RHD1 model. Nearly all of the best fitting models are from faster rotating cores. None of the best models are from the first collapse stage, indicating that the source is more evolved than a dense `starless' core, however there is an equal proportion of models from FHSC phase and the second collapse so we cannot constrain the evolutionary stage any further. None of the best rotating models have a \ang{90} inclination, since the SED peak is at a comparatively short wavelength of between \SIrange{150}{200}{\micro\metre} even though the source remains undetected at \SI{24}{\micro\metre}. The majority of best models used a maximum grain size of \SI{10}{\micro\metre} and the rest used \SI{1}{\micro\metre}, all of the silicate grain type. The optimum scale factor of the best fitting model was \num{4.1} from which we estimate the mass contribution to be \SI{5}{\solarmass} at \SI{260}{\parsec} or \SI{15}{\solarmass} at \SI{430}{\parsec}. The \SI{850}{\micro\metre} dust opacity was \SI{3.9e-3}{\centi\metre\squared\per\gram}. The best fitting model SEDs have been scaled beyond a factor of two which means the models are likely to be brighter at \SI{70}{\micro\metre} than a similar source of a higher mass. Given that the scaled low-mass model SEDs were not quite able to fit the \SI{70}{\micro\metre} point along with the other observations, we are unable to place strong constraints on this source. However, it seems likely that Aqu-MM2 is at least as evolved as an FHSC. It is also likely to have a low inclination, therefore a resolved observation would be likely to show rotational structures.

An outflow has been observed at this source \citep{dunham2014} with a maximum velocity of \SI{9}{\kilo\metre\per\second}, which is faster than expected from an FHSC. It looks likely that Aqu-MM2 is more evolved than an FHSC and, if so, this demonstrates an overlap in the SED properties of some FHSCs and protostars.

\subsubsection{SerpS-MM22}
SerpS-MM22 is detected at all of the observed wavelengths and is brighter at \SI{24}{\micro\metre} and \SI{70}{\micro\metre} than the other sources even though the peak still appears to lie between \SI{200}{\micro\metre} and \SI{300}{\micro\metre}, which suggests that it is unlikely to be an FHSC. The smallest $\chi^2$ value was \num{4.02} and 124 models fell within a factor of two of this. None of the models could fit the \SI{24}{\micro\meter} flux and it should be noted that the model with the minimum $\chi^2$ may in fact provide a poorer fit than models that are brighter at \SI{24}{\micro\meter} but fit the remaining points less well.

Since none of the models could replicate the broad shape of the SED of SerpS-MM22 and none came close to fitting the \SI{24}{\micro\metre} flux it would not be meaningful to draw conclusions regarding the properties of the source. As expected, there was little consistency among the `best fitting' models, except that most were from RHD2 and all but one used grains of ${a_\mathrm{max}}<$~\SI{100}{\micro\metre}. Given the variation of properties among the best fitting models and the poor quality of the fits, we are not able to constrain the nature of this source, but it is likely to be at least as evolved as an FHSC, and may have a stellar core.

\subsubsection{SerpS-MM19}
SerpS-MM19 is undetected at \SI{24}{\micro\metre} and \SI{70}{\micro\metre} but the peak is between \SI{150}{\micro\metre} and \SI{200}{\micro\metre}. This source is fitted reasonably well by our models with a minimum $\chi^2$ of \num{0.19} and 11 model SED are defined as providing good fits. The optimum scale factor of the best fitting model is \num{1.2}, which gives a mass of \SI{1.3}{\solarmass} at \SI{260}{\parsec} or \SI{4.2}{\solarmass} at \SI{430}{\parsec}. The \SI{850}{\micro\metre} dust opacity of this model was \SI{4.0e-3}{\centi\metre\squared\per\gram}. All of the best fitting models are from the RHD2 model and used silicate dust grains with ${a_\mathrm{max}}=$~\SI{10}{\micro\metre}. The best fitting models favour a more slowly rotating core, with all but one having $\beta=0$ or $\beta=0.01$. All the selected models are also from late FHSC or second collapse, except for the one fast rotating model, which is early FHSC. The models from rotating cores are nearly all at moderate or high inclination. These results allow us to place constraints on the nature of the source. Although there was no detectable flux even at \SI{70}{\micro\metre}, the peak is at a sufficiently short wavelength to suggest this source is an evolved FHSC or in the second collapse and stellar core formation phase. It looks likely that the object is rotating slowly and inclined, such that much of the short wavelength radiation from the hot core is absorbed by the oblate distribution of material surrounding the core.

\subsubsection{Aqu-MM1}
For Aqu-MM1 we obtained nine good fits with the minimum $\chi^2$ value of \num{0.07}. All but one were from the RHD2 model and all used amorphous carbon dust grains with ${a_\mathrm{max}}=$~\SI{1}{\micro\metre}. Most models were rotating quickly and at high inclination. As for SerpS-MM19, the combination of a quickly rotating object at high inclination allows the peak to be at $\lambda\sim$~\SI{200}{\micro\metre} without the source being bright at \SI{70}{\micro\metre}. The optimum scale factor was $\sim$~\num{0.6}, which gives a mass of \SI{0.4}{\solarmass} at \SI{260}{\parsec} or \SI{1.9}{\solarmass} at \SI{430}{\parsec}. The \SI{850}{\micro\metre} dust opacity was \SI{1.3e-2}{\centi\metre\squared\per\gram}.The best models were mostly from the FHSC phase. This source is therefore likely to be an FHSC and to show strong rotational signatures, being most likely edge-on.

\subsubsection{K242}
For K242, the smallest $\chi^2$ value was 0.01 and seven models fitted well, all of which were from the RHD1 model and used amorphous carbon grains of ${a_\mathrm{max}}=$~\SI{1}{\micro\metre}. Only one of the best fitting models was not rotating, but the remaining models are shared equally between the three sampled rotation rates. Of the rotating models, all but one of these are inclined at \ang{90}. The optimum scale factor was 0.35 which gives a mass of $<$\SI{0.3}{\solarmass} at \SI{260}{\parsec} or \SI{0.9}{\solarmass} at \SI{430}{\parsec}. The \SI{850}{\micro\metre} dust opacity was \SI{1.3}{\centi\metre\squared\per\gram}. Crucially, all of the best fitting models are from the FHSC phase, skewed towards early in the phase. This source is then likely to be an FHSC in a low mass core and may show some rotation.

\section{Discussion}
We have performed 3D RHD simulations of the collapse of pre-stellar cores with a range of properties and then modelled the SEDs of these FHSCs and their envelopes using a radiative transfer code. The properties of the FHSCs themselves are very similar, the only real variation being in the increase in radius and oblateness with rotation. These simulations have shown that differences in the core properties lead to variations in the SED and that these variations are due to differences in the temperature structure and morphology of the collapsing envelope rather than the FHSC itself.

Many properties have similar effects on the SED, for example a higher mass, lower ISRF exposure, slower rotation and higher inclination all lead to a steeper drop in flux in the far-infrared. Differences in the initial rotation rate and core radius (with a constant mass and Bonnor-Ebert density profile) gave rise to the greatest SED variation. Many factors have a significant effect on the SED but only models with faster rotation can produce detectable flux at \SI{24}{\micro\metre}. This may be a property that can be constrained with the SED and can also be verified through molecular line observations. We find a shift of the SED peak to shorter wavelengths and an increase in far infrared flux as the FHSC evolves, which is consistent with the findings of others, including \mbox{\citet{omukai2007}} and \mbox{\cite{tomida2010dec}}. We also find, however, this effect is considerably reduced by a higher initial temperature or a larger initial radius, to the extent that for $T_{\mathrm{init}}=$~\SI{15}{\kelvin} or  $r_{\mathrm{init}}=$~\SI{1.5e17}{\centi\metre} (\SI{10000}{au}) there is no SED evolution.

In Section~\ref{sec:resultsgrainsize} we described how changing the maximum dust grain size shifted the SED peak by up to \SI{200}{\micro\metre} and can cause the object to appear nearly an order of magnitude brighter at submillimeter wavelengths. The choice of dust grain model can therefore have a significant impact on SED modelling. We chose to include models with very large grains in our selection for fitting to observations because it appears to be the only way to reproduce both the \SI{70}{\micro\metre} flux and a peak at $\lambda >$~\SI{200}{\micro\metre} that we see in several observations of FHSC candidates.
There is evidence for dust grains in dense molecular cores with $a_{\mathrm{max}} > $~\SI{1.5}{\micro\metre} from coreshine observations \citep{steinacker2015}, of micron sized grains in the ISM from detectors on solar system spacecraft (e.g. \citealt{gruen1994,sterken2014}) and of $\sim$~\SI{40}{\micro\metre} sized Earth-impacting meteoroids (e.g. \citealt{baggaley2000}). \citet{ormel2009} suggests that if molecular cloud lifetimes are lengthened by long-term support mechanisms, dust aggregates of $\sim$~\SI{100}{\micro\metre} may be produced. From measurements of the spectral index of the millimetre dust opacity, \citet{ricci2010} also suggest that millimetre-sized grains may already exist in class 0 YSOs so it is perhaps not unreasonable to expect grain sizes of over \SI{10}{\micro\metre} in pre-stellar cores although we note that there are also protostellar cores for which there is no evidence of grain growth (e.g. \citealt{IHsiuLi2017}). The sources that were fitted best by our model SEDs, except Per-Bolo 58, were fitted well exclusively by models with $a_{\mathrm{max}} = $~\SI{1}{\micro\metre} or $a_{\mathrm{max}} = $~\SI{10}{\micro\metre}, possibly indicating that the remainder would be better fitted by models with a different combination of other parameters rather than with very large grains.
 The alternative is that sources best fit by models with large grains are more evolved but embedded objects. Indeed, the SEDs of Per-Bolo 58 and Cha-MMS1 appear similar to the SEDs of more evolved objects, most likely accreting stellar cores, as shown by \citet{young2005} and \citet{dunham2010}.

We set out to ascertain whether an SED is useful for constraining the nature of these sources by fitting the model SEDs to observations. We note that the set of models covers a limited range of parameters: for example, we show in \ref{sec:scaling} that these models are less likely to be valid for sources outside a mass range of $0.5\lesssim M \lesssim 2 ~\mathrm{M}_{\sun}$ and we only use an initial radius of \SI{7e16}{\centi\metre}. Despite similarities in the SEDs of some models, such as between inclined, rotating FHSCs and young cores in first collapse phase, we find it is possible to constrain the nature of the sources in some cases. The model SEDs are distinct enough that each source is fitted well by a limited number of models, which have mostly consistent properties. Some sources are better constrained than others. None of our model SEDs fitted Aqu-MM2 and SerpS-MM22 well and so these are equally well fitted by large numbers of models. These sources are more luminous than the other Serpens South candidates and, even at the close distance, the models required scaling beyond the range of masses for which the SEDs are valid. It is possible that these sources could be fitted by using a combination of parameters we have not considered here, such as a different radius, greater mass and a fast rotation rate. It is more likely, however, that Aqu-MM2 and SerpS-MM22 are more evolved young protostars, especially given the bolometric luminosities of 0.95~$L_{\odot}$ and 0.2~$L_{\odot}$ \citep{Maury:2011aa}.

The results of fitting models for SerpS-MM19, Aqu-MM1 and K242 provided some insight into their nature because a small number of models fitted well. Interestingly, where there was a strong preference for models from the early FHSC phase, no models were selected for the earlier first collapse stage, suggesting that these two stages are observationally distinguishable. After first collapse the \SI{160}{\micro\metre} flux rises quickly relative to the longer wavelengths and the peak shifts. The cold, dim first collapse SEDs could not be reproduced by a more evolved object, even at high inclination, meaning that it does seem possible to differentiate between empty `starless' cores and cores containing an FHSC. Similarly, \mbox{\citet{commercon2012a}} suggest that a starless cores and an FHSC are distinguishable at far-infrared wavelengths. CB17-MMS was fitted by both first collapse SEDs and SEDs from more evolved cores at high inclination, so there is some level of degeneracy between those parameters. However, we can see from Fig.~\ref{fig:CB17MMS} that none of the models is able to fit the \SI{160}{\micro\metre} flux and the location of the SED peak is not well constrained. This is likely to explain the spread of best fitting models, since the location of the peak is different for very young prestellar cores and more evolved cores at high inclination.

On the other hand, it is more difficult to distinguish between an FHSC and more evolved object observationally. Where the SED fitting shows a strong preference for the late FHSC stage, for example SerpS-MM19, second collapse SEDs also provide good fits. Differences in the SED are caused by changes in the temperature structure and opacity of the envelope and infalling  material. These changes are more significant between first collapse and FHSC formation as the object transitions from the isothermal to adiabatic phase. Second collapse takes place far quicker than the first and, although the temperature rises quickly, it does so primarily within the radius of the FHSC. The SED only allows observers to probe down to the region surrounding the FHSC. This region is slow to heat and so there are unlikely to be significant differences between the SEDs of the FHSC, second collapse and early stellar core phases.

\citet{omukai2007} suggests that the optical depth is reduced enough in the vertical direction of a rotating core that FHSC radiation may be visible at low inclinations. However, like \citet{saigo2011}, we find that the FHSC heats the region immediately surrounding it and that it is from this region that most of the observed radiation is emitted. Even for a core with a higher rotation rate viewed face-on the FHSC is unlikely to be directly visible at wavelengths $<$~\SI{850}{\micro\metre}.

 Currently, a detection at \SI{70}{\micro\metre} is used to distinguish protostellar (class 0) sources from starless cores after radiative transfer modelling showed that the \SI{70}{\micro\metre} flux is proportional to the internal luminosity \citep{dunham2008} and from previous attempts to model FHSC SEDs. With a realistic density structure, we find that the FHSC is obscured which significantly reduces the \SI{70}{\micro\metre} flux (see also \citealt{saigo2011,commercon2012a}). The \SI{70}{\micro\metre} flux is highly dependent upon the structure of the centre of the core and the extent of the envelope. With a photometric sensitivity of a few \si{\milli\jansky} at \SI{70}{\micro\metre}, \textit{Herschel} may be able to detect some FHSCs, depending on the geometry of the FHSC and its envelope but others will be too faint.

RHD2 includes a more physical treatment of radiation than RHD1 and we do see a strong preference for SEDs produced from the RHD2 model in the SED fitting. Only three of the FHSC candidates here, B1-bN, CB17-MMS and K242, were fitted best by RHD1 SEDs and these are the only sources to have SED peaks at $\lambda>$~\SI{250}{\micro\metre}. The SEDs produced from RHD2 models all peaked at shorter wavelengths than this, except when large dust grains were used, due to heating from the ISRF. It is interesting to note that K242 appears to lie outside the main filament (see Fig.~\ref{fig:s2_maps}), which means the ISRF and extinction could be different for this source. When the incident ISRF was reduced, the SED peaked at a longer wavelength. This suggests that the preference for RHD1 SEDs could simply be due to local differences in the ISRF and we have not included model SEDs of cores with different levels of exposure to the ISRF in the fitting procedure. We attempted to fit these three sources with only RHD2 SEDs. We could not obtain a good fit for CB17-MMS because the RHD2 SEDs are too `warm': a lower envelope temperature is required to reproduce the position of the SED peak and submillimetre brightness. B1-bN was only fitted well by RHD2 SEDs with a maximum grain size of \SI{200}{\micro\metre}, which is probably unrealistic. For K242 we obtained reasonable fits with RHD2 SEDs. The properties derived from this fitting are very similar to those from the RHD1 fits except that the RHD2 fits require the source to be younger.

Finally, the selection of the best fitting SEDs is sensitive to small differences in the observed fluxes. Consequently, any errors in the observation could lead to a different interpretation of the properties of the core, although these are not expected to affect the possibility of distinguishing starless cores and FHSCs.

\section{Conclusions}
We have produced RHD simulations of the collapse of pre-stellar cores with different initial properties. Snapshots from these models at the key stages of FHSC formation and evolution were then used to simulate SEDs using a 3D radiative transfer code. We compared the SEDs of cores at different stages in their evolution and of cores with different properties. Secondly, we fitted the observed SEDs of several FHSC candidates with model SEDs from a set of 1440.

Differences in the temperature structure of the core affect the shape of the SED even though the FHSC itself is not observable directly at wavelengths $<$~\SI{850}{\micro\meter}. Most notably, fast rotation leads to significant flux at \SI{24}{\micro\metre} during FHSC phase and so FHSC candidates with observed flux at this wavelength should not necessarily be ruled out. If the core is initially warmer (e.g. \SI{15}{\kelvin}) or initially less compact (e.g. $r_{\mathrm{ini}}=$~\SI{1.5e17}{\centi\metre}) we find no evolution of the SED as the core evolves, which means a younger core could appear more evolved.

Some of the FHSC candidates were fitted well by just a small number of model SEDs with consistent properties, which allowed us to make suggestions as to their nature. We found that the SED can be used to distinguish between cores undergoing first collapse and cores containing an FHSC but not between the FHSC and second collapse stages. Although FHSC SEDs may appear featureless, they can nonetheless be useful in characterising sources.

Of the FHSC candidates that we have fitted with model SEDs, B1-bN, CB17-MMS, Aqu-MM1 and K242 are most likely to be FHSCs. Aqu-MM2 is likely to be rotating and at a low inclination. SerpS-MM19 may be more evolved than an FHSC but is probably rotating moderately quickly and oriented at a high inclination. Aqu-MM1 is also likely to have a high rotation rate and to have an edge-on inclination and CB17-MMS could be an FHSC embedded in an edge-on disc. From our results, we consider Chamaeleon-MMS1 and Per-Bolo 58 to be more evolved and they have probably undergone stellar core formation.

\section*{Acknowledgements}
The groundwork for this paper began as an MPhys project completed by Alison Young and Jemma Holloway at the University of Exeter. We are grateful to David Acreman for assistance with running TORUS. We would also like to thank Nathan Mayne and Tim Naylor for helpful discussions regarding SED fitting and photometry. Finally, we thank the anonymous referee and Michael Dunham for comments that helped to improve the paper. Some figures were produced using the publically available SPLASH visualization software \citep{price2007}.

This work was supported by the European Research Council under the European Commission's Seventh Framework Programme (FP7/2007-2013 Grant Agreement No. 339248).  The calculations discussed in this paper were performed on the University of Exeter Supercomputer, a DiRAC Facility jointly funded by STFC, the Large Facilities Capital Fund of BIS, and the University of Exeter. This work used the DiRAC Complexity system, operated by the University of Leicester IT Services, which forms part of the STFC DiRAC HPC Facility (www.dirac.ac.uk). This equipment is funded by BIS National E-Infrastructure capital grant ST/K000373/1 and STFC DiRAC Operations grant ST/K0003259/1. DiRAC is part of the National E-Infrastructure. 

The JCMT has historically been operated by the Joint Astronomy Centre on behalf of the Science and Technology Facilities Council of the United Kingdom, the National Research Council of Canada and the Netherlands Organisation for Scientific Research. Additional funds for the construction of SCUBA-2 were provided by the Canada Foundation for Innovation. Chris Mowat is supported by an STFC studentship. 

This research has made use of NASA's Astrophysics Data System. This research used the services of the Canadian Advanced Network for Astronomy Research (CANFAR) which is supported by CANARIE, Compute Canada, University of Victoria, the National Research Council of Canada, and the Canadian Space Agency. This research used the facilities of the Canadian Astronomy Data Centre operated by the National Research Council of Canada with the support of the Canadian Space Agency.

Starlink software \citep{Currie:2014aa} is supported by the East Asian Observatory. Matplotlib is a 2D graphics package used for Python for application development, interactive scripting, and publication-quality image generation across user interfaces and operating systems. This research made use of APLpy \citep{Robitaille:2012aa}, an open-source plotting package for Python hosted at http://aplpy.github.com.

Herschel is an ESA space observatory with science instruments provided by European-led Principal Investigator consortia and with important participation from NASA. This research has made use of data from the Herschel Gould Belt survey (HGBS) project (http://gouldbelt-herschel.cea.fr). The HGBS is a Herschel Key Programme jointly carried out by SPIRE Specialist Astronomy Group 3 (SAG 3), scientists of several institutes in the PACS Consortium (CEA Saclay, INAF-IFSI Rome and INAF-Arcetri, KU Leuven, MPIA Heidelberg), and scientists of the Herschel Science Center (HSC).

The authors wish to recognize and acknowledge the very significant cultural role and reverence that the summit of Maunakea has always had within the indigenous Hawaiian community. We are most fortunate to have the opportunity to conduct observations from this mountain.




\bibliographystyle{mnras}
\bibliography{paper_refs} 



\appendix

\section{First Hydrostatic Core Candidates in Serpens South}
\label{sec:SSobs}

First hydrostatic core candidates in Serpens South were identified on the basis of their submillimeter, far-infrared, and mid-infrared fluxes. These fluxes were extracted from observations by SCUBA-2 \citep{Holland:2013aa}, \textit{Herschel} \citep{Pilbratt:2010aa}, \textit{Spitzer} MIPS \citep{Rieke:2004aa}, and WISE \citep{Wright:2010aa}.

There is some uncertainty as to the distance of Serpens South, ranging from 225~$\pm$~55~pc derived by \citet{Straizys:2003aa} to the commonly adopted value of \SI{260}{\parsec} \citep{Gutermuth:2008aa,Bontemps:2010ab,Maury:2011aa,Konyves:2015aa} and the more recent suggestion that it is physically associated with Serpens Main at 415~$\pm$~5~pc \citep{Dzib:2010aa}. The most up-to-date measurement is now 436.0~$\pm$~9.2~pc \citep{ortizleon2017}. While fitting the observations, we scale the model SEDs and discuss the results considering both distances of \SI{260}{\parsec} and \SI{430}{\parsec}.

\subsection{Observations}

\begin{figure}
	\centering
		\includegraphics[width=1.0\columnwidth,trim = 8cm 3cm 9cm 1.5cm, clip]{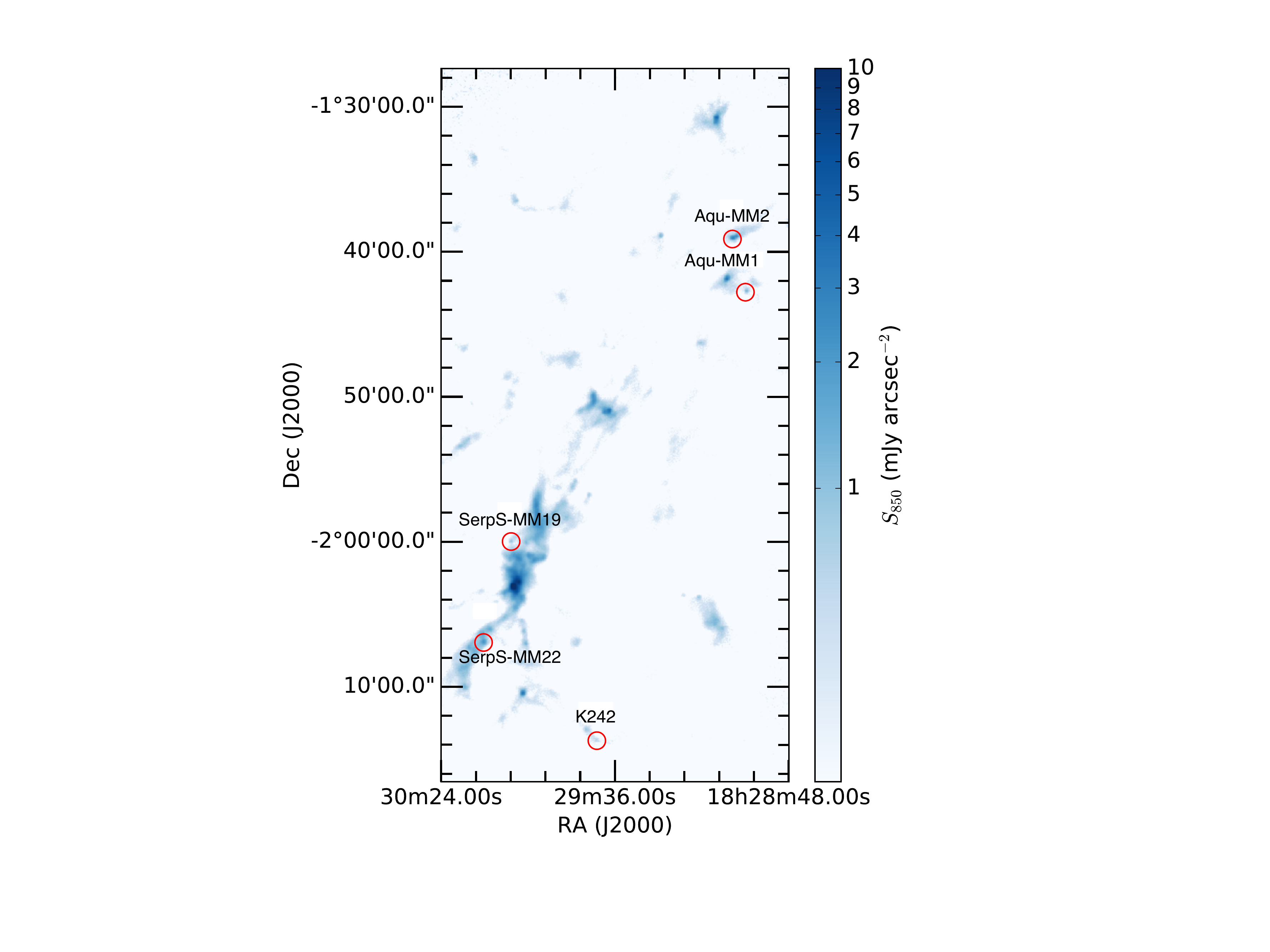}
	\caption{SCUBA-2 maps of the Serpens South region at 850~\si{\micro\metre}. FHSC candidates are labelled, and identified using red circles.}
	\label{fig:s2_maps}
\end{figure}

The Aquila region, containing Serpens South, was observed with SCUBA-2 \citep{Holland:2013aa} as part of the James Clerk Maxwell Telescope Gould Belt Survey (JCMT GBS, \citealt{Ward-Thompson:2007aa}). The full map was first presented in \citet{Rumble:2016aa}, though the map presented in this work was produced using an improved data reduction procedure. Figure~\ref{fig:s2_maps} shows the Serpens South region of the SCUBA-2 Aquila map at 850~\si{\micro\metre}. The FHSC candidates in Serpens South are labelled. 

The Multiband Imaging Photometer for \textit{Spitzer} (MIPS) observed Serpens South as part of the \textit{Spitzer} Gould Belt Legacy Survey, described in \citet{Dunham:2015aa}. The MIPS 1 maps containing Serpens South were used to obtain photometry for FHSC candidates at 24~\si{\micro\metre}.

The \textit{Herschel} Space Observatory was also used to observe Serpens South as part of the \textit{Herschel} Gould Belt Survey \citep{Andre:2010aa}. The observations were performed with both PACS \citep{Poglitsch:2010aa} and SPIRE \citep{Griffin:2010aa}. Photometry for FHSC candidates in the Serpens South region was obtained at 70 and 160~\si{\micro\metre} (observed with PACS), and 250~\si{\micro\metre} (observed with SPIRE).

The Wide-field Infrared Survey Explorer (WISE) mapped the whole sky in four wavelength bands. Band W4 has an effective wavelength of 22~\si{\micro\metre}, and so the W4 map in the vicinity of Serpens South was used, in conjunction with the \textit{Spitzer} data, in the identification of FHSC candidates (detailed below, in Section~\ref{sec:identification})

\subsection{Identification}\label{sec:identification}
FHSC candidates were identified using a five-step procedure. Initially, the FellWalker clump-finding algorithm \citep{Berry:2015aa} was used to identify peaks in the SCUBA-2 450~\si{\micro\metre} emission. Next, centrally-concentrated cores were selected as collapse candidates, with concentration values of $\> 0.3$ accepted, following \citet{Enoch:2007aa}. Infrared-bright cores were then rejected by cross-matching with sources in the WISE \citep{Cutri:2012aa}, and \textit{Spitzer} YSO \citep{Dunham:2015aa}, catalogues, at 22 and 24~\si{\micro\metre} respectively. Additionally, potential FHSCs from 450~\si{\micro\metre} were overlaid on the WISE 22~\si{\micro\metre} and \textit{Spitzer} 24~\si{\micro\metre} maps. Then, \textit{Herschel} non-detections at 160~\si{\micro\metre} (based on a preliminary flux extraction) were rejected, as it would be expected for the SED of a cold source such as an FHSC to peak near this wavelength. Finally, FHSC candidate SEDs were compared with FHSC models, and those with the highest 24:450~\si{\micro\metre} flux ratios were rejected. This procedure found a total of five compact submillimetre infrared-faint sources considered to be FHSC candidates in Serpens South.

\begin{figure*}
	\centering
		\includegraphics[width=0.9\textwidth, trim= 0.5cm 4.5cm 0 3cm, clip ]{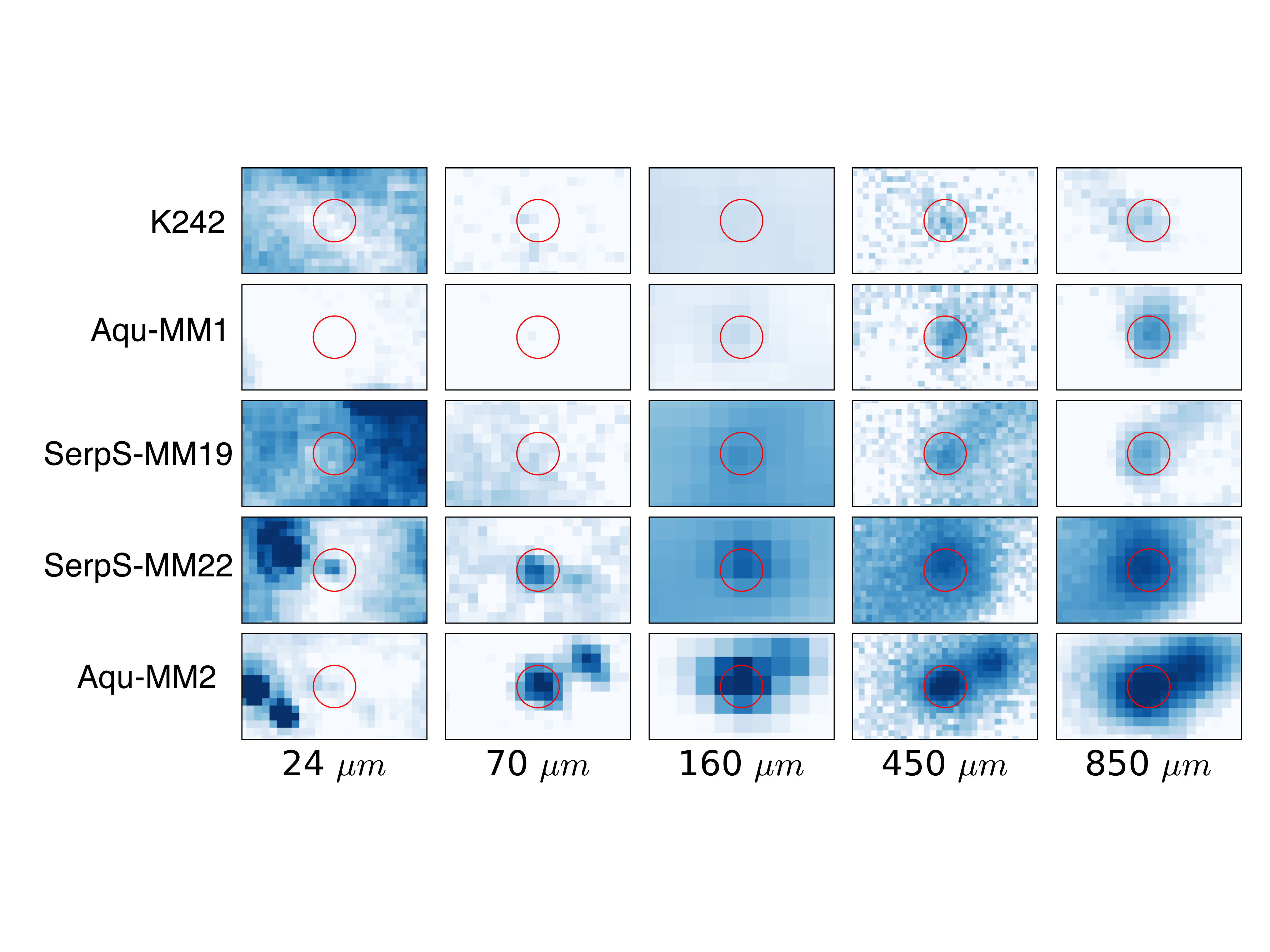}
	\caption{Candidate FHSCs in Serpens South, with their positions shown at five wavelengths. Red circles denote the positions of the candidates in each image, and their size represents the size of the aperture used in the photometry (3.5$''$, 15.0$''$, 15.0$''$, 10.0$''$, and 15.0$''$, respectively). We note that we used the photometry from \citet{Konyves:2015aa} for the \SI{70}{\micro\metre} detection of Aqu-MM1 instead. The x-axis direction is East to West and the y-axis direction is South to North. These are centred on the peak flux position when the candidate is detected, and the 450~\si{\micro\metre} peak position when it is not. The colour scales are logarithmic and run from 4.57 to 4.87~mJy/pixel for 24~\si{\micro\metre} except for SerpS-MM22 and SerpS-MM19 which range from 4.72 to 5.02~mJ/pixel because they are in a different map; from 10.59 to 105.9 for 70~\si{\micro\metre}; from 21.18 to 211.8~mJy/pixel for 160\si{\micro\metre}; from 4.0 to 80.0~mJy/pixel for 450~\si{\micro\metre}; and from 2.7 to 22.5~mJy/pixel for 850~\si{\micro\metre}.}
	\label{fig:fhsc_images}
\end{figure*}

Images of these five FHSC candidates are shown in Figure~\ref{fig:fhsc_images}. Five images per FHSC candidate are shown, at 24~\si{\micro\metre} (MIPS 1), 70 and 160~\si{\micro\metre} (\textit{Herschel} PACS), and 450 and 850~\si{\micro\metre} (SCUBA-2). Pixel sizes are 2.54, 3.0, 3.0, 2.0, and 3.0 arcseconds, at 24, 70, 160, 450, and 850~\si{\micro\metre}, respectively.
\begin{table*}
\centering
\caption{Fluxes and upper limits for the five candidate FHSCs in Serpens South. Fluxes are given at 24~\si{\micro\metre} (\textit{Spitzer} MIPS 1), 70 and 160~\si{\micro\metre} (\textit{Herschel} PACS), and 450 and 850~\si{\micro\metre} (SCUBA-2). The positions given are the coordinates of the peak of flux at 450~\si{\micro\metre}. Upper limits used are $2\sigma$ limits. The apertures used are \SI{3.5}{\arcsec}, \SI{15.0}{\arcsec}, \SI{15.0}{\arcsec}, \SI{10.0}{\arcsec} and \SI{15.0}{\arcsec} for the \SI{24}{\micro\metre}, \SI{70}{\micro\metre}, \SI{160}{\micro\metre}, \SI{450}{\micro\metre} and \SI{850}{\micro\metre} fluxes respectively. $^{\mathrm{a}}$From \citet{Konyves:2015aa}.}
\label{tab:fhsc_fluxes}
\begin{tabular}{ l c c c c c c c}
	\hline \\ [-1.5ex]
	 ID & Position & $S_{24}$ & $S_{70}$ & $S_{160}$ & $S_{250}$ & $S_{450}$ & $S_{850}$ \\
	 & (J2000) & (mJy) & (mJy) & (mJy) & (mJy) & (mJy) & (mJy) \\[+0.8ex]
	\hline \\ [-1.5ex]
	Aqu-MM2 & 18:29:03.6 -01:39:07 & $\le$ 1.0 & 1410 $\pm$ 220 & 15700 $\pm$ 2400 & 15400 $\pm$ 800 & 3995 $\pm$ 353 & 1076 $\pm$ 42 \\
	SerpS-MM22 & 18:30:12.3 -02:06:57 & 2.3 $\pm$ 0.9 & 403 $\pm$ 83 & 4940 $\pm$ 790 & 6940 $\pm$ 450 & 1923 $\pm$ 306 & 639 $\pm$ 39 \\
	SerpS-MM19 & 18:30:04.7 -01:59:59 & $\le$ 1.1 & $\le$ 104 & 2530 $\pm$ 440 & 2450 $\pm$ 470 & 810 $\pm$ 271 & 216 $\pm$ 33 \\
	Aqu-MM1 & 18:29:00.0 -01:42:47 & $\le$ 0.9 & 31 $\pm$ 28$^{\mathrm{a}}$ & 764 $\pm$ 155 & 1370 $\pm$ 460 & 916 $\pm$ 358 & 304 $\pm$ 43 \\
	K242 & 18:29:41.0 -02:13:43 & $\le$ 1.1 & $\le$ 37 & 279 $\pm$ 125 & 718 $\pm$ 135 & 581 $\pm$ 358 & 197 $\pm$ 43 \\[+0.8ex]
	\hline \\ 
\end{tabular}
\end{table*}

\subsection{Fluxes}
\subsubsection{SCUBA-2 Fluxes}
Flux extraction from the SCUBA-2 450 and 850~\si{\micro\metre} maps of Aquila used 10 and 15$''$ aperture radii, respectively. In this instance, aperture corrections of 1/0.72 and 1/0.85 were used at the respective wavelengths \citep{Dempsey:2013aa}. Uncertainties were extracted from the error maps at the positions of each candidate, and then combined with SCUBA-2 calibration uncertainties of five and ten per cent at 850 and 450~\si{\micro\metre}, respectively. The use of a 10$''$ aperture radius for 450~\si{\micro\metre} photometry differs from the 15$''$ radius used to extract 70, 160, and 850~\si{\micro\metre} fluxes. However, this smaller aperture size means overall lower uncertainty on the 450~\si{\micro\metre} fluxes, and less background contamination (visible in Figure~\ref{fig:fhsc_images}, particularly for Aqu-MM2 and SerpS-MM22). A 10$''$ aperture radius was also tested on the 70~\si{\micro\metre} map, but it was found to produce less consistent fluxes than a 15$''$ aperture radius, so an aperture radius of 15$''$ was maintained. The SCUBA-2 fluxes are given, along with the MIPS and \textit{Herschel} fluxes, in Table~\ref{tab:fhsc_fluxes}.

\subsubsection{24~\si{\micro\metre} fluxes}
Fluxes for each of the five FHSC candidates were extracted from the MIPS 1 maps of Aquila using aperture photometry. Each aperture had a radius $r$ =  3.5$''$, with background annuli between 1.71 and 2.29 $r$. This aperture encircles half of the first dark ring of the beam. Uncertainties were extracted from the corresponding error maps, using apertures of the same size, but without annuli. Fluxes from these apertures were corrected using a factor of 2.78, and a four per cent calibration uncertainty was added \citep{Engelbracht:2007aa}. These apertures were tested on known sources in the \citet{Dunham:2015aa} catalogue, and all fluxes were within ten per cent of the literature values. These fluxes are given in Table~\ref{tab:fhsc_fluxes}.

\subsubsection{\textit{Herschel} Fluxes}\label{sec:herschel_fluxes}
Aperture photometry was also used to find fluxes for each candidate from the 70 and 160~\si{\micro\metre} maps. Apertures had 15$''$ radii, with annuli between 1.5 and 2.0 $r$, giving aperture corrections of 1/0.829 (70~\si{\micro\metre}) and 1/0.729 (160~\si{\micro\metre}, \citealt{Balog:2014aa}). Photometric uncertainties were then combined with a conservative calibration uncertainty of 15 per cent, following \citet{Kelly:2012aa} and \citet{Konyves:2015aa}. The resulting fluxes are provided in Table~\ref{tab:fhsc_fluxes}.

Fluxes for each of the candidates were also obtained at 250~\si{\micro\metre}. Aperture photometry is not recommended for use with SPIRE maps for point sources, so 250~\si{\micro\metre} fluxes were found by cross-matching FHSC candidate positions with the source catalogue of \citet{Konyves:2015aa}. These fluxes are also given in Table~\ref{tab:fhsc_fluxes}. We do not use the SPIRE fluxes for 350 or 500~\si{\micro\metre} because the angular resolution of those maps is significantly lower.


\bsp	
\label{lastpage}
\end{document}